\documentclass[12pt]{article}
\usepackage{geometry}                
\usepackage{graphicx}
\usepackage{amssymb} 
\usepackage{epstopdf}
\usepackage[dvips]{epsfig}
\usepackage{rotate}
\newcommand{\Rp}{\mbox{$\not \hspace{-0.15cm} R_p$}}
\def\GeV{\hbox{$\;\hbox{\rm GeV}$}}
\def\TeV{\hbox{$\;\hbox{\rm TeV}$}}

\newcommand{\fb}{\mbox{{\rm ~fb}}}

\newcommand{\bs}{\overline{s}}
\newcommand{\bu}{\overline{u}}

\newcommand{\bU}{\overline{U}}
\newcommand{\bD}{\overline{D}}

\newcommand{\xpom}{x_{\PO}}
\newcommand{\PO}{I\!\!P}
\newlength{\dinwidth}                                                          
\newlength{\dinmargin}                                                         
\begin{document}
\begin{titlepage}                                                    
\begin{flushright}
%
DESY 06-006 \\
Cockcroft-06-05 \\
JINST 1 P10001
\end{flushright}

\vspace*{1.cm} 

\begin{center}                                                                 

\begin{LARGE}  
{\bf Deep Inelastic Electron-Nucleon Scattering 
 at the LHC}
\\                                                             
\vspace*{1.5cm}   
\end{LARGE}                                                              
\begin{Large}
J.\,B.\,Dainton$^{1}$, M.\,Klein$^{2}$, P.\,Newman$^3$, E.\,Perez$^4$, F.\,Willeke$^2$                                                         
\end{Large}
\vspace{0.5cm} \\
$^1$ Cockcroft Institute of Accelerator Science and Technology, \\
Daresbury International Science Park, UK \\
$^2$ DESY,  Hamburg and Zeuthen, Germany \\
$^3$ School of Physics and Astronomy, University of Birmingham, UK \\
$^4$ CE Saclay, DSM/DAPNIA/Spp,  Gif-sur-Yvette, France
\vspace{0.5cm}
\\
\vspace*{1.2cm} {\bf Abstract  }
\begin{quotation}               
\noindent
The physics, and a design, of a Large Hadron Electron Collider (LHeC) are
sketched. With high luminosity, $10^{33}$cm$^{-2}$s$^{-1}$,
and high energy, $\sqrt{s}=1.4$\,TeV, such a collider can be built
in which a 70\,GeV electron (positron) beam in the LHC tunnel 
is in collision with one of the LHC hadron beams and which 
operates simultaneously with the LHC. The LHeC makes possible
deep-inelastic lepton-hadron ($ep$, $eD$ and $eA$) scattering  for
momentum transfers $Q^2$
beyond $10^6$\,GeV$^2$ and for Bjorken $x$ down to the $10^{-6}$.
New sensitivity to the existence of new states of matter, primarily in
the lepton-quark sector and in dense partonic systems,
is  achieved.  The precision possible with an electron-hadron experiment
brings in addition crucial accuracy in the determination of hadron structure,
as described in Quantum Chromodynamics,
and of parton dynamics at the TeV energy scale. The LHeC thus 
complements the proton-proton and ion programmes,
adds substantial new discovery potential to them, and is important
for a full understanding of physics in the LHC energy range. 
\end{quotation}
\vspace*{1.2cm} 
\end{center}                                                                
\cleardoublepage                     
\end{titlepage}
%

%
%
\newpage
\section{Introduction}
Deep-inelastic lepton-hadron scattering (DIS) has  long been \cite{feyn,oldies,nobelslac,nobel}
the most accurate means of exploring the substructure of matter at short distances.
Nowadays, such physics is concerned with electron-quark interactions at the
highest possible energy and momentum transfer. An $ep$ collider 
 allows new states coupling to leptons 
and quarks to be produced and their quantum numbers to be determined.
It is also concerned with the exploration of new forms of hadronic
matter, which may be manifest in very high parton densities at very 
low Bjorken-$x$.  With an $ep$ collider the parton dynamics and
the momentum distributions of quarks and gluons,
which are crucial for the discovery and interpretation of new physics,
are most accurately determined.
\begin{figure}[h]
   \centering
     \centerline{\includegraphics[width=.6\textwidth]{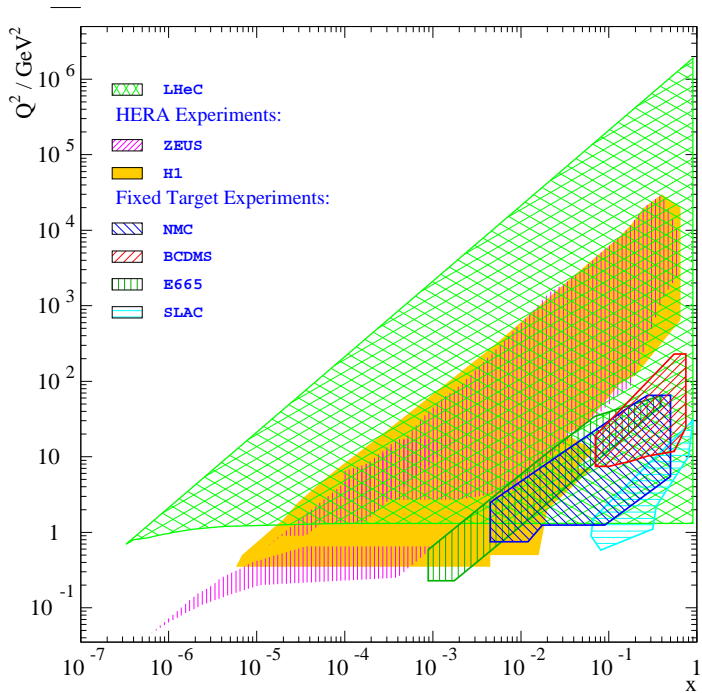}}
\caption{\it Kinematic regions in Bjorken-$x$ and momentum transfers $Q^2$
 covered by fixed target unpolarised  lepton-proton scattering experiments, 
 the H1 and ZEUS experiments at HERA and   
 the proposed electron-proton collider LHeC.}
   \label{fig:epkine}
\end{figure}

The Large Hadron Collider (LHC) will explore a new range of energy and mass. An
$ep$ facility is essential to complete this exploration, both through its
unique sensitivity to some possible new states of matter and since it
provides complementary information which will be needed to resolve puzzles
thrown up by $pp$ and $AA$ data.
The most attractive proposition for an  $ep$ collider operating in this energy domain
is to make use of the 7\,TeV LHC $p$ beam by colliding it with an intense 
 electron or positron beam stored in a ring mounted above the LHC,
a Large Hadron Electron Collider (LHeC).
Compared with  linac-ring solutions
recently considered, such as the ILC with HERA~\cite{THERA} or
a CLIC prototype with the LHC \cite{explorer}, this proposition
increases the luminosity by  two orders of magnitude.
An LHeC, in which for example 70\,GeV electrons collide with 7\,TeV protons,
will substantially extend the phase space explored hitherto 
in deep-inelastic lepton-hadron scattering (Fig.\,\ref{fig:epkine}).

\begin{figure}[h]
   \centering
    \centerline{\includegraphics[width=0.65\textwidth]{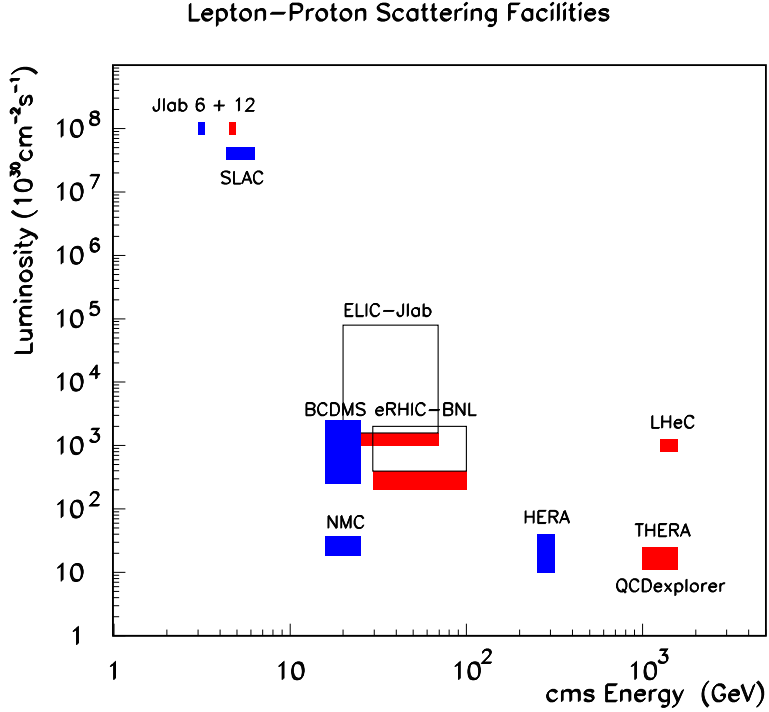}}
\caption{\it Summary of existing (dark, blue boxes) and proposed (grey, red boxes)
facilities for unpolarised lepton-proton deep-inelastic scattering investigations
in terms of the luminosity and centre-of mass energy.
The Jlab fixed target programme is directed to
high statistics physics at low $Q^2$ and very large~$x$. The SLAC box indicates
the luminosity of the classic $ep$ experiment at the 2\,mile linear accelerator.
BCDMS and NMC have provided the most accurate DIS muon-proton
structure function data using 30\,m and 3\,m long unpolarised hydrogen targets, respectively.
The  large luminosities envisaged at eRHIC and ELIC (hollow boxes),  which is desirable for
polarised $ep$ physics, are based on  energy-recovery
linac technology.  HERA  has reached peak luminosities
of up to $5 \cdot 10^{31}$cm$^{-2}$s$^{-1}$
with a luminosity upgrade. The linac-ring accelerator designs
of THERA (TESLA/ILC-HERA) and the QCD explorer (CLIC-LHC) barely 
provide luminosity above
$10^{31}$cm$^{-2}$s$^{-1}$. The LHeC is designed for the highest energy
at  the largest luminosity. 
 }
   \label{fig:lumiep}
\end{figure}

Deep-inelastic lepton-hadron physics at the TeV
scale has been considered previously  at the LEP-LHC workshop
in 1990 \cite{joel,rueckl}  and as part of the TDR
for  TESLA  in 2001 (THERA)~\cite{therabook}.  
A ring-ring $ep$ collider using the LHC has been considered based 
on LEP  \cite{verdier,bartel,keil}. 
This paper is concerned with a new evaluation  taking advantage
of the experience gained at HERA.
A feasibility study  for an $ep$ collider at the LHC using
an electron ring of  energy $E_e=70$\,GeV leads to an estimated
luminosity of about $10^{33}$\,cm$^{-2}$s$^{-1}$, which 
corresponds to an annual integrated luminosity of about 10\,fb$^{-1}$,
at a center of mass (cms) energy of $\sqrt{s}=2 \sqrt{E_eE_p}$ of 1.4\,TeV. 
This places the LHeC very favourably in the luminosity-energy map of
DIS facilites (Fig.\,\ref{fig:lumiep}). 

This paper was first prepared  for inclusion in the deliberations of the 
European Strategy Group of the CERN Council \cite{CERNCOUNCIL} during 2006.
It highlights LHeC physics with examples and it presents a feasibility
study for the machine. The scale of the LHeC is that of an upgrade of 
the LHC. The next steps are to complete a full evaluation of the 
feasibility of the LHeC, including injection, and to develop 
further the optimisation of an experiment and its interface
to the machine. The first data at the LHC and the
completion of the physics programme at HERA with its
final data sample will also have a direct bearing on this optimisation.

The paper is organised as follows: In Section 2 examples are given of
new physics  at high masses and  at very low Bjorken-$x$. Section\,3
is concerned with exploiting the precision of DIS to quantify and test
QCD at a new level of accuracy. Section\,4 highlights the opportunities
at the LHeC to probe QCD in more complex hadronic environments
making use of the wide range of ion beams at the LHC. There follows
a brief consideration of the kinematic reach and its implications
for a detector design in Section\,5. Finally, Section\,6 is concerned with
luminosity prospects based on an initial consideration of the LHeC 
machine design.
\begin{figure}[h]
   \centering
   \centerline{\includegraphics[width=.6\textwidth]{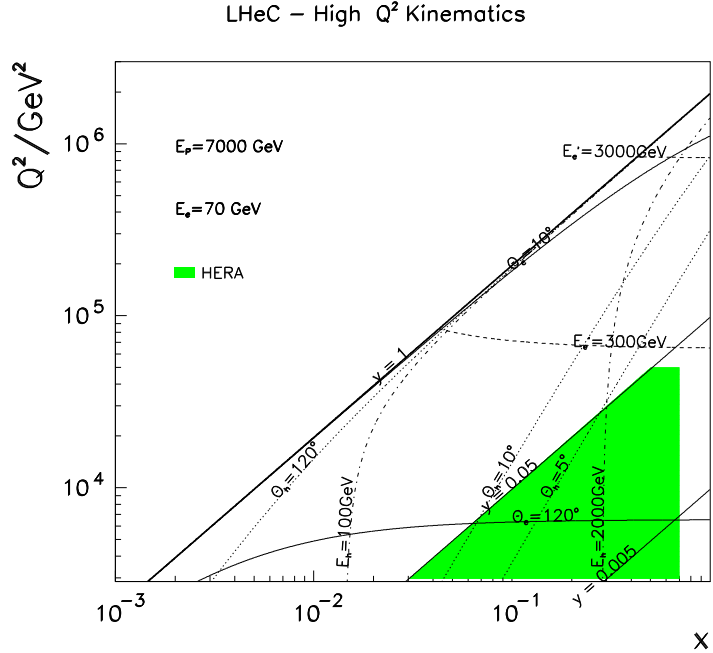}}
\caption{\it Kinematics of $ep$ scattering at the LHeC
 at high $Q^2$. Solid (dotted) curves correspond to constant polar
 angles $\theta_e$ ($\theta_h$) of the scattered electron (hadronic final
 state). The polar angle is defined with respect to the
 proton beam direction. 
 Dashed (dashed-dotted) curves
 correspond to constant energies $E_e'$ ($E_h$) of the scattered
 electron (hadronic final state). The iso-angle and iso-energy 
 lines are derived from Eq.\,4, Section~5.
 The shaded area illustrates
 the region of kinematic coverage in neutral current (NC) scattering at HERA. 
 In $ep$ scattering electron-quark resonances can be formed with
 mass $M = \sqrt{xs}$.  Due to luminosity and energy range, the
 search limit at HERA has been at about  290\,GeV while the 
 LHeC extends to  large $M$ values of about 1300\,GeV.}
   \label{fig:kineq}
\end{figure}
%
   
%
\section{New Physics at the LHeC}
\subsection{Physics Beyond the Standard Model}

In the kinematic domain in which the new physics underpinning
the Standard Model (SM)  is manifest, new electron-quark and positron-quark dynamics could
be observable,  revealing a relationship between the quark and lepton sectors of the SM. 
This sensitivity to lepton-quark physics at the highest energy and shortest distance  is
at the root of  the importance of  the  LHeC.

The high energy of the LHeC extends the kinematic range of DIS physics 
to much higher values of  electron-quark mass $M=\sqrt{sx}$
(Fig.\,\ref{fig:kineq}). An $ep$ collider, providing both baryonic and leptonic quantum
numbers in the initial state, is ideally suited to a study of the properties
of new bosons possessing couplings to an electron-quark pair in this new mass range.
Such particles can be squarks in supersymmetry with $R$-parity violation (\Rp),
or first-generation leptoquark (LQ) bosons which appear naturally
in various unifying theories beyond the Standard Model (SM).
They are produced as  single $s-$channel resonances via the fusion of 
incoming electrons with  quarks in the proton.  They are generically
referred to as ``leptoquarks" in what follows.
\begin{figure}[h]
\begin{center}
  \includegraphics[height=7cm]{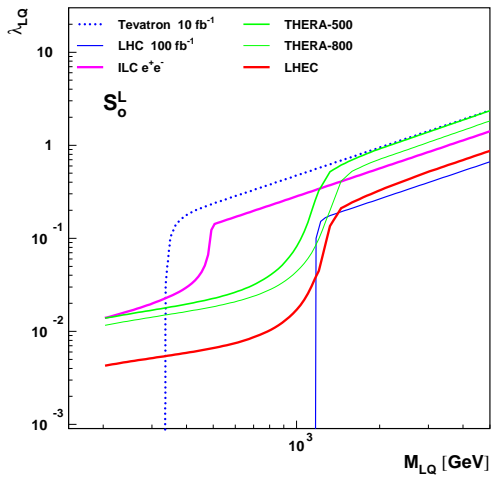} 
  \caption{\label{fig:lq_lambda}
    \it  Mass-dependent upper bounds on the LQ coupling
     $\lambda$ as expected at LHeC for a luminosity of $10 \fb^{-1}$ 
     (full red curve) and at the LHC for $100 \fb^{-1}$ (full blue curve) \cite{filip}.
     These are shown for an example scalar LQ coupling to $e^- u$.}
\end{center}
\end{figure}

Fig.~\ref{fig:lq_lambda} shows the expected sensitivity \cite{filip} of the LHC and LHeC colliders
for scalar leptoquark production. The single LQ production cross section depends on
the unknown coupling $\lambda$ of the LQ to the electron-quark pair, and means 
for a coupling $\lambda$ of ${\cal{O}}(0.1)$, that LQ masses up to about $1 \TeV$ could
be probed at the LHeC. In $pp$ interactions at the LHC such leptoquarks would be
mainly produced via pair production or singly  \cite{belaev} 
with a much reduced cross section \footnote{Single leptoquark production via
photon splitting into an $e^+e^-$ pair \cite{peterz}, which for masses of about 
1\,TeV has a lower cross section than the quark-gluon induced processes,
is not considered here.}.
In $ep$ collisions LQ production can be probed in detail, taking advantage of the formation and
decay of systems which can be observed directly as a combination of jet and lepton invariant mass in
the final state. It will thereby be possible at the LHeC to probe directly
and  with high precision  the
perhaps complex structures which will result in the lepton-jet system
and to determine the quantum numbers of new states.
Examples of the sensitivity of high energy $ep$ collisions to the properties
of LQ production follow. \\

\begin{figure}[h]
\begin{center}
 \begin{tabular}{cc}
  \includegraphics[height=7.3cm]{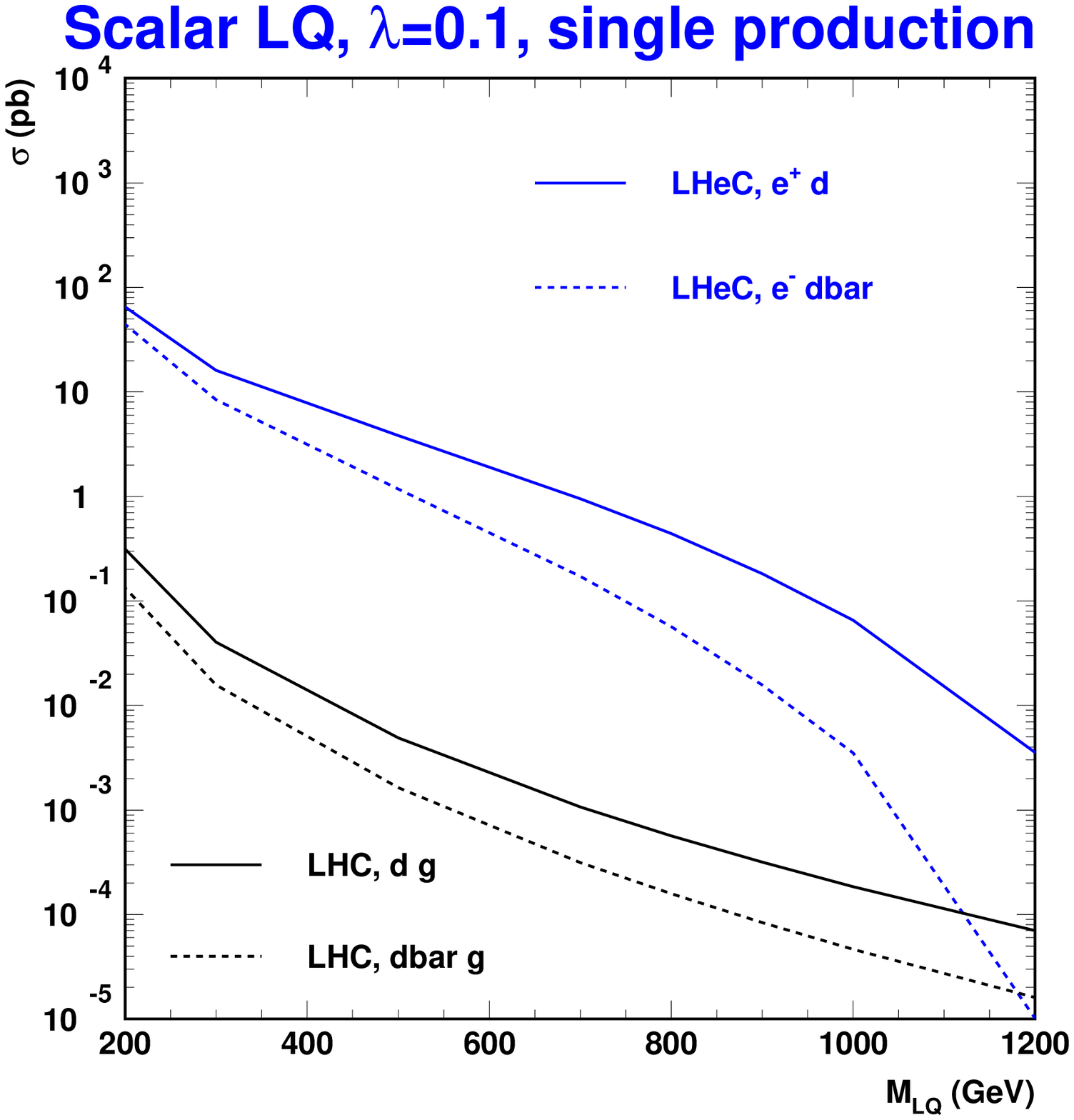} &
  \includegraphics[height=7.3cm]{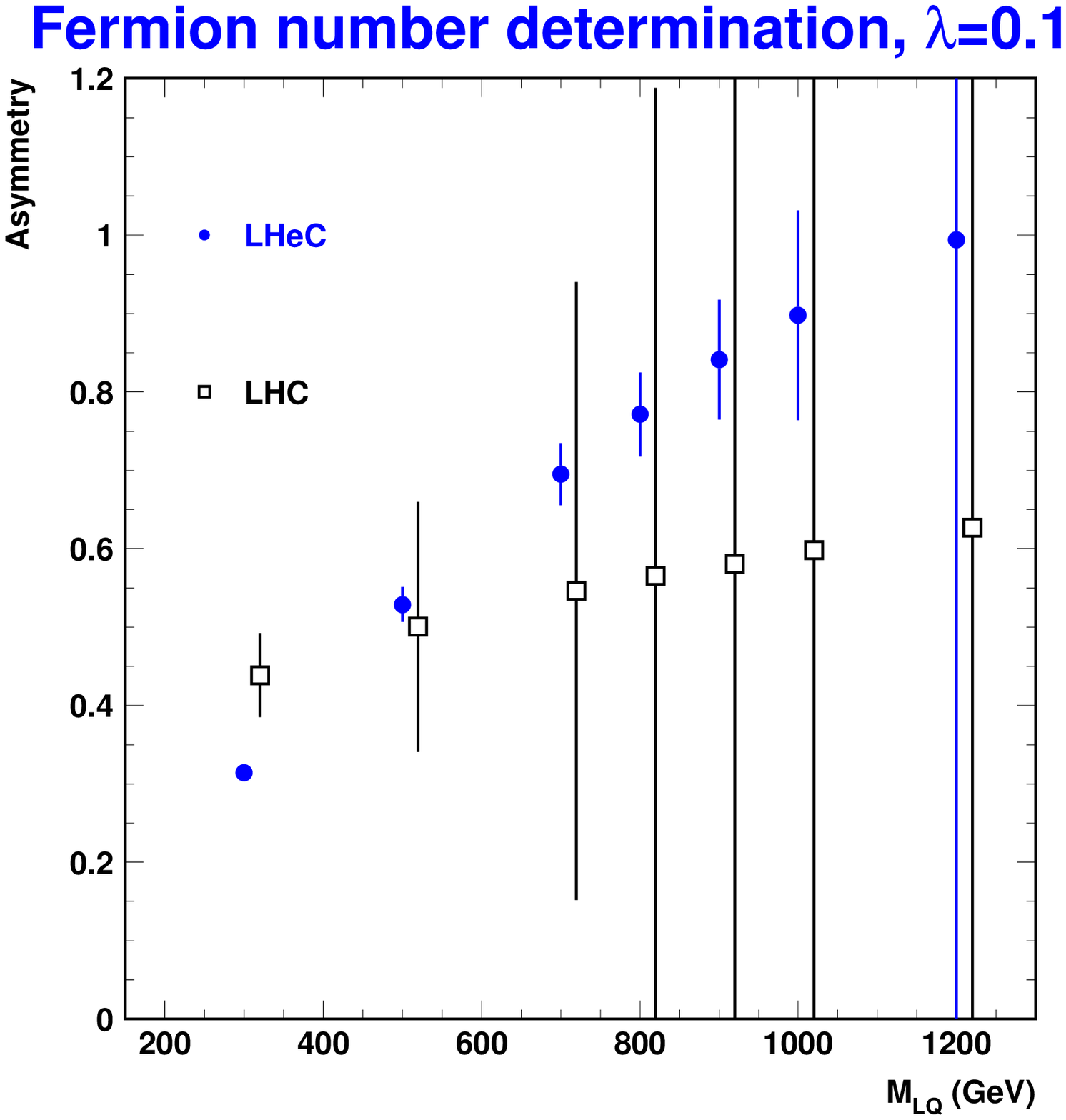}
 \end{tabular}
  \caption{\label{fig:lq_single} \it
     (Left) Single LQ production cross section at the LHeC (top) and LHC (bottom), for
     a scalar LQ coupling to $e^+ d$ with a coupling $\lambda=0.1$;
      (Right) Asymmetries which would determine the fermion number of
     such a LQ,
     the sign of the asymmetry being the relevant quantity.
     An integrated luminosity of $100 \fb^{-1}$ ($10 \fb^{-1}$ per lepton charge) has been
     assumed for the LHC (LHeC).
     }
\end{center}
\end{figure}

{\bf Fermion number ($F$)} :
 Since the parton densities for $u$ and $d$ at high $x$ are much larger than
 those for $\bar{u}$ and $\bar{d}$, the production cross section at LHeC of an
 $F=0$ ($F=2$) LQ is much larger in $e^+ p$ ($e^- p$) than in
 $e^- p$ ($e^+ p$) collisions. A measurement of the asymmetry between the $e^+p$
 and $e^- p$ LQ cross sections thus determines the fermion number
 of the produced leptoquark. 
 Pair production of first generation LQs at the LHC will not allow
 this determination.  Single LQ production at the LHC, followed
 by the LQ decay into  $e^{\pm}$ and  $q$ or $\bar{q}$, could determine $F$
 by comparing the signal cross sections with an $e^+$
 and an $e^-$ coming from the resonant  state.
 However, the single  LQ production cross section
 at the LHC is two orders of magnitude lower than at the LHeC (Fig.\,\ref{fig:lq_single}a),
 so that the asymmetry measured at the LHC may suffer from statistics
 in a large part of the parameter space. 
 For a coupling $\lambda = 0.1$, no information on $F$ can be extracted from
 the LHC data for a LQ mass above $\sim 700 \GeV$, while the LHeC can determine 
 $F$ for LQ masses up to $1 \TeV$ (Fig.\,\ref{fig:lq_single}b). 
%

 {\bf Spin}: At the LHeC, the angular distribution of the
 LQ decay products is unambiguously related to its spin.
 This determination will be much more complicated,
 even possibly ambiguous, if only the LHC
 leptoquark pair production data are available.  
 Angular distributions for vector LQs depend 
 strongly on the structure of the $g \, LQ \, {\overline{LQ}}$ coupling, 
 i.e. on possible anomalous couplings.
 For a structure similar to that of the $\gamma W W$ vertex, vector LQs produced via
 $q \bar{q}$ fusion are unpolarised and, because both LQs are produced with the
 same helicity, the distribution of the LQ production angle will be similar to that
 of a scalar LQ. The study of LQ spin via single LQ production at the LHC
 will suffer from the relatively low rates and more complicated backgrounds.

{\bf Neutrino decay modes}:
 At the LHeC, there is similar sensitivity 
 for LQ decay into both $eq$ and $\nu q$.  At the LHC, in $pp$ collisions,
 LQ decay into neutrino-quark final states is plagued by huge QCD
background. At the LHeC, production through $eq$ fusion with subsequent
$ \nu q$ decay is thus very important if the complete pattern of LQ decay couplings
is to be determined. 

{\bf Coupling $\lambda$}: At the LHeC there is large sensitivity down to small values of the
 coupling $\lambda$. With less sensitivity, in $pp$ interactions at the LHC, information can be 
 obtained from single LQ production  and also from dilepton production via
 the $t$-channel LQ exchange.  Since the single LQ
 production cross sections depend on both $\lambda$ and the flavour of the quark 
 to which the LQ couples, determining $\lambda$ and this flavour  requires $pp$ and
 $ep$ data. 

{\bf Chiral structure of the LQ coupling}:  Chirality is central to the SM Lagrangian.
 Polarised electron and positron  beams\footnote{Whether it is possible
 to achieve longitudinal polarisation
 in a 70\,GeV $e^{\pm}$ beam in the LHC tunnel remains to be clarified.}
 at the LHeC will shed light on the chiral structure of the 
 LQ-e-q couplings. Measurements of a similar nature at LHC are impossible.

Table~\ref{tab:lqqn} summarises the observables \cite{lqvirey} which could disentangle
the various scalar LQ species at an $ep$ machine. A similar discrimination power exists
for vector LQs.

\begin{table}[htb]
 \begin{center}
  \begin{tabular}{|c|cccc|ccc|}
      \hline
      &  $S_{0,L}$  & $S_{1,L}$  & $\tilde{S}_{0,R}$ & $S_{0,R}$
      &  $S_{1/2,L}$   & $\tilde{S}_{1/2,L}$  & $S_{1/2,R}$ \\
         \hline    \hline

      \begin{tabular}{l}
          $S_{0,L}$ \\
          $S_{1,L}$ \\
          $\tilde{S}_{0,R}$ \\
          $S_{0,R}$ \\
       \end{tabular}
   &
     \begin{tabular}{l}
          \\
        $\beta_{\nu}$ \\
        $P_e$  \\
        $P_e$ \\
     \end{tabular}
   &
     \begin{tabular}{l}
       $\beta_{\nu}$ \\
          \\
       $P_e$  \\
       $P_e$  \\
     \end{tabular}
    &
     \begin{tabular}{l}
       $P_e$ \\
       $P_e$ \\
         \\
      $ - $ \\
     \end{tabular}
    &
     \begin{tabular}{l}
       $P_e$ \\
       $P_e$ \\
      $ - $ \\
         \\
     \end{tabular}
    &
  \multicolumn{3}{|c|}{ $e^+ / e^-$ } \\
  \hline  \hline
   \begin{tabular}{l}
     $S_{1/2, L}$ \\
     $\tilde{S}_{1/2,L}$ \\
     $S_{1/2,R}$ \\
   \end{tabular}
  &
   \multicolumn{4}{|c|}{ $e^+ / e^-$ }
  &
   \begin{tabular}{l}
        \\
     $ - $  \\
     $P_e$  \\
   \end{tabular}
   &
   \begin{tabular}{l}
      $ - $ \\
        \\
      $P_e$ \\
    \end{tabular}
   &
   \begin{tabular}{l}
      $P_e$ \\
      $P_e$ \\
        \\
    \end{tabular} \\
  \hline
 \end{tabular}
  \caption[]
          {\label{tab:lqqn}
          \it  Discrimination between LQs with different
           quantum numbers in $e^{\pm}p$ scattering with an electron beam 
           polarisation $P_e$.
           The nomenclature of~\cite{LQNAME} has been used
           to label the different scalar LQ species described by
           the model of Buchm\"uller, R\"uckl and Wyler~\cite{BRW}, in which the branching ratio
           $\beta_{\nu}$ of the LQs to decay into $\nu + q$
           is known. }
  \end{center}
 \end{table}

\begin{figure}[h]
\begin{center}
\includegraphics[height=13cm]{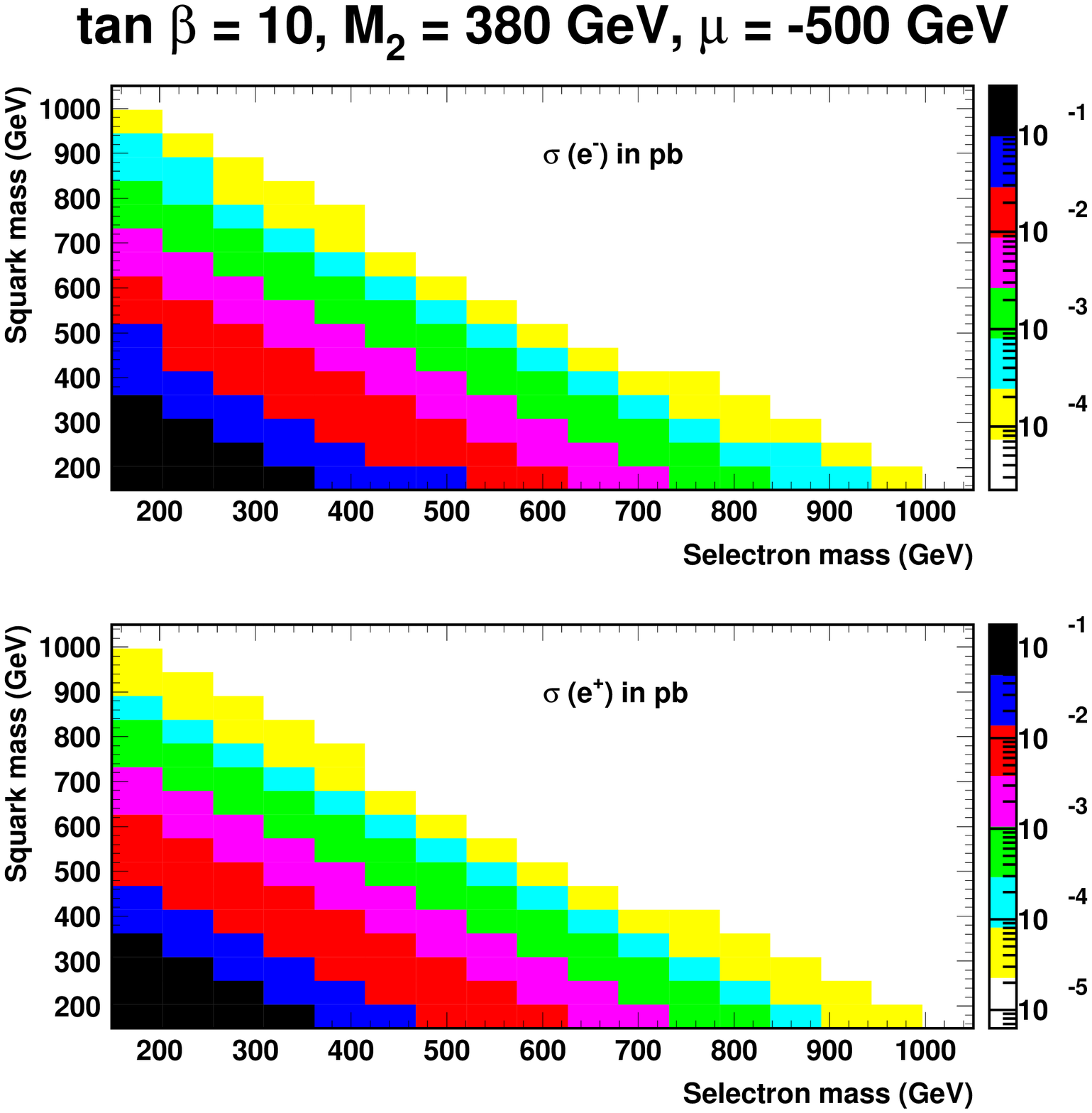} 
  \caption{\label{fig:susysig} \it
     Production cross section (right scale, in pb) of a selectron-squark pair at the LHeC,
     for example values of MSSM parameters.
     It is assumed that $\tilde{u}_L$, $\tilde{u}_R$, $\tilde{d}_L$ and
     $\tilde{d}_R$ are degenerate, as are $\tilde{e}_R$ and $\tilde{e}_L$.
     }
\end{center}
\end{figure}

If Supersymmetry is manifest at TeV energy,
in $ep$ interactions the associated production of squarks and 
sleptons ($\tilde{e}$ and $\tilde{\nu_e}$) proceeds via the
$t$-channel exchange of a neutralino or chargino.
 Fig.~\ref{fig:susysig} shows that the rates can be sizeable
at the LHeC when the sum of the squark and slepton masses is below
$\sim 1 \TeV$. If squarks are relatively light, $\sim 500 \GeV$,
selectron masses up to about $500 \GeV$ could be probed at the LHeC.
This sensitivity may extend somewhat beyond the discovery reach for selectrons 
in $pp$ scattering at the LHC.
\begin{figure}[h]
\begin{center}
 \includegraphics[height=9cm]{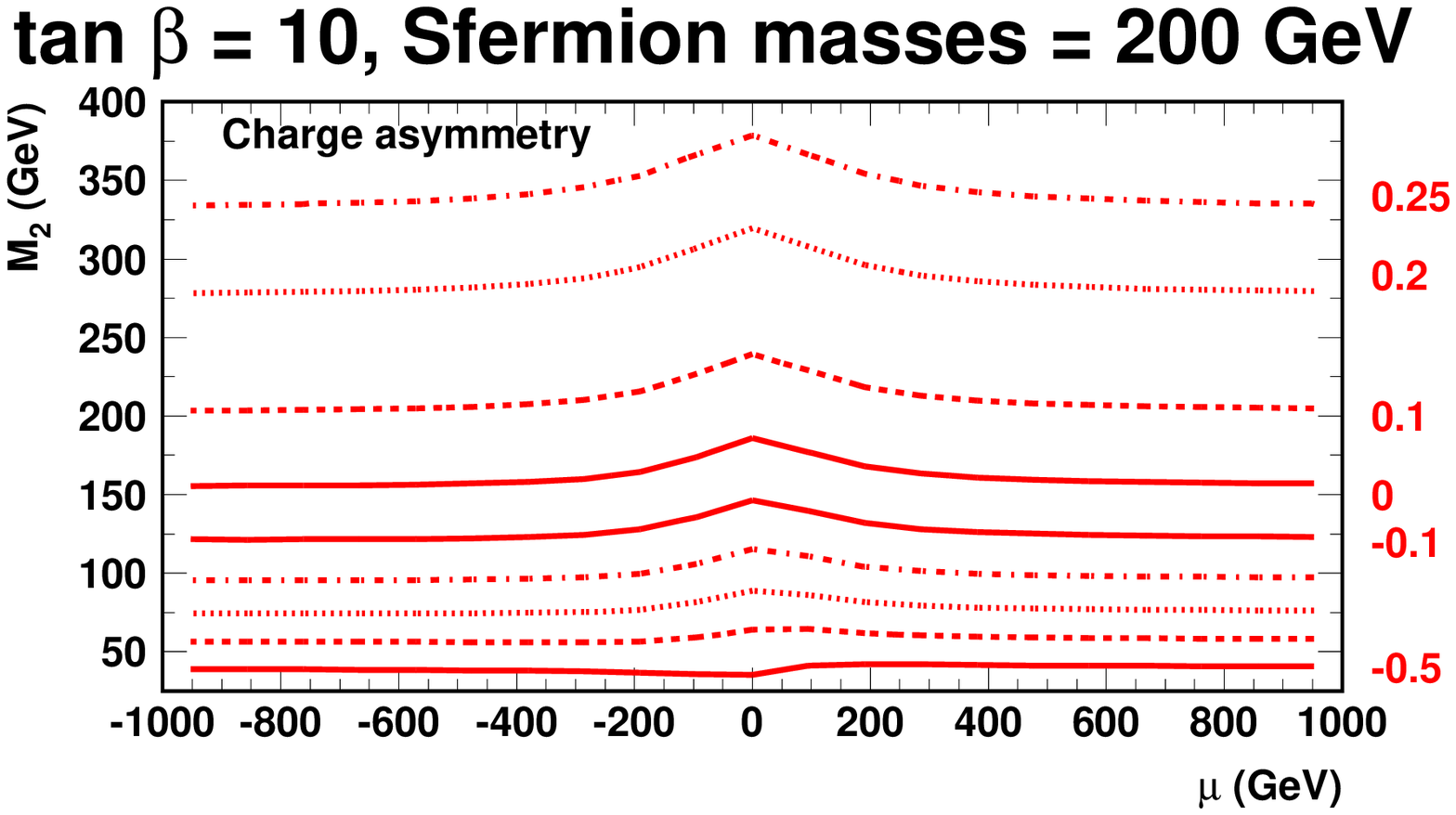}
  \caption{\label{fig:susyasy}
\it  Charge   asymmetry of the $\tilde{e} \tilde{q}$
     production cross section at LHeC in the $M_2-\mu$ plane in
     the MSSM. Light sfermion masses are assumed. The numerical values on the
     right of the plot correspond to the asymmetry values of the contours.
     }
\end{center}
\end{figure}
%


Charge asymmetries and, possibly, polarisation asymmetries could
provide additional information on the couplings of the exchanged
gauginos and on the mass difference between $\tilde{e}_L$ and $\tilde{e}_R$.
Examples of such asymmetries are shown in Fig.~\ref{fig:susyasy}. 
Light sfermions are assumed and the framework of the Minimal Supersymmetric
Standard Model is used.

%
%
\subsection{Physics of High Parton Densities (Low $x$)}
The observations at HERA of the rise of the structure function
$F_2(x,Q^2)$ and of its derivative $\partial  F_2 /\partial \ln
Q^2$ as  $x$  decreases  \cite{rise} 
imply that  at low $x$ the sea quark  and the 
gluon distributions, respectively, of the proton increase dramatically. 
At low $x$  proton structure is thus driven by gluons and the
formation  of quark-antiquark pairs. While the charge of the proton
is determined by its valence quarks, the kinetic and potential
energy of gluons determines its mass. An understanding of quark-gluon
dynamics is thus a key to the mass of the universe \cite{wilczek}.

The sharp rise  of the density of the gluons \cite{alfash1,alfaszeus} 
in the proton
leads to the possibility of non-linear parton interaction effects \cite{glr}.
A new dense state of parton matter is likely to exist, sometimes referred
to as a Colour Glass Condensate \cite{cgc},  which is characterised
by a high parton density and small coupling constant. 
\begin{figure}[h]
   \centering
   \centerline{\includegraphics[width=.6\textwidth]{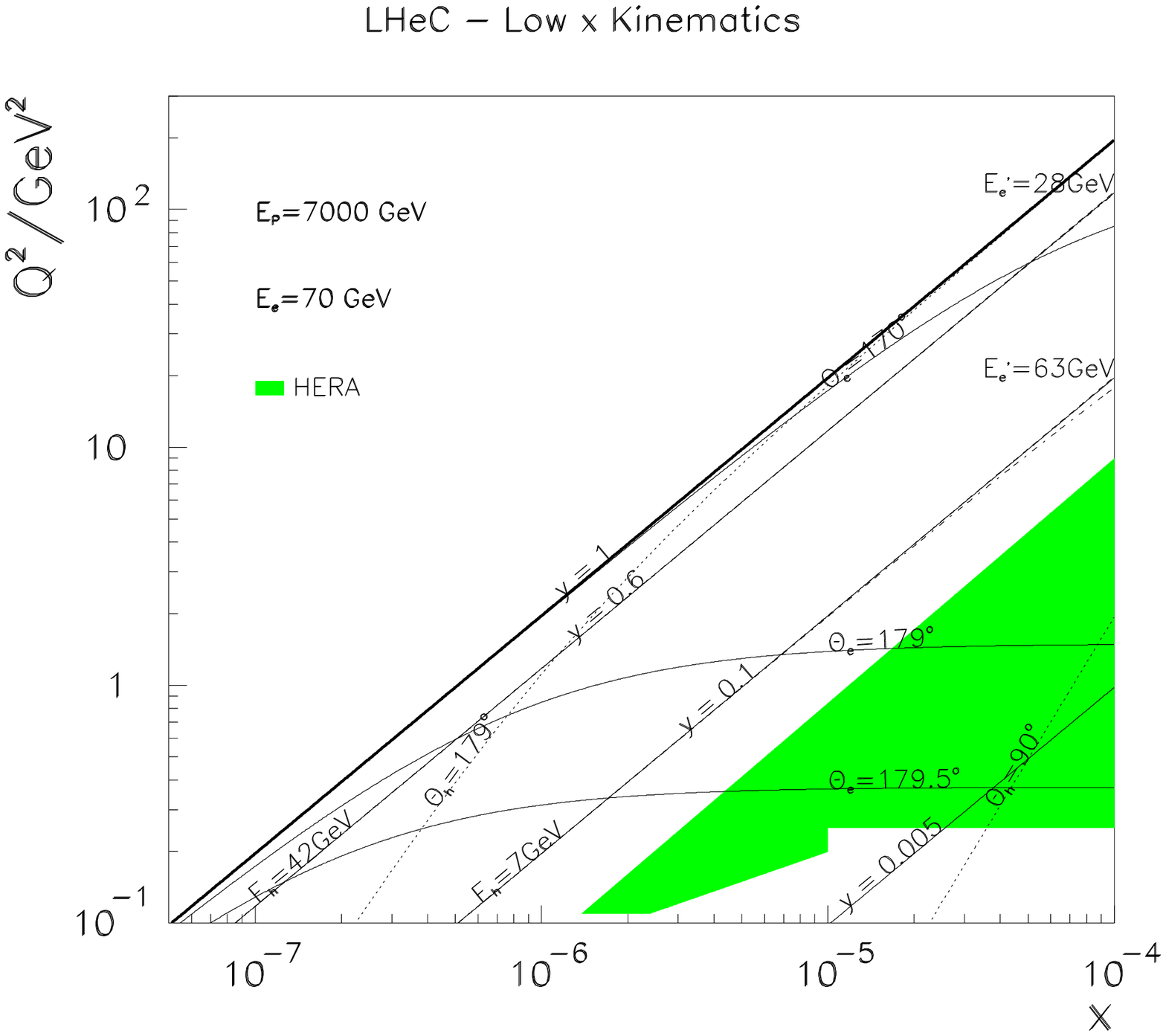}}
\caption{\it Kinematics of $ep$ scattering at the LHeC
  at low $x$.  Access to the $Q^2$ region down to 1\,GeV$^2$
  requires a dedicated interaction region with acceptances extended
  to about 179$^{\circ}$, with respect to the proton beam direction,
  i.e. an extended ``backward detector".
   The shaded area illustrates
 the region of kinematic coverage in NC scattering at HERA, which 
 has been extended below
 $x \sim 10^{-4}$  with special techniques, namely a detector
 attached to the beam pipe backward of the main apparatus and also
 by shifting the interaction vertex into the proton beam direction.}
   \label{fig:kinex}
\end{figure}

Much theoretical development  in low $x$ physics is concerned with 
evolution equations which
are the most appropriate approximation to a full solution to QCD 
(such as the BFKL \cite{bfkl},
the CCFM  \cite{ccfm},
and the Balitzky Kovchegov equations \cite{balkov}), and the incorporation of
small $x$ resummation combining the classic DGLAP approach \cite{dglap}
with BFKL evolution \cite{altforte}.   
DGLAP theory has been calculated to NNLO \cite{vermaseren}.
Physics at low $x$ has also been formulated using the
colour dipole approach \cite{dipole}. However,
there is   still  no universally accepted formulation of QCD at low $x$. 
This situation can be traced in large part to the fact that
the kinematic reach of HERA is insufficient to establish 
unambiguously the existence
of non-linear parton interaction effects and parton saturation phenomena
 \cite{kgbdiff,bartelsat,levin}.
%
%
%

Although the derivative $\partial F_2 / \partial \ln x$  shows
no evidence for a damping of the growth of the sea quark 
density with decreasing $x$, some effects observed at HERA
in forward jet production \cite{fwdjets} and in azimuthal (de)correlations
\cite{azimu}  seem to indicate departures from 
the conventional radiation pattern in QCD. 
Future progress requires the substantial increase of the kinematic 
reach to low $x$  (Fig.\,\ref{fig:kinex}) which is made possible by the LHeC.

\begin{figure}[htbp]
    \centerline{\includegraphics[width=0.55\textwidth]{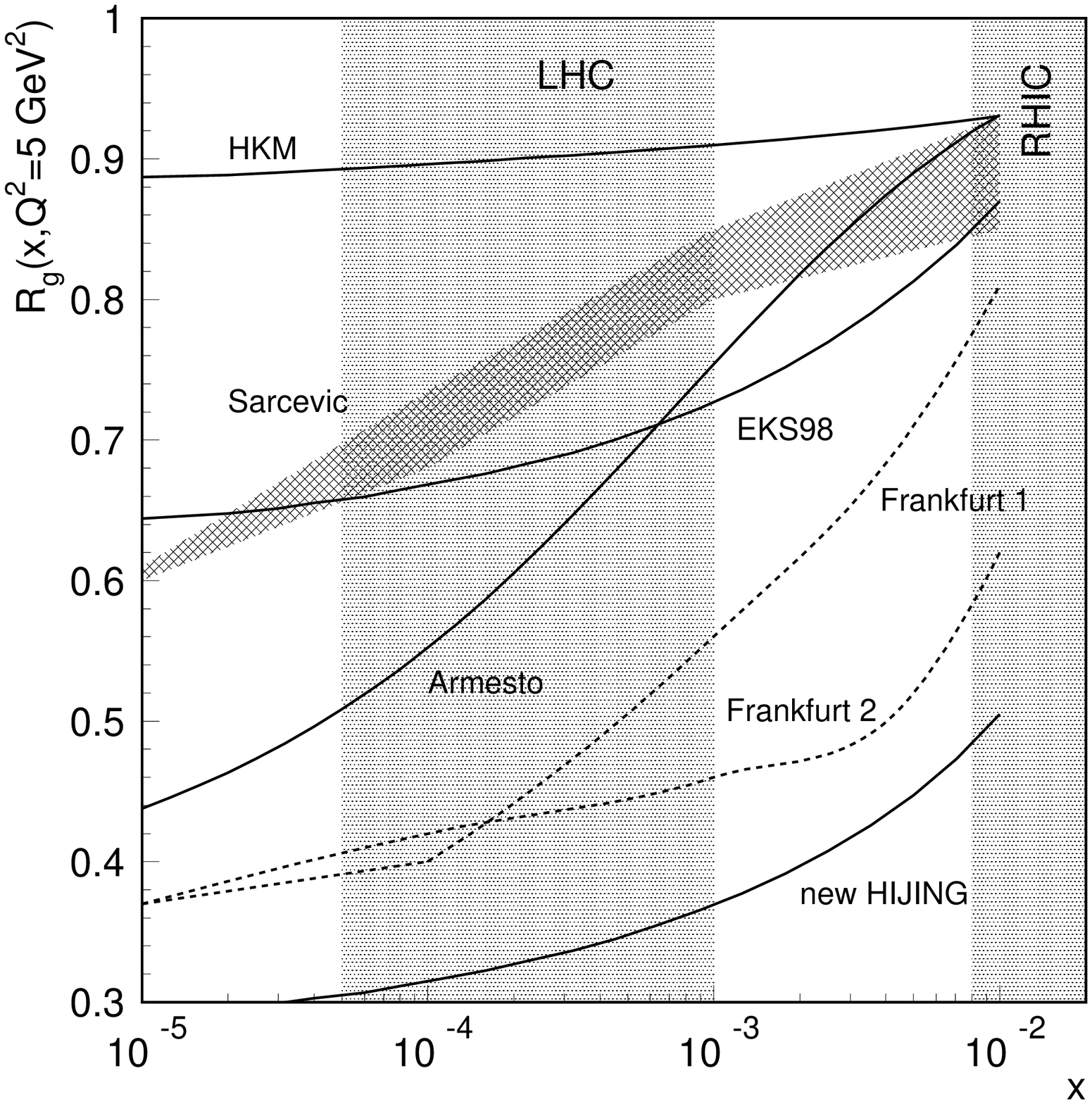}}
\caption{\it Ratios of gluon distribution functions in lead (Pb) relative to
the gluon distribution in the proton using different models,
at $Q^2 = 5$\,GeV$^2$.}
   \label{fig:glurat}
\end{figure}

At the LHC measurements will be performed of the interaction
of nuclei, in for example Pb-Pb collisions, with the aim of investigating
a high density parton phase (Quark Gluon Plasma).  An unambiguous determination
of the nuclear parton distributions in the domain accessed with
these measurements, which is possible in $eA$ scattering at the 
LHeC, will be important if
the equilibrium of the nucleon (colour singlet) and parton 
(colour non-singlet, QGP) phases are to be understood.
A complete unknown is the gluon distribution in nuclei at low $x$ 
 (Fig.\,\ref{fig:glurat}) \cite{aliceRG}.
%


If, as one expects, the evolution at low $x$ proceeds differently from
DGLAP based predictions,
there are important consequences for high energy physics, which reach
beyond the intrinsic questions of low $x$ theory. These may include 
\begin{itemize}
\item{ the  observation of the black disc limit of high energy scattering \cite{blackdisc},
i.e. of the unitarity limit, which causes cross sections at high energy to saturate
resulting in }
\item{
 new, possibly radically different, predictions for high energy
cosmic ray and neutrino physics, which currently are based on large 
extrapolations \cite{engel}, and also }
\item{  
 a different perspective on physics in the forward region of $pp$, $pA$ and $AA$
  interactions at the LHC.}
\end{itemize}
%
\section{Precision Quantum Chromodynamics}
QCD is one of the cornerstones of our understanding of the physics of the 
universe, which is encapsulated in the Standard Model (SM).  Like the electroweak
sector of the SM, its predictive power depends on the accuracy with which its
gauge coupling is known and on our ability to make predictions and compare
them with experiment for a variety of hadronic phenomena.  Lepton-hadron
interactions, in which hadronic matter is probed deeply in a well 
understood and point-like manner,
is an extremely powerful means of achieving these aims. 

There follow a few
examples, in which at the LHeC,
because of the precision of an $ep$ experiment and because
of kinematic reach, one can foresee a major improvement in the accuracy,
with which QCD can be tested, quantified and developed further.
\subsection{Structure Functions and Partons for the LHC}
\begin{figure}[htbp]
   \centering
   \centerline{\includegraphics[width=0.65\textwidth,angle=90.]{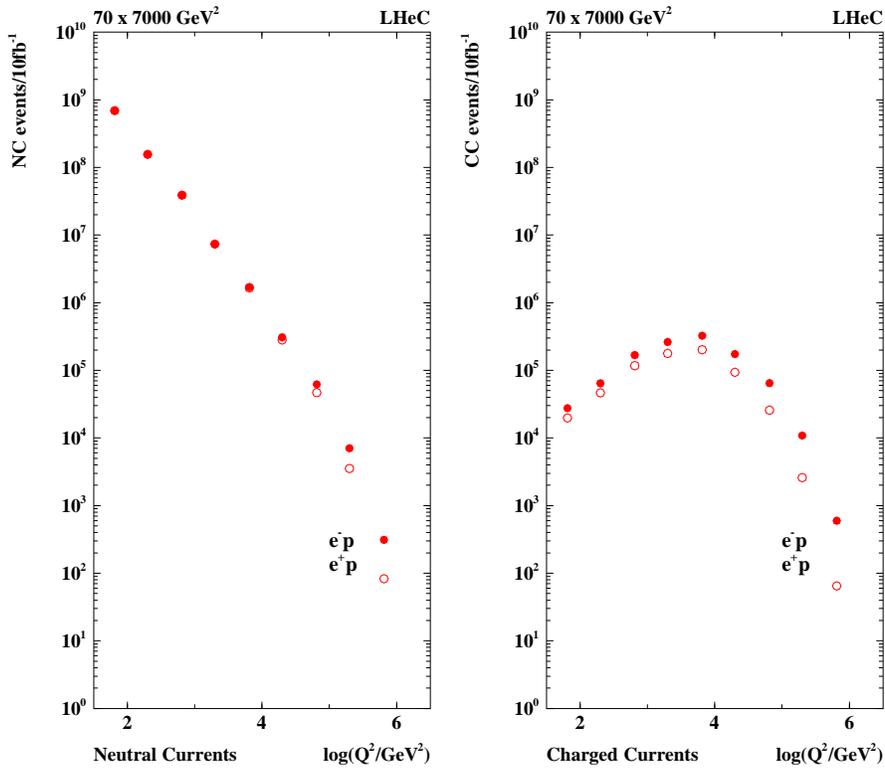}}
\caption{\it Event rates per 10\,fb$^{-1}$ in neutral (left) and charged (right)
  current $e^{\pm}p$ scattering at the LHeC for the high lumnosity
   acceptance, Section\,5. At design luminosity such statistical
  sensitivity  may be accumulated within one year of operating the LHeC.}
   \label{fig:rates}
\end{figure}

\begin{figure}[h]
 \centering
  \centerline{\includegraphics[width=.6\textwidth,angle=90.]{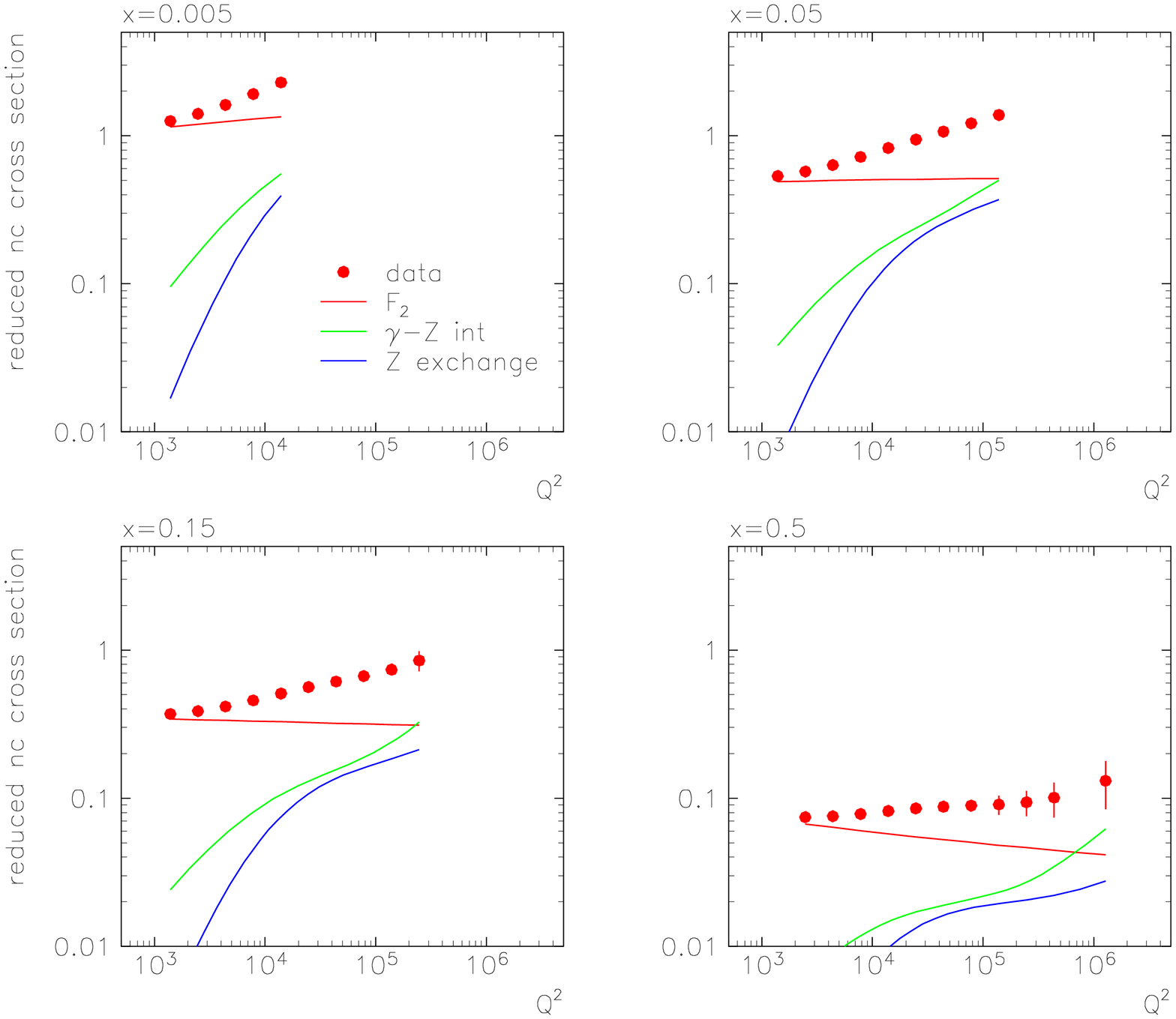}}
\caption{\it Simulation of a measurement in unpolarised $e^-p$ scattering
of the reduced NC cross section, 
$\sigma_r= d\sigma/dxdQ^2 \cdot Q^4 x/2 \pi \alpha^2 (1+(1-y)^2)$.
In the one-photon exchange approximation $\sigma_r = F_2$ for $F_L=0$.
With rising $Q^2$ the effects due to the exchange of weak bosons
become sizeable such that, for example, the reduced cross section
does not decrease at large $x$ any more, as would be
the case for pure photon exchange, solid (red) line. The luminosity used here
is only 200\,pb$^{-1}$. Accurate measurements at the LHeC in the rapidity plateau region
of the LHC can thus be made using data from a very short running period, namely
less than two weeks at design luminosity.}
   \label{fig:nccross}
\end{figure}

\begin{figure}[htbp]
 \centering
   \centerline{\includegraphics[width=.6\textwidth,angle=90.]{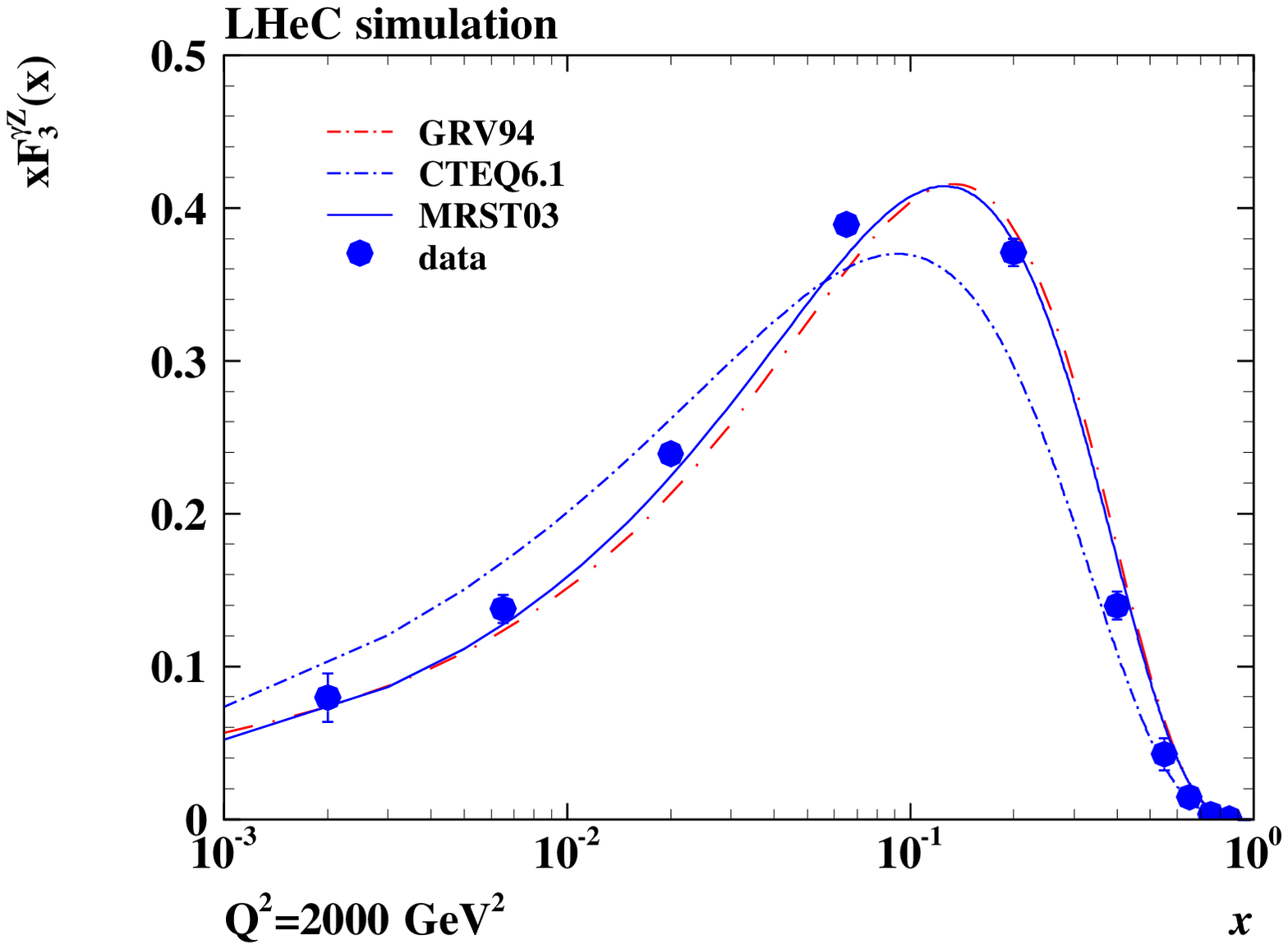}}
\caption{\it Simulation of a measurement of the neutral current
interference structure function $xG_3=xF_3^{\gamma Z}$
from an $e^{\pm}p$ NC charge asymmetry measurement assuming 
a luminosity of 10\,fb$^{-1}$ per beam setting. This function is defined as \cite{klrie}
$xG_3 = 2x [a_u e_u (U-\bU) + a_d e_d (D-\bD)]$, where
$U (D)$ denotes the sum of the up (down) quark distributions in the proton. 
For the first time one thus has direct access to the antiquark-sea (a?)symmetry at low $x$
in the deep inelastic region. It is usually assumed that antiquarks and
sea quarks are equal to each other, e.g. $\bu = u_s$ and $s = \bs$.
This assumption is common to all pdf parameterisations used here,
so that $xG_3$ tends to zero at low $x$.
Recent analyses of the NuTeV anomaly, however, suggest a possible difference
between strange and anti-strange quark distributions\,\cite{cteqanomaly}.
Symmetry implies that $xG_3 =(2u_v + d_v)/3$, which is expected to drop to zero at
low $x$ and to be roughly independent of $Q^2$. The data points are thus
projected to an average $Q^2$ to display the accuracy
of the measurement. From the very high $Q^2$ region one will be able to 
derive a measurement over 3 orders of magnitude in $x$. Note that at HERA this function  can 
only be measured down to $x \simeq 0.02$ and with much less accuracy since
the $Q^2$ values involved are  smaller, i.e. electroweak effects weaker,
and the luminosity is inferior to what is projected for the  LHeC.
}
   \label{fig:xg3}
\end{figure}

\begin{figure}[htbp]
   \centering
     \centerline{\includegraphics[width=.9\textwidth]{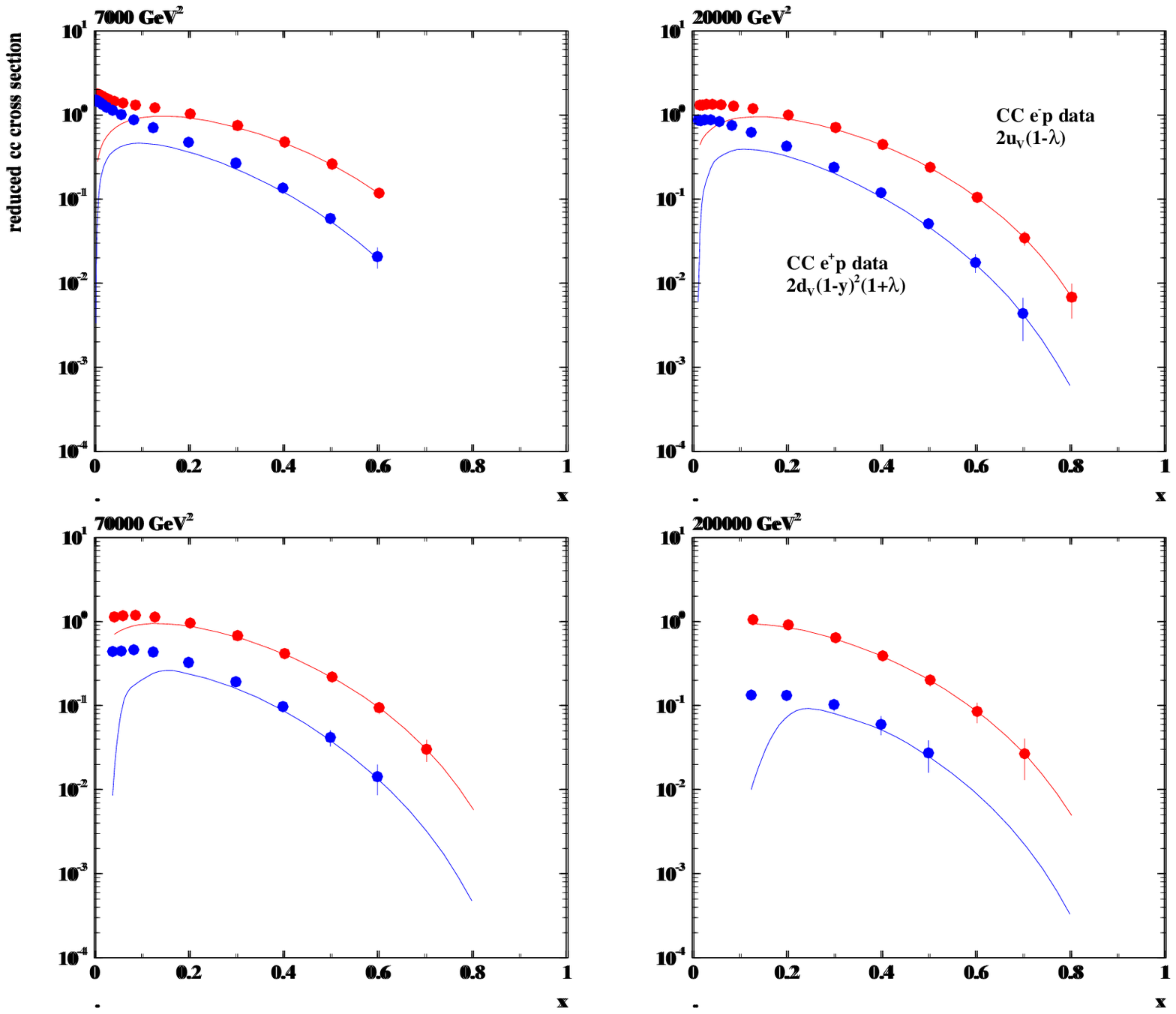}}
\caption{\it Reduced charged current cross sections
 with statistical uncertainties corresponding to
   1\,fb$^{-1}$ in  
  electron (top data points, red) and positron (lower data points, blue)
  proton scattering at the LHeC, simulated for the low $x$ detector configuration,
   see Section\,5.
  The curves are determined by the dominant valence quark distributions, $u_v$ for $e^-p$
  and $d_v$ for $e^+p$. In the simulation
   the lepton polarisation $\lambda$ is
  taken to be zero. The valence quark approximation of the reduced cross section
  is seen to hold already at $x \simeq 0.2$ and a rather accurate determination
  of the $u/d$ ratio up to large $x$ appears to be feasible at very high $Q^2$.}
   \label{fig:uoverd}
\end{figure}

Precision measurements at the LHeC of the neutral current (NC) and charged current (CC) 
deep-inelastic scattering cross sections (Fig.\,\ref{fig:rates})
are pivotal in building a successful physics programme at the LHC.
Large electroweak effects, which add quark
flavour and matter-antimatter 
sensitivity, are present (Fig.\,\ref{fig:nccross}).
The primary purpose of these measurements is to extract a number of important 
structure functions of the proton, and
hence to  determine a comprehensive set of parton 
density functions (pdf) for the nucleon at the LHC energy scale,
thereby avoiding the uncertainties introduced in the evolution of measurements
from the HERA energy scale.  Important examples where new constraints
are needed are  the behaviour of the sea and valence quarks at
low $x$, which can be obtained
from a measurement of the $\gamma Z$ interference structure
function $xG_3$ (Fig.\,\ref{fig:xg3}), and the limit of  the $u/d$ ratio at large $x$,
obtainable
from a measurement of charged current $e^{\pm}p$ scattering (Fig.\,\ref{fig:uoverd}). 

The accuracy with which the parton distributions
can be determined contributes directly to the sensitivity to new physics at both the LHC
and the LHeC.  An example is the envisaged  measurement to 1\%
of the luminosity in $pp$ interactions at the LHC using $W$ and $Z$ production \cite{dittmar}.
So as to cover the rapidity plateau region at the  LHC, it is necessary to know the
pdfs of quarks and gluons \cite{stirling} in the Bjorken-$x$ range $10^{-4}$ to 0.1 at
$Q^2 =M_{Z,W}^2 \simeq 10^4$ \,GeV$^2$ with adequate precision.
This range  is covered  directly by the
LHeC. If HERA data alone are to be used, a large extrapolation in $Q^2$ is necessary,
 which requires an accurate understanding of perturbative QCD.
The theory underpinning the extrapolation is subtle and far from unambiguous 
\cite{forteheralhc}. At large $x$, resummation effects are
important. At low $x$ the usual assumption of DGLAP evolution must
ultimately break down.
Furthermore, the parton dynamics may be non-linear (see Section 2.2), and multiple
interactions may have to be taken into account when extrapolating to high energy
\cite{martinwatt}.

\begin{figure}[h]
 \centering
  \centerline{\includegraphics[height=8cm,width=13.5cm]{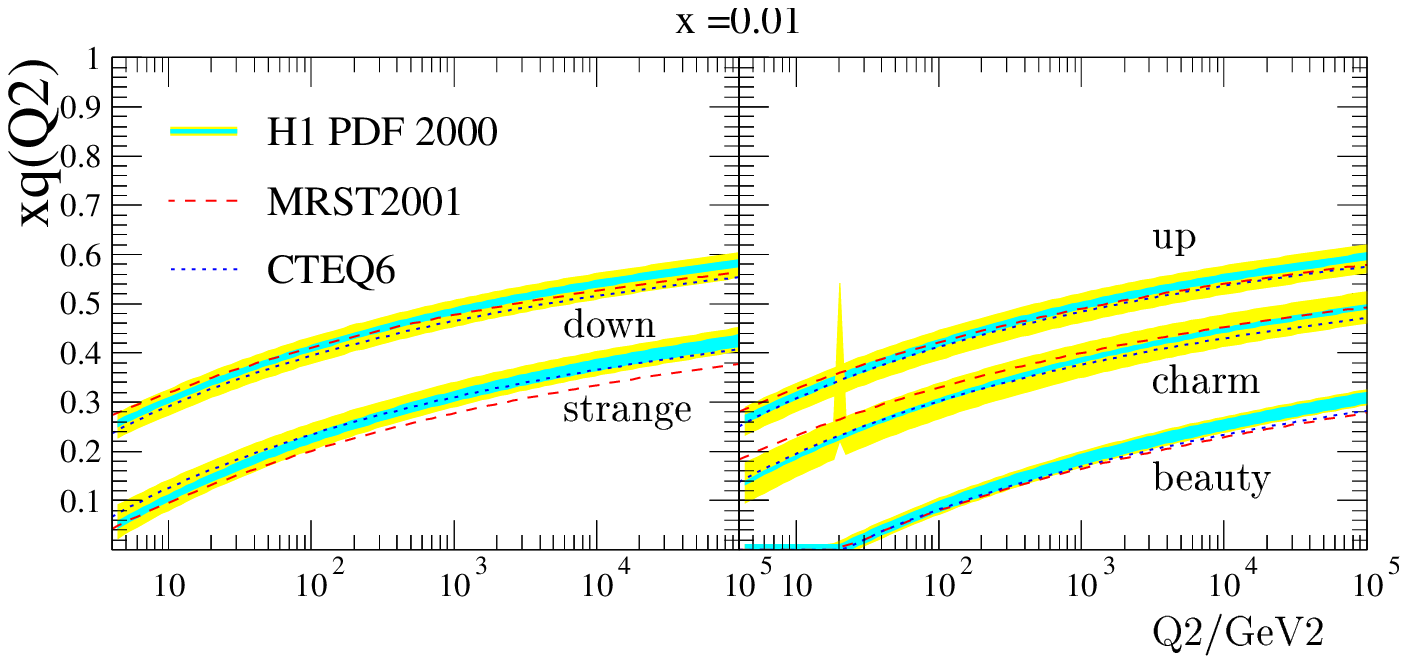}}
\caption{\it The $Q^2$ evolution of the sea quark distributions, for $x$ within the
rapidity plateau at the LHC, as predicted using the NLO DGLAP pdf analyses of H1, MRST and CTEQ.
At $x=0.01$, the LHeC will cover a $Q^2$ range up to $10^4$\,GeV$^2$, by which 
point the
heavy quark distributions become similar in size to the light quarks.
The complete determination to high accuracy of the parton distributions
is essential for searches for new physics at the LHC and also for
measuring its ``partonic" luminosity from the $W, Z$ boson production rates.
}
   \label{fig:quarks}
\end{figure}

At LHC, heavy quarks play an essential role in QCD dynamics, and  their
contributions to proton structure must therefore be understood well. In 
particular the bottom quark distribution needs to be known rather accurately
because $b$ quarks contribute substantially to the production mechanisms
for new physics.
With rising $Q^2$ the fraction
of the heavy quark contributions increases (Fig.\,\ref{fig:quarks}), 
for $b$ quarks from a few per mil
near threshold at HERA \cite{bh1} to about 5\% at the LHeC. 
Thus with silicon vertex detectors, and taking advantage
of the small beam spot size ($15 \cdot 35$\,$\mu $m$^2$),
very accurate measurements of the beauty and the charm quark densities 
are possible in a very wide range of $Q^2$ and $x$.  The greatly enhanced
kinematic reach at the LHeC, together with vertex flavour tagging,
will   make possible 
measurements of high $x$ strange and beauty quark densities in the proton
using the couplings  $s W \rightarrow c$ and $b W \rightarrow t$ respectively.
%

Thus, at the LHeC determinations at a new level of precision 
and to previously unexplored $x$ values
of all quark distributions in the proton can be anticipated, 
and, with the combination of the LHeC and HERA structure function measurements, 
a determination of the gluon distribution in the proton 
with unprecedented accuracy over an extended range of $x$ 
and of momentum transfer will also result.

%
\subsection{Strong Coupling Constant}
\begin{figure}[htbp]
 \centering
     \centerline{\includegraphics[width=.5\textwidth]{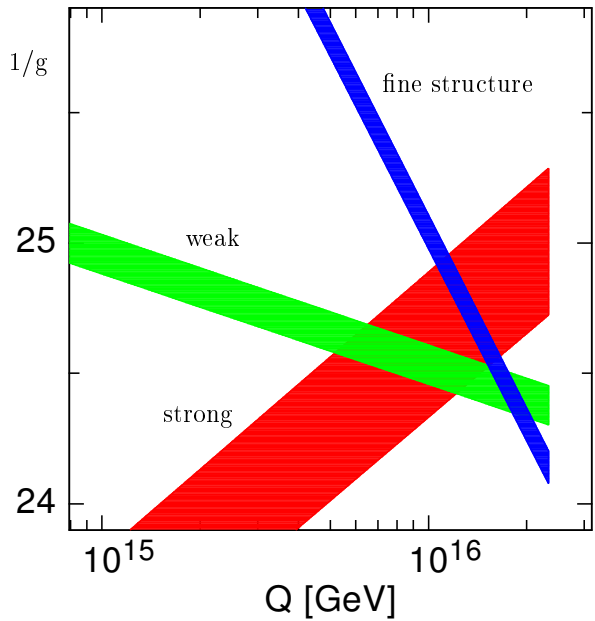}}
\caption{\it Two-loop extrapolation of the inverse coupling constants of $U(1)$, $SU(2)$ and
$SU(3)$ to the unification point, defined as $1/g_1 = 1/g_2$, in the 
MSSM model using the $\overline{DR}$ scheme, where $g_1 (g_2)$ denotes
the electromagnetic (weak) coupling constant. The uncertainty is dominated 
by the rather moderate knowledge of the strong coupling constant.
Improvements on the measurement of $\alpha_s$ are expected also from the
Giga $Z$ mode at the ILC \cite{ulim}.}
   \label{fig:alphasusy}
\end{figure}

The strong coupling constant $\alpha_s$ is currently known to 1-2\%
experimental error. This is many orders of magnitude worse than the determination of the
fine structure constant and the Fermi constant, which are known
at the level of $10^{-9}$ and  $10^{-5}$, respectively. The gravitational
constant is known to  0.1\%. 
In unified  theories the electromagnetic, weak and strong couplings
are expected to approach a common limit at some large unification scale.
Presently, the accuracy of such extrapolations is limited by the uncertainty with
which $\alpha_s$ is known (Fig.\,\ref{fig:alphasusy} \cite{ulim}).   
Precision tests of QCD and comparisons
with lattice QCD calculations \cite{alfalattice} require a significant improvement in the
knowledge of $\alpha_s$.

Deep-inelastic scattering is a well defined process theoretically \cite{feyn,vannervendis}, 
and has recently been calculated to NNLO \cite{vermaseren}.  The determination of
the strong coupling constant in DIS requires the simultaneous determination
of the gluon distribution, $xg$, and the quark distributions.
At HERA $\alpha_s$ is determined to within an experimental
accuracy of about 1\% \cite{glasman}. With the inclusion of the LHeC data,
the experimental accuracy is expected to reach a few per mil. 

At such a high level of experimental 
accuracy, it will be necessary to reassess
many theoretical and phenomenological problems 
 of crucial importance to our understanding of QCD and 
how to use it, for example,  the treatment of the renormalisation scale  ($\mu_r$) uncertainty.
By  convention\footnote{Such a large variation of the renormalisation scale
is already not supported in NLO QCD analyses of the HERA structure function data
of H1 \cite{alfash1} and ZEUS \cite{alfaszeus} and the prescription
for estimating the resulting  theoretical uncertainty  of $\alpha_s$ 
needs to be reconsidered.}, one  still varies
$\mu_r ^2$ between $Q^2/4$ and $4Q^2$,
which at NLO introduces an uncertainty on $\alpha_s$ of about
5\% and at  NNLO is estimated to be about 1\% \cite{vermaseren}.
Further examples of theoretical
issues which appear at the new level of accuracy with LHeC data are the treatment of heavy
flavours in QCD evolution \cite{wukiprivate} and the 
limits to the validity of the DGLAP approximation  in deep-inelastic scattering \cite{mrstconserv}. 

\subsection{Hard Diffraction}
As discussed in the previous sections, a detailed 
understanding of physics in the LHC energy range will require substantial 
developments in the knowledge of  Quantum
Chromodynamics in the high  density, low $x$, environment. Low 
$x$ studies at HERA and the Tevatron have shown clearly 
that diffraction has to be an 
integral component of any successful low $x$ theory.
\begin{figure}[h] \unitlength 1mm
 \begin{center}
 \begin{picture}(90,115)
    \put(-15,67){\epsfig{file=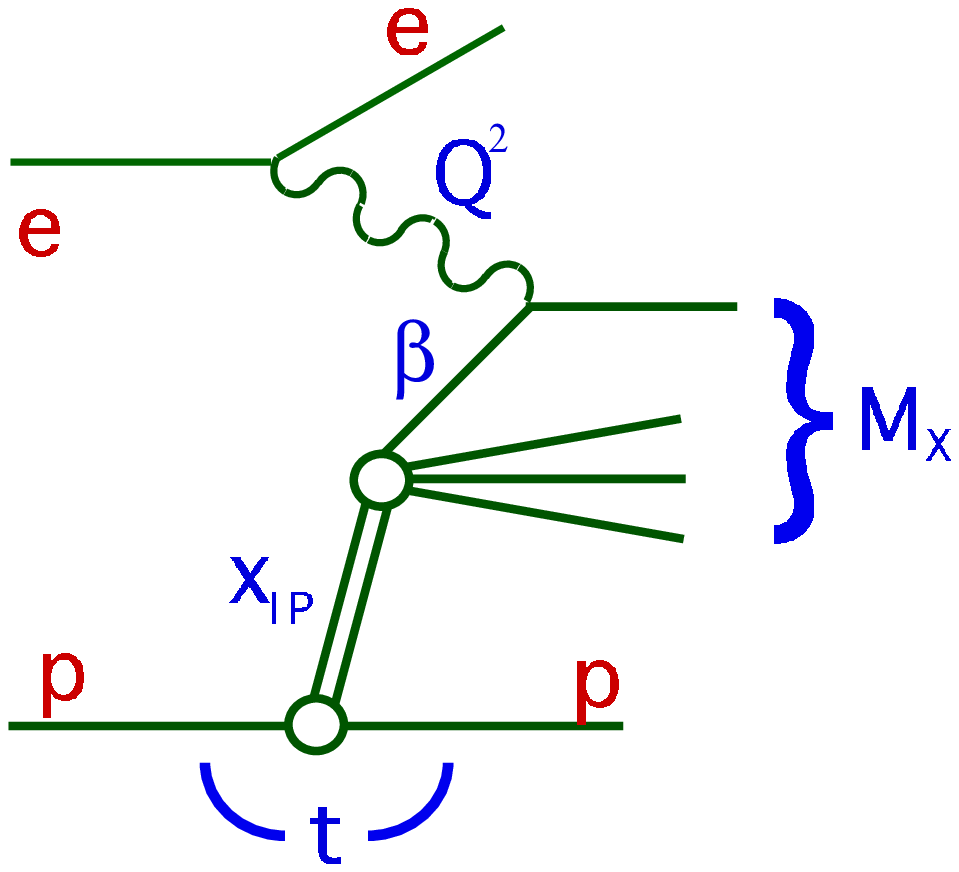,width=0.35\textwidth}}
  \put(45,35){\epsfig{file=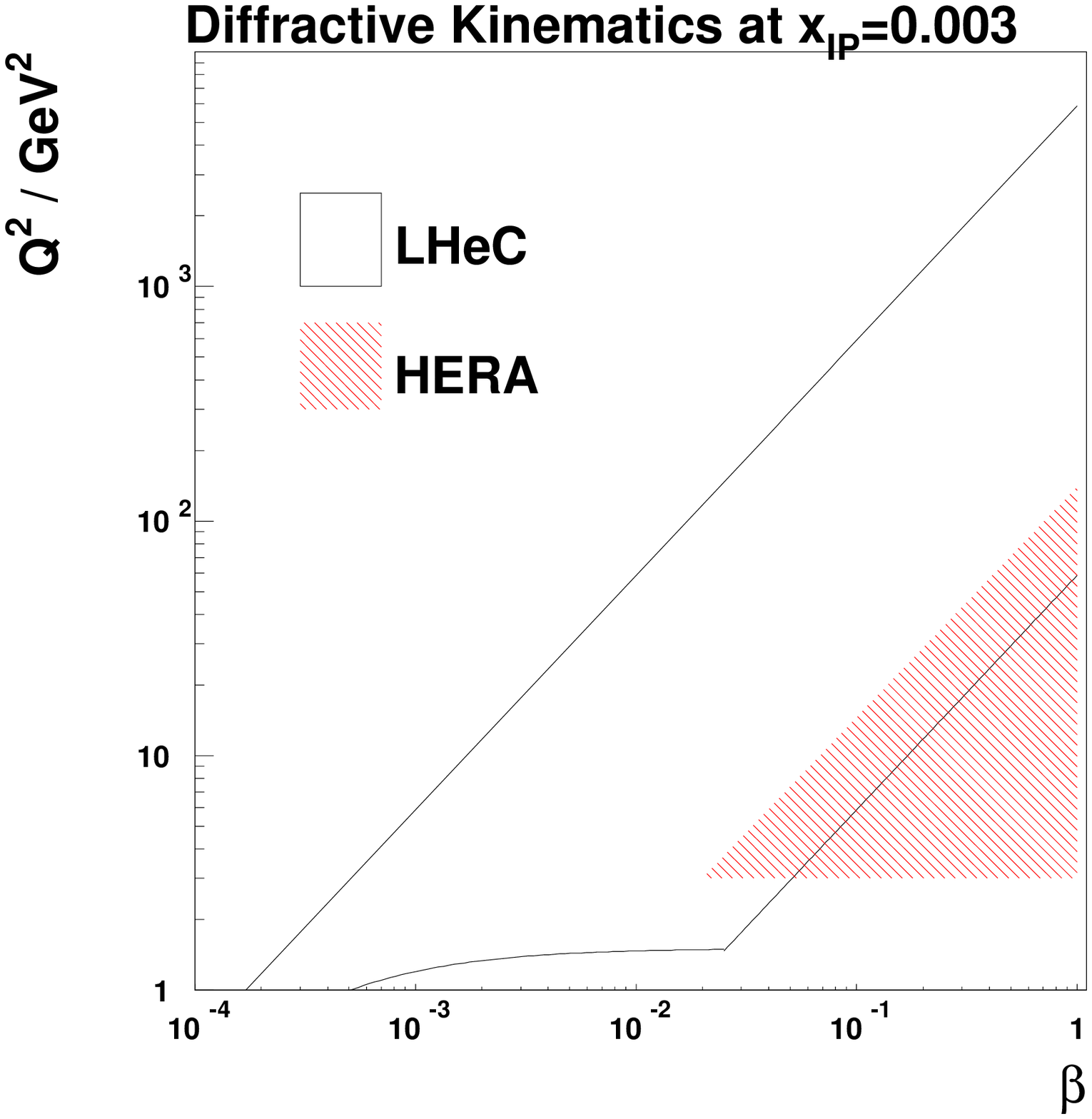,width=0.55\textwidth}}
  \put(-30,-10){\epsfig{file=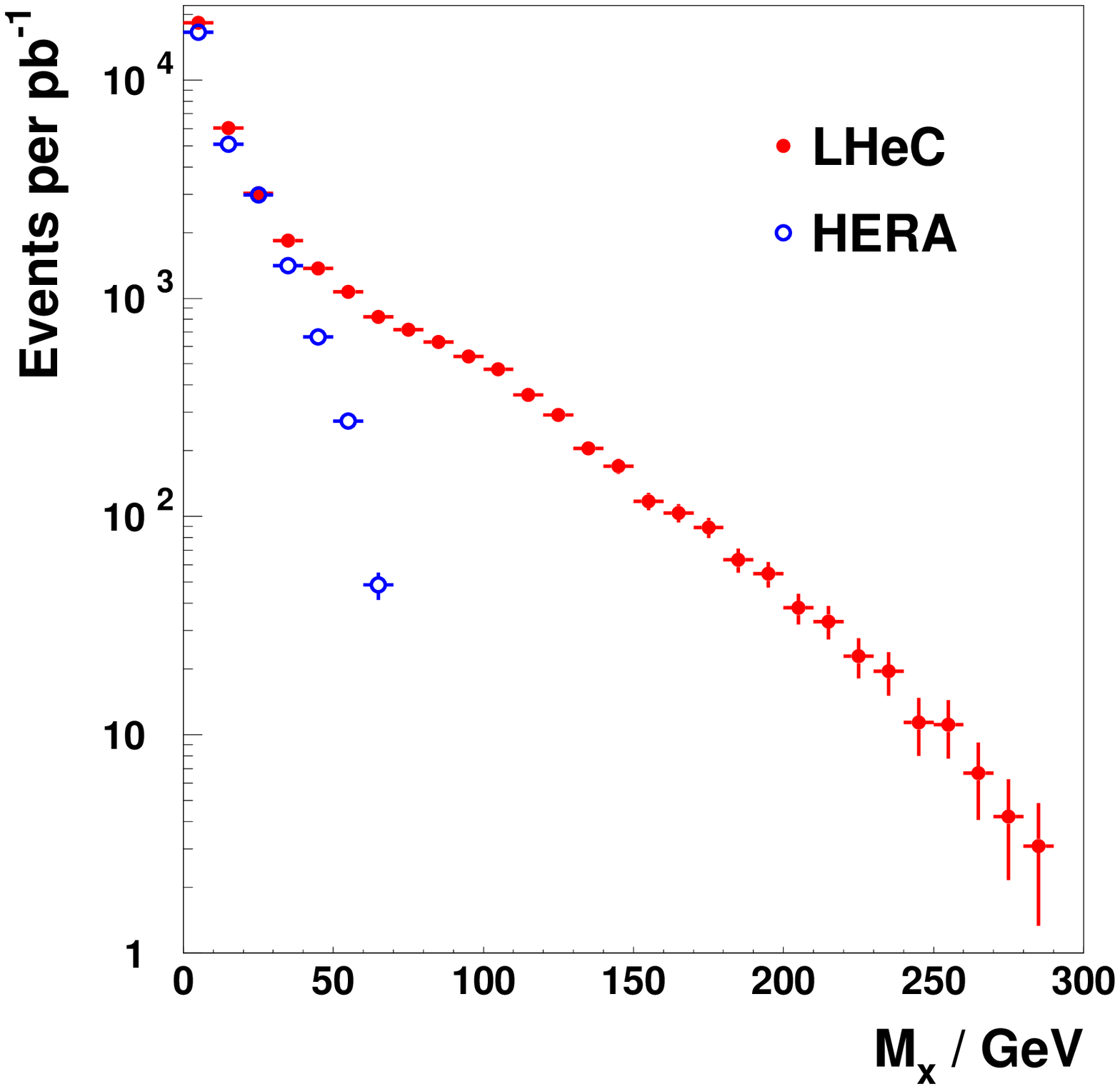,width=0.55\textwidth}}
  \put(-17,95){\Large{\bf (a)}}
  \put(97,102){\Large{\bf (b)}} 
  \put(27,42){\Large{\bf (c)}}
 \end{picture}
 \end{center}
 \caption{\it 
(a) Feynman diagram for diffractive DIS, indicating the kinematic
variables used in the text. (b) Comparison of the kinematic region in $\beta$
and $Q^2$ accessible at HERA and LHeC for an example $\xpom = 0.003$. The
LHeC region corresponds to $0.01 < y < 1$ and $\theta_e^\prime <179^\circ$.
Similar extensions are available
over the full range of $\xpom < 0.05$ which is relevant to diffraction.
(c) Monte Carlo simulation, comparing the diffractive mass distributions 
at HERA and the LHeC
available for studies of the hadronic final state with the selection
$Q^2 \geq 3$ GeV$^2$, $0.1 < y < 0.9$ and $\xpom < 0.05$.}
\label{difpic1}
\end{figure}
%
Furthermore, 
the contrast between non-diffractive  DIS, 
where the proton is rather violently broken up, and diffractive DIS where
the proton remains intact, offers a rare experimental window on the mechanism 
which confines quarks within hadrons.
%

One of the biggest successes of HERA has been the development of the
understanding of diffraction in QCD through the study of the diffractive
DIS process (Fig.~\ref{difpic1}a)  $ep \rightarrow eXp$ \cite{obsdiff}.
In this process, the proton remains intact, but loses a fraction $\xpom$ of
its longitudinal momentum in the form of some net colourless partonic system.
The virtual photon probes this colourless system, coupling to a quark with
a fractional momentum $\beta$, in much the same way as the whole proton is 
probed in inclusive DIS, and producing a diffractive system of mass 
$M_{\rm X}$. The new kinematic regions and very large luminosities 
possible at the LHeC make it ideal 
for  further precision study of hard diffraction. 
Figures~\ref{difpic1}b and~\ref{difpic1}c illustrate the 
dramatic improvements
in diffractive kinematic coverage at the LHeC compared with HERA. 

A general framework  to describe diffractive DIS is provided by a QCD factorisation 
theorem \cite{collins},  which allows ``diffractive parton distributions" (dpdfs) to
be defined. Such dpdfs have been extracted from the diffractive
structure function $F_2^D$ and its scaling violations at HERA \cite{h1f2d94}. 
However, unlike the case of inclusive
scattering, the dpdfs are only applicable to DIS processes.
Large and very poorly constrained 
``survival" factors are needed before dpdf's can be used
to predict $pp$ scattering processes. Testing the factorisation properties
and dpdfs extracted
at HERA has therefore only been possible thus far by predicting
final state DIS observables such as jet or charm cross sections. 
This approach is heavily
limited by the relatively low accessible $M_{\rm X}$ 
values at HERA (Fig.~\ref{difpic1}c),
which imply that jets can be studied only at uncomfortably low $p_t$ 
values ($\leq M_{\rm X} / 2$)
and that charm cross sections are frustratingly small. The much 
larger invariant masses accessible at the LHeC would circumvent these
problems and would open up new and complementary channels to study, such as
diffractive beauty production and diffractive electroweak gauge boson production. 

The only way of making significant further 
constraints and extensions to the dpdfs and studying their QCD evolution 
is at a higher energy facility such as the LHeC. 
With the high available luminosities, it will 
be possible to make measurements 
in the HERA $\beta$ range, but at larger $Q^2$ (Fig.~\ref{difpic1}b), 
which would provide a 
unique opportunity to test the applicability of DGLAP 
evolution to diffraction. 

Fig.~\ref{difpic1}b also shows
a substantial extension towards lower $\beta$
at the LHeC, similar to that available to lower $x$
in the inclusive case. As discussed above,
it is almost certain that new QCD physics associated with
the taming of the growth of the gluon density will be
observed in this region.
The diffractive exchange is often modelled as being derived from a pair 
of gluons, in which case any novel effects observed in the gluon density 
should be amplified in the diffraction case. 
Such measurements of the QCD structure of diffraction are fundamental,
amounting to direct experimental measurements of the partonic structure
of the vacuum and its quantification in terms of distribution functions.

%
\subsection{Final State  Physics}
\begin{figure}[htbp]
   \centering
     \centerline{\includegraphics[width=0.6\textwidth]{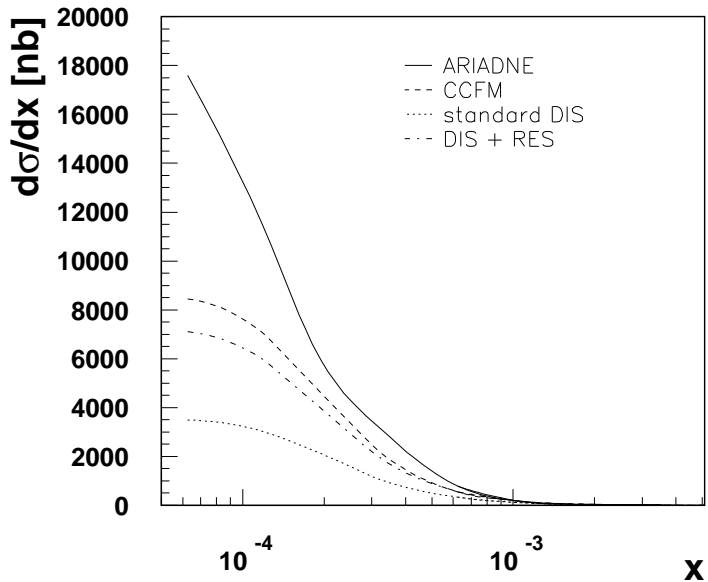}}
   \caption{\it Forward jet cross section as a function of $x$ predicted by  different models
    for the pattern of parton radiation, for $0.5 < p_T^2/Q^2 < 2$ and  minimum
    polar jet angle of $1^{\circ}$ as is envisaged for the low $x$ interaction region at the LHeC.
    The measurements at HERA are limited to $x \geq 2 \cdot 10^{-3}$,
    a range in which      the resolved photon effects may still mimic non-$k_t$
    ordered parton emission. The predictions of ARIADNE and the CCFM
    equation become largely different only at  lower $x$ values,  below the kinematic 
    range accessed by HERA, see \cite{therabook}. }
   \label{fig:fwdjets}
\end{figure}

The hadronic final state in deep-inelastic interactions probes 
the {\bf QCD cascade}, and therefore is a unique and important probe
of the chromodynamics of confinement. While predictions
based on the DGLAP evolution equations provide a good description
of the current inclusive DIS data, there are indications that
this approach might not be sufficient. For example, HERA data on
forward jet production at low $x$ show  that non $k_t$-ordered 
contributions play an important role. 
At the LHeC the kinematic domain is extended to much
lower $x$ values,  providing much greater sensitivity to the topological
features of the cascade. With LHeC data various approaches
for modeling QCD cascades could thus be
distinguished for the first time (Fig.\,\ref{fig:fwdjets})
and constraints on the unintegrated
parton densities \cite{unintegr} could be obtained.
In addition, the increase of the centre of mass energy with respect 
to HERA will lead to a much larger cross section for heavy flavor 
production. This would allow the $b$ and $c$ photoproduction cross sections
to be measured up to large transverse momentum, providing a much larger
level-arm when comparing the measurements to the QCD predictions of
the collinear or $k_t$-factorisation approach \cite{therabook}. 

%
%
The {\bf hadronic structure of the photon} and its interaction with hadronic matter have long
been something of a conundrum. The advent of both $e^+e^-$ and $ep$ colliders in the last two 
decades, and the huge increases in kinematic reach which have resulted,
have meant that it is now possible to see how QCD contributes to the interplay of photon
structure and photon interaction dynamics. As a result there is now a phenomenology of 
hadronic photoproduction ($ep$) and two-photon production ($e^+e^-$). 
A  distinction between ``resolved" and ``direct" interactions can be made by 
means of a Bjorken-$x$ like variable $x_{\gamma}$, which specifies the fraction of the 
photon momentum participating in the interaction. Resolved photon interactions
($x_{\gamma}<1$) depend  on the structure of the photon and direct photon interactions ($x_{\gamma}=1$) correspond to the  direct coupling of the photon in a hadronic interaction.

Photoproduction measurements at HERA of  hard QCD 
phenomena (final state transverse
momentum squared $p_T^2 > Q^2$) are used to extract structure functions of the photon
over a substantial 
range of Bjorken-$x$ and  of photon dimension
($\propto 1/Q$).  Features of photon structure, 
which derive from its underlying
$\gamma \rightarrow q \overline{q}$ splitting,
are now  established, and the subtle interplay between
 the hard scales ($Q^2, p_T^2$, and when involved a
heavy quark mass) are emerging as the photon virtuality, that is its size, changes. 

As a result, there now exists a rather comprehensive picture of photon-hadron coupling in 
terms of QCD for values of $x_{\gamma}$
down to about 0.1.  Furthermore, the ability to ``tune" the nature of the photon interaction 
between direct and
resolved (hadron-like)  now begins to throw light on issues in hard hadronic 
interactions  such as multiple parton interactions which would break QCD factorisation.
Measurements at LHeC will extend hadronic photoproduction 
measurements to $x_{\gamma} \simeq 0.01$
at  large momentum transfer, and thus will begin to reveal the properties of the gluon-dominated
region of photon structure \cite{therabook}, that is of a flavour singlet object
of variable dimension ($Q^2$). \\

{\bf Prompt photon production}, $ep \rightarrow \gamma X$, the
deep inelastic Compton scattering process, is also sensitive to QCD
and photon structure. In the foward region, $\eta_{\gamma}  > 0$,
this process is dominated by the reaction $gq \rightarrow \gamma q$
with a cross section which is an oder of magnitude higher at LHeC  than at HERA and
which extends to larger $p_t$ of the photon. Prompt photon production
at the LHeC will thus explore the gluon content of the photon \cite{maria}. \\

A new way of probing hadronic matter involves the physics of {\bf deeply virtual Compton
scattering (DVCS)}, $ep \rightarrow ep \gamma$ \cite{dvcs}. Here a
parton in the proton absorbs the virtual photon, emits a real photon
and the proton ground state is restored. In this process, two gluons
at low $x$ (at collider experiments) or two quarks
at larger $x$ (as in fixed target experiments),
carry different fractions of the initial proton momentum. 
The DVCS process thus measures generalised parton distributions (GPDs) \cite{gpds}
which depend on two momentum fractions $x$ and $\xi$, as well as
on $Q^2$ and the squared four-momentum transfer $t$ at the proton vertex.
The DVCS process interferes with Bethe-Heitler scattering, which
allows parton scattering amplitudes to be measured,  rather than just cross
sections in the form of standard pdfs. 

The huge increases in kinematic reach and luminosity with the  LHeC will make it 
possible to build an understanding of proton structure in terms of both transverse 
(impact parameter or $p_T$) and longitudinal (rapidity or Bjorken-$x$)
dimensions. Such proton ``tomography" \cite{belmuel} will thereby provide 
a deeper understanding  of proton structure. 
The possibility that DVCS at the LHeC may also be measured in charged current
 interactions would, as in inclusive scattering, 
shed light on the variation of flavour structure  across the
transverse profile of the proton.

GPDs can also be accessed in the production of vector mesons \cite{gpdvec}.
 Measurements of the $t$ and $W$ dependence of light ($\rho$) 
 and heavy ($J/\psi, \Upsilon$) vector
meson  production will be important for the determination of 
proton structure and saturation effects at low $x$ at the LHeC.

\section{Electron-Nucleus Scattering}
\subsection{Nuclei}
\begin{figure}[htbp]
   \centering
      \centerline{\includegraphics[width=0.55\textwidth]{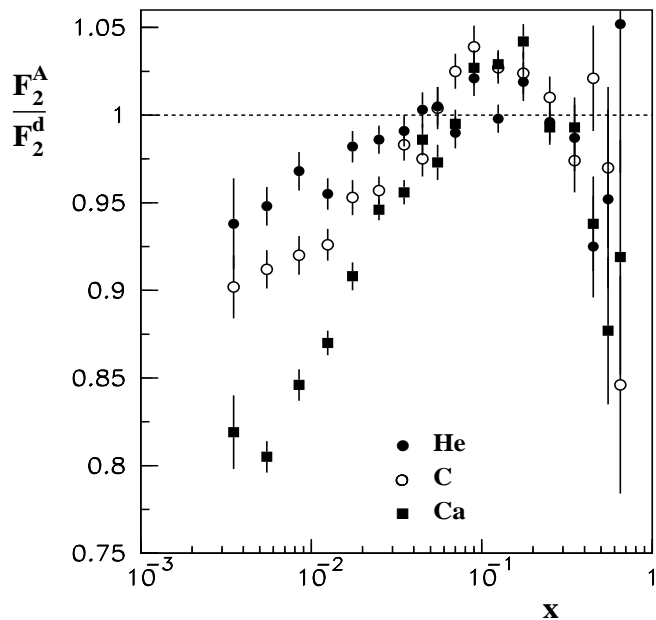}}
\caption{\it Ratio of $F_2$ structure functions in nuclei and the  deuteron as measured in fixed target
 muon scattering by the NMC Collaboration. At the LHeC this measurement can be extended 
 down to nearly $x = 10^{-6}$ in the deep inelastic region. Note that the $Q^2$ values at the
 smallest $x \simeq 3 \cdot 10^{-3}$ in this plot are below 1\,GeV$^2$ for the NMC but 
 reach a few 1000\,GeV$^2$ at the LHeC.}
   \label{fig:F2eA}
\end{figure}

Current knowledge of the partonic structure of nuclei is
limited to a very small range of $x$ and $Q^2$ (Fig.\,\ref{fig:F2eA}),
and is likely to remain so for the foreseeable future now
that there are no plans to develop HERA as an electron-ion collider  \cite{eAHERA,eDLoIH1}. 
The possibilities for $eA$ collisions in the future at
lower energy ($\sqrt{s} \leq 100$\,GeV) 
at JLab  \cite{ent} and at BNL \cite{eRHIC,milner} will not access the low-$x$ 
region of nuclear structure, which is crucial to a
full understanding of phase equilibria in nuclear matter, 
that is to the possible existence of a Quark-Gluon Plasma (QGP).
Realisation of the LHeC will in contrast put in place a unique tool in the hunt for 
an understanding of chromodynamic phase
equilibria. 

\begin{figure}[h]
   \centering
       \centerline{\includegraphics[width=0.4\textwidth]{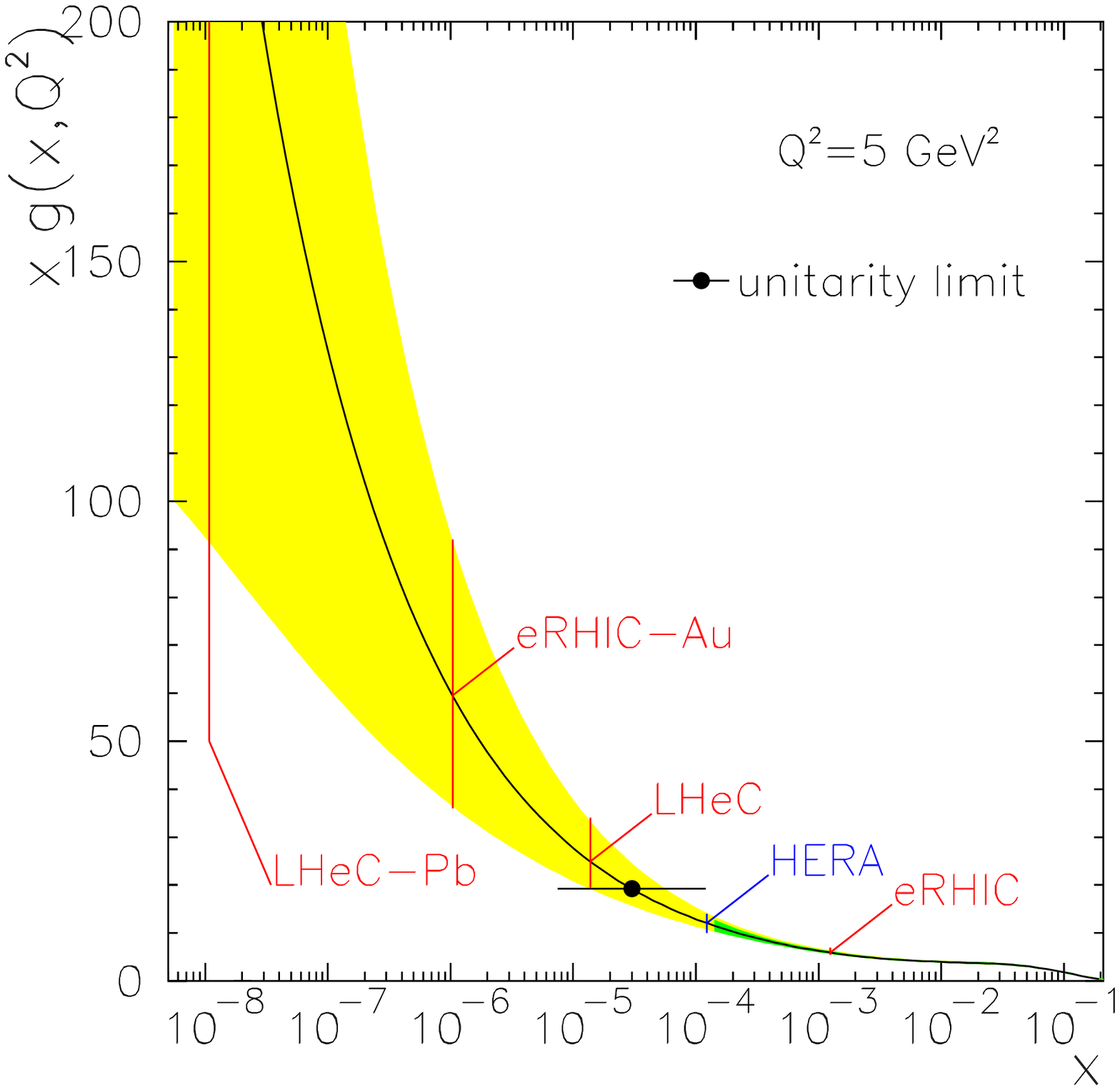},
    \includegraphics[width=0.4\textwidth]{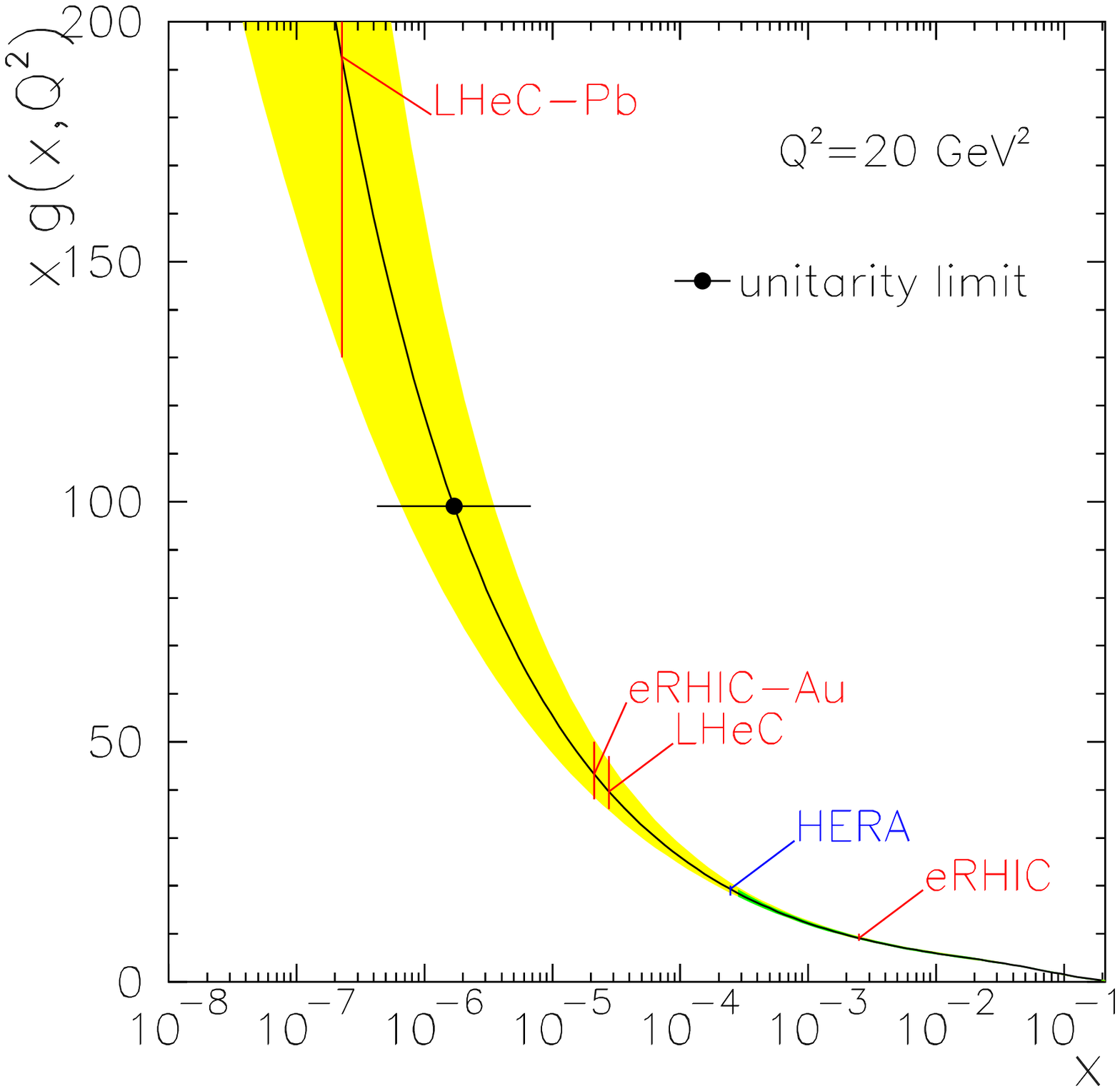}}
\caption{\it The gluon distribution as determined in an NLO DGLAP QCD analysis
 of H1 data extrapolated from the accessed region above $10^{-4}$ to much lower
values of Bjorken $x$, for $Q^2=5$\,GeV$^2$ (left) and $Q^2=20$\,GeV$^2$ (right).
The gluon distribution is expected not to rise strongly beyond the unitarity limit, which is estimated
to be of the order of $Q^2/\alpha_s(Q^2)$ (solid points). The extension of the kinematic range
by the LHeC leads beyond the unitarity limit at lower $Q^2$ in $ep$ scattering 
and also for larger $Q^2$ in $eA$ scattering,
 when the increase of the gluon density in nuclei $\propto A^{1/3}$
is taken into account, considering here $ePb$ scattering. 
Due to the much higher energy, the LHeC explores a region of much smaller  $x$
than is accessible by  eRHIC in $eAu$ scattering. A clear
observation of saturation 
in $ep$ scattering at the LHeC will be crucial in distinguishing saturation phenomena
from nuclear effects in $eA$ and $AA$ interactions.
}
   \label{fig:gluonex}
\end{figure}

The electron-nucleus scattering experiments
at the LHeC will  have a tremendous impact on the understanding
of  partonic matter in nuclei and on basic questions of QCD regarding the
(de)confinement of quarks \cite{deconf}, the existence of a saturated
gluon state,  the Colour Glass 
Condensate \cite{cgc}, and the relationship of nuclear Gribov-Glauber shadowing
to hard diffraction \cite{shado}.  
At  a certain scale, due to unitarity, the rise of the gluon
density  towards low $x$ has to be tamed, and may not exceed a limit
which is estimated to be roughly of the order of $Q^2/\alpha_s^2$ \cite{therabook}, with $Q^2$ in GeV$^2$ and a factor of order 1
or smaller which depends on the dipole type considered in such estimates \cite{mark}.
Such a limit,  illustrated in Fig.\ref{fig:gluonex}, may have been close to, but not
yet reached, at HERA. The limit is likely to be observed in $ep$ scattering at the LHeC
and its effects  in $eA$ scattering will be large, because  the gluon density 
in nuclei is amplified $\propto A^{1/3}$. 


In the regime of very high parton densities, the interaction of  small colour singlets with
hadronic targets becomes of comparable strength to the geometric cross section
$\pi R_A^2$ and thus approaches  the black disk limit or limit of 
opacity (absorption). In this limit of large, fixed $Q^2$ and decreasing $x$,
striking observations are thereby expected in $eA$ DIS   \cite{martinbb}:
\begin{itemize}
\item{
a  change in the  $Q^2,x$  dependence of the  structure function, 
$F_2 \rightarrow Q^2/\ln{(1/x)}$,}
\item{
an increased diffractive scattering component approaching  50\%
of the scattering cross section and dominated by   dijet production, and}
\item{
 a much  less rapid decrease of the  dependence of
 the cross sections  on $Q^2$ for  exclusive  vector meson production.}
 \end{itemize}
%

The energy densities achieved in an $AA$ interaction at the LHC are immense, 
and to fully explore the nature of the
interactions will require comparable data in $pA$, $pp$, and $eA$ collisions. 
LHeC and the LHC will thus constitute an
experimental tool unparalleled in the history of hadron physics in that nowhere else
has there ever been such a range of
possible measurements at such an energy scale. Given the importance attached to an
understanding of the existence and nature
of a QGP, and given the complexities of this understanding, 
establishing $eA$ physics with the LHeC is of primary value.
\subsection{Deuterons}
Electron-deuteron scattering complements $ep$ scattering in that it
makes possible accurate measurements of neutron structure
in the new kinematic range accessed by the LHeC.  In a collider
configuration, in which the hadron ``target"  has momentum
much larger than the lepton probe,
the spectator proton can be tagged and its momentum measured
with high resolution  \cite{eDLoIH1}. The resulting neutron structure
function data are then free of  nuclear corrections.
For the first time, since diffraction is related to shadowing,
one will be able to control the shadowing corrections at the per cent level 
of accuracy \cite{mark}. 

Accurate $en$ cross section measurements will resolve the quark 
flavour decomposition of the
sea, i.e.  via isospin symmetry, unfolding $\bar{u}$ from $\bar{d}$
contributions to the rise of $F_2^p \propto x(4 \bar{u} + \bar{d})$
towards low $x$, and, from the full set of $e^{\pm}p$ and $e^{\pm}n$ 
charged current cross section data, a full unfolding of the flavour content of the nucleon. 
For the study of the parton evolution with $Q^2$, the measurement
of $F_2^N=(F_2^p+F_2^n)/2$ is crucial \cite{forteprivate} since it 
disentangles the evolution of the non-singlet and the singlet contributions. 
This provides additional accuracy in the determination of the strong coupling 
constant \cite{botje}.
\section{Kinematics and Detector Requirements}
The kinematics of inclusive $ep$ scattering are determined from
the scattered electron with energy $E_e'$ and polar angle $\theta_e$ 
and from the hadronic final state of energy $E_h$ and scattering angle $\theta_h$.
The negative four-momentum            
transfer $Q^2$, the relative energy transfer $y$ and the Bjorken                
variable $x$ can be calculated from the scattered electron kinematics as                                                                                         
\begin{eqnarray} \label{QYe}                                                     
 Q^2_e & = & 4 E_e E_e' \cos^2(\frac{\theta_e}{2}) \nonumber\\                             
 y_e~~ & = & 1 - \frac{E_e'}{E_e} \sin^2(\frac{\theta_e}{2})                                      
\end{eqnarray}                                                                  
and from the hadronic final state kinematics as
\begin{eqnarray} \label{QYh}                                                                              
 Q^2_h & = & \frac{1}{1-y_h} \cdot E_h^2 \sin^2(\theta_h ) \nonumber\\                             
 y_h~~ & = & \frac{E_h}{E_e} \sin^2(\frac{\theta_h}{2}) ,                                         
\end{eqnarray}                                                                  
where $x$ is given as $Q^2 / sy$. Note that the angles                          
$\theta$ are defined between the directions of the                              
outgoing electron and the proton beam ($\theta_e$) and                        
between the hadronic final state and the proton beam ($\theta_h$).
%
The inclusive DIS cross section depends on two variables, besides the
centre-of mass energy squared $s=4E_eE_p=Q^2/xy$. 
The kinematic reconstruction in neutral current scattering  therefore
contains high
redundancy, which is one reason why DIS experiments at $ep$ colliders are
precise\footnote{An important example is the
calibration of the electromagnetic energy scale from the measurements
of the electron and the hadron scattering angles. This leads to
energy calibration accuracies for $E_e'$ at the per mil level at HERA, since in
a large part of the phase space, around $x=E_e/E_p$, the scattered
electron energy is approximately equal to the beam energy, $E_e' \simeq E_e$,
which causes a large ``kinematic  peak" in the scattered electron energy distribution.
The hadronic energy scale can be obtained from the transverse 
momentum balance in neutral current scattering, $p_t^e \simeq p_t^h$. It 
is determined to around 1\% at HERA.}.

Following Eq.\ref{QYh}, in charged current scattering the kinematics is
reconstructed from the transverse and longitudinal momenta and energy
of the final state particles according to \cite{blondel}
\begin{eqnarray} \label{QYjb}                                                                              
 Q^2_h & = & \frac{1}{1-y_h}\sum{p_t^2} \nonumber\\                             
 y_h~~ & = & \frac{1}{2E_e} \sum{(E-p_z)}.                                         
\end{eqnarray}   
%
%
\begin{figure}[htbp]
      \centerline{\includegraphics[width=0.75\textwidth]{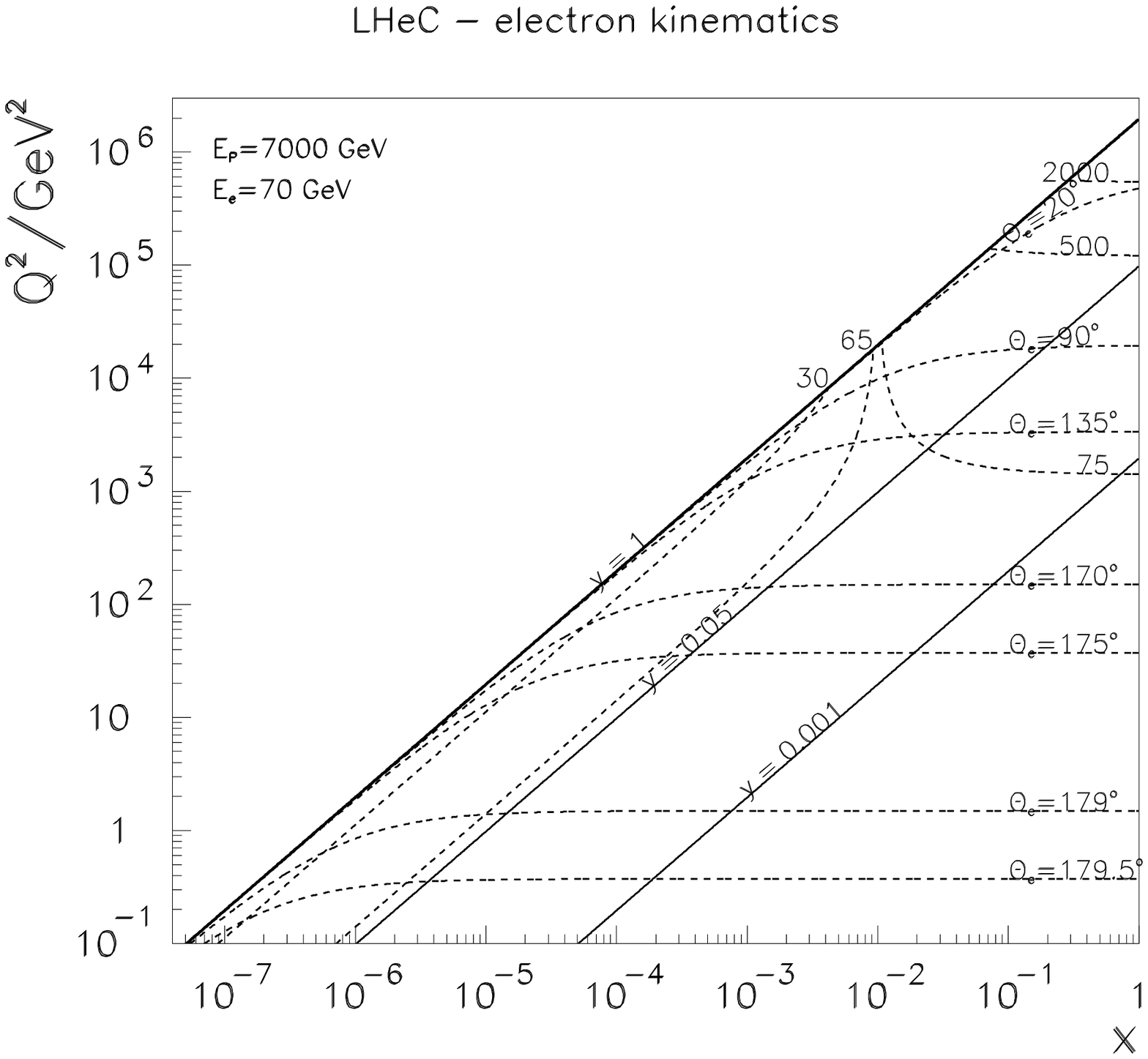}}
   \centerline{\includegraphics[width=0.75\textwidth]{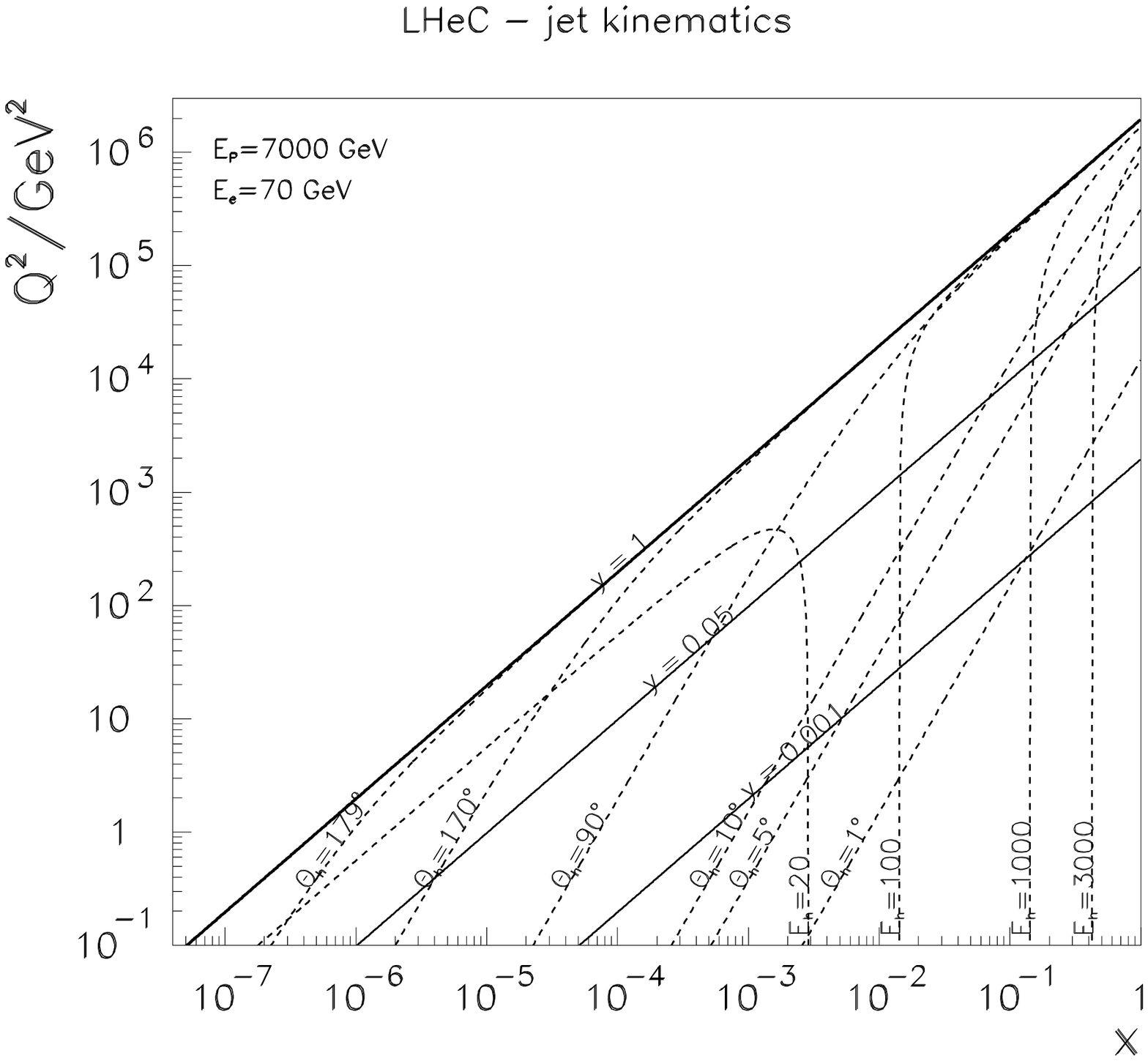}}
\caption{Kinematics of electron (top) and hadronic final state, ``jet" (bottom) detection at the LHeC.}
   \label{fig:kine}
\end{figure}

The kinematics of $ep$ scattering at the LHeC are illustrated in Fig.\,\ref{fig:kine}.  
Lines of constant energy and angle of the scattered electron                    
and the current jet are  located  in the $(Q^2,x)$ plane according to the                       
relations:                                                                      
\begin{eqnarray} \label{QX}                                                     
 Q^2(x,E_e) &=& s x ( 1 - E_e'/E_e  ) / [ 1 - x E_p/E_e  ] \nonumber\\           
 Q^2(x,E_j) &=& s x ( 1 - E_j/xE_p ) / [ 1 - E_e/(x E_p) ] \nonumber\\          
 Q^2~(x,\theta_e) &=& s x / [ 1 + x E_p \cot^2(\theta_e /2)/E_e ]                
                                                          \nonumber\\           
 Q^2~(x,\theta_j) &=& s x / [ 1 +   E_e\tan^2(\theta_j /2)/xE_p].                
\end{eqnarray}                                                                  
The  accessible kinematic range is limited by the
position and dimension of the focussing magnets, see Section 6.5.
Limitation of the scattered electron angle to 
a value $\theta_{e,min}$ defines, for not too small $x$,
a constant minimum $Q^2$ which is independent 
of $E_p$ and which is a linear function of $E_e^2$,                                                                    
$Q^2(x,\theta_{e,min}) \simeq [2 E_e \tan (\theta_e /2)]^2$.
Thus the LHeC, with a rather low electron beam energy
of 70\,GeV, allows the low $Q^2$ region
to be accessed more easily than THERA \cite{therabook}. 
As is illustrated in   Fig.\,\ref{fig:kine},
roughly a $1^{\circ}$ angular cut corresponds to a minimum $Q^2$
of about 1\,GeV$^2$.  To achieve  high luminosity,
focussing magnets need to be placed close to the interaction 
region (IR). In the present design an aperture limit of $10^{\circ}$
is used. Thus for  high luminosity operation, $Q^2$
reconstructed with electrons is larger than 100\,GeV$^2$.

For THERA a  ``backward" electron spectrometer was sketched \cite{therabook}
using a calorimeter to reconstruct high energies of several hundreds of GeV
and a silicon tracker telescope to measure the 
electron scattering angle down to $179.5^{\circ}$
and to identify the electron. The requirements at 
the LHeC are more modest, because $E_e'$ is lower and the minimum
$\theta_e$  is larger.
At low $x$ the kinematics are such that
the hadronic final state and the scattered electron energy
add up to approximately the electron beam energy, $E_e' + E_h \simeq E_e$. 
Final state physics at low $x$ requires access to the region 
within a few degrees of the beam pipe (Fig.\,\ref{fig:kine}).
Building a  spectrometer for low $x$
physics with a $179(1)^{\circ}$ degree acceptance limit for $\theta_e (\theta_h)$
is challenging.

The increased asymmetry between the electron and proton beam energies at 
LHeC means that  diffractively produced systems are strongly boosted in
the outgoing proton direction. The rapidity gap-based
diffractive selection, usually employed at HERA, will thus only be effective
over a wide $\xpom$ range
if there is substantial instrumentation at large rapidities 
(up to at least $\eta = 5$). Another intriguing possibility
is the use of Roman pots, which are planned for the LHC at the TOTEM
experiment \cite{totem} and which are proposed in connection 
with ATLAS and CMS in the FP420 project \cite{FP420}.

At larger $Q^2$ the electron is scattered to angles between
$170^{\circ}$ and $10^{\circ}$ with energies between 
about 10\,GeV and a few TeV.   
A minimum angle cut $\theta_{h,min}$ in the forward region, the direction
of the proton beam, excludes the large $x$ region
from the hadronic final state acceptance (Fig.\,\ref{fig:kine}), along a line
 $Q^2~(x,\theta_{h,min})\simeq [2 E_p x \tan^2(\theta_{h,min} /2)]^2$,
which is linear in the $\log Q^2$, $\log x$ plot and depends on 
$E_p$ only. Thus at                
$E_p=7$\,TeV the minimum $Q^2$ is roughly $(1000[500] x)^2$ at a minimum                  
angle of $10[5]^{\circ}$. Since this dependence is quadratic with $E_p$,
lowering the proton beam energy may be of interest for reaching 
the highest possible $x$.     As is seen in Fig.\,\ref{fig:kine}, the
high luminosity IR ($\theta_h < 170^{\circ}$) 
will have full coverage of the final state up to the largest
$y$ for $Q^2$ above 100\,GeV$^2$.
The hadronic final state energies vary between a few GeV  at low $x$
and  a few TeV at large $x$. The central detector
thus needs a strong solenoidal field, silicon and gaseous tracking
detectors and a calorimeter capable of reconstructing electromagnetic
and hadronic energy from a few tens of GeV up to a few TeV.
Accurate muon identification and momentum measurement as needed
for example for $b$ physics complement such an $ep$ detector.

Issues on the interaction region (IR) and
detector requirements were discussed previously at the LEP-LHC workshop
\cite{bartel,joel}, which was held prior  to the startup of HERA.
The consideration of the machine-detector interface which is described
in the next section is very preliminary. There has to be much
further consideration on the detector and on the IR,
 with direct experience of the first operational LHC beam,
of the optimisation for luminosity and physics scope than
has been achieved at this stage.
%
%
\clearpage
\section{Luminosity Prospects for the LHeC}
\subsection{Introduction}
In the following  the possibilities are explored for achieving a high 
luminosity of lepton-proton collisions using the existing LHC proton beam 
together with a new lepton storage ring in the LHC tunnel while maintaining 
the existing facility for proton-proton collisions. Earlier studies \cite{ferdi1,therabook} 
of possible lepton-proton colliders, which considered a configuration of a 
proton storage ring and a lepton linac, yielded a relatively small 
luminosity. Using the expressions derived in \cite{ferdi1,therabook} for the case 
of collisions between the LHC proton beam and a high energy lepton beam 
produced by a linear accelerator, the luminosity is given by 
\begin{equation}
\label{eq1}
L=4.8\times 10^{30}\cdot cm^{-2}s^{-1}\frac{N_p }{10^{11}}\cdot 
\frac{10^{-6}m}{\epsilon _{pN} }\cdot \frac{\gamma _p }{1066}\cdot 
\frac{10cm}{\beta _p^\ast }\cdot \frac{P_e }{22.6MW}\cdot \frac{250GeV}{E_e 
}
\end{equation}
Assuming the so-called ``ultimate'' LHC parameters \cite{ferdi3} with the number of 
protons per bunch $N_{p }$= 1.67 $\cdot $10$^{11}$, the normalized proton 
transverse emittances of \textit{$\epsilon $}$_{pN }= 3.75 \cdot 10^{-6}$m,
 a proton beta function 
\textit{$\beta $}$_{p}^*$ of 0.5\,m at the interaction point, a proton beam energy of 7\,TeV, 
(corresponding to a Lorentz-factor of \textit{$\gamma $}$_{p }$= 7460) and assuming a lepton 
energy of $E_{e}= $70 GeV and an available electron beam power of $P_{e}=50$\,MW 
one expects peak luminosities of  $L = 2.4 \cdot 10^{31}$cm$^{-2}$s$^{-1}$
for a ring-linac solution. The luminosity could be increased somewhat by 
reducing the proton emittance by using transverse cooling. The cooling of 
the bright and very high energy proton beams of the LHC is considered to be 
very challenging as experimental verification of proton beam cooling at 
these high energies has yet to be demonstrated. 
It thus seems worthwhile to reconsider 
the luminosity which could be achieved in a ring-ring configuration, using 
the LHC to accommodate an additional lepton storage ring, which would be 
installed above the existing superconducting magnets of the LHC and which would be 
brought  into collisions with the LHC proton beam at one interaction point. 

Similar estimates have been performed previously \cite{verdier,keil}. The present estimate 
makes explicit use of the available experience of high luminosity operation 
at the circular lepton-proton collider HERA to derive a realistic scaling of 
the achievable luminosity of a ring-ring-based lepton proton collider in the 
LHC tunnel. 

The aim of the present study is to investigate the parameters required for a 
realistic LHeC with a peak luminosity of $L = $10$^{33}$cm$^{-2}$s$^{-1}$ 
while 
maintaining the option for continued operations with proton-proton 
collisions. 

The present study does not purport to be exhaustive in that detailed designs 
of the lattice, investigations of beam stability, and conceptual designs of 
accelerator components have not been performed. However, no limitations 
associated with these issues are foreseen as similar systems have been 
demonstrated in existing accelerators or will be demonstrated soon. In 
particular there is no detailed design of the magnets in the lepton-proton 
interaction region (IR) or the layout of other accelerator hardware 
systems. The injector system is assumed to be identical to the LEP injector 
system. 


\subsection{ General Requirements}

The present study of the LHeC is based on the following general requirements 
and assumptions:
\begin{itemize}
\item A high luminosity of the order of $L=$10$^{33}$cm$^{-2}$s$^{-1}$ is assumed necessary to perform the physics programme in a reasonable amount of time.
\item The electron beam energy should be as high as possible to obtain a large centre of mass energy which exceeds the HERA centre of mass energy of 318\,GeV by a significant factor. The choice of the centre of mass energy is motivated by the physics and is a sensitive parameter for the achievable luminosity. For this study values of around $E_{cm }= $1.4\,TeV were explored corresponding to an electron beam energy of 70\,GeV given a proton beam energy of 7\,TeV. 
\item It appears to be highly desirable to find a solution which allows for proton-proton collisions in the interaction regions (IR) of the LHC denoted by IR\,1 and IR\,5 during lepton proton collision operation.  This implies that the proton beam parameters for LHeC are already predetermined by the design for the LHC. The parameters used in this study are listed in Tab.\,\ref{tabfer1} from \cite{ferdi3}. 
\item In order to provide sufficient space and acceptance for the experimental detector, a magnet-free space of at least 2.4\,m is required in the interaction region. The aperture of the first accelerator lattice elements should allow for a detector acceptance angle of 10$^{\circ}$.
\item The electric power consumption of the LHeC should be in a reasonable range compared to the CERN overall power consumption.
\end{itemize}

\begin{table}[h]
\caption{\it LHC  proton beam parameters used in this study}
\label{tabfer1}
\begin{center}
\begin{tabular}{|l|c|c|}
\hline
Proton Beam Energy                                & TeV           & 7                       \\
Circumference                                      & m             & 26658.883               \\
Number of Protons per bunch                        &$10^{11}$          &1.67                    \\
Normalized transverse emittance                    & $\mu $m            & 3.75                    \\
Bunch length                                       & cm            & 7.55                    \\
Bunch spacing                                      & ns            & 25    \\                  
\hline
\end{tabular}
\end{center}
\label{tablhc}
\end{table}%

\subsection{Overall Parameters of the LHeC}
The luminosity of a high energy proton-lepton circular collider, expressed 
in terms of the limiting parameters, is given by 
\begin{equation}
\label{eq2}
L=\frac{I_e \cdot N_p \cdot \gamma _p }{4\cdot \pi \cdot e\cdot \epsilon 
_{pN} \sqrt {\beta _{xp} \beta _{yp} } }.
\end{equation}
In this equation, $I_{e}$ is the total lepton beam current, $N_{p}$ is the 
number of protons per bunch, \textit{$\gamma $}$_{p}$ is the Lorentz factor of the protons which depends on
the proton beam energy, \textit{$\epsilon $}$_{pN}$ is the normalized proton 
transverse beam emittance (which is assumed to be equal in both planes) and\textit{ $\beta $}$_{xp}$ 
and \textit{$\beta $}$_{yp}$ are the values of the proton beam amplitude function 
($\beta $-function) at the interaction point (IP). Implicit in this formula 
are the requirements that the beam cross sections $\sigma $ of the proton 
and lepton beams at the IP be equal, $\sigma _{xp}=\sigma 
_{xe}$, $\sigma _{ye }=\sigma _{ye}$, which is consistent with 
experience at the SPS and HERA \cite{ferdi5}, and that the beam-beam tune shift 
parameters are in a tolerable range, which is particularly relevant for the 
lepton beam.

The proton beam parameters are assumed to be identical to the so-called 
``ultimate'' LHC proton beam parameters with the beam energy of 7 TeV, the 
number of protons per bunch $N_{p }$= 1.67$\cdot $10$^{11}$, and the 
normalized emittances of \textit{$\epsilon $}$_{pN}= 3.75\cdot10^{-6}$m. Given the 
aforementioned constraints, to achieve a luminosity of $L$ = 1 $\cdot 
$10$^{33}$cm$^{-2}$s$^{-1}$, only the total lepton beam current $I_{e}$ and 
the beta-functions at the IP remain as free parameters.

The maximum lepton beam current in a high energy lepton storage ring is 
limited by the available RF power. Other potential limitations, such as the 
dissipation of this power around the accelerator or localized heating due to 
higher-order modes, can be avoided by an appropriate design of accelerator 
components. The total output RF power is typically limited by cost: an RF 
output power of P$_{rf }$= 50\,MW (which is approximately 86{\%} of the LEP 
power consumption or 28{\%} of the 1999 CERN site power consumption of 
910\,GWh \cite{ferdi6} assuming a yearly operating time of 5000\,h) 
is considered here (somewhat arbitrarily) to be 
an upper limit. Since the basic ring parameters can only be varied slightly, 
the maximum lepton beam current is then determined by the lepton beam 
energy. A lepton which travels with an energy of $E_{e}$ around the ring with 
a bend radius of \textit{$\rho $ =}2997 m  incurs an energy loss per turn of 
\textit{e$\Delta $U = C}$_{g}E_{e}^{4 
}$\textit{/ (e$\rho )$} = 707 MeV where $C_{g }= $4/3\textit{ $\pi $r}$_{0}E_{0}^{-3 }= $8.821 $\cdot $10$^{-5}$ GeV$^{-3}$m \cite{ferdi7}. 
For a given available power, the total beam current is
\begin{equation}
I_e = 0.351 mA \cdot ( P_{rf} / MW ) \cdot  (100 GeV / E_e)^4,     
\label{eq:power}
\end{equation}
with $E_e= 70$\,GeV, $P_{rf} =50$\,MeV 
and assuming quasi loss-free superconducting RF resonators, the total beam current is $I_e=71$\,mA.

With the proton beam parameters, the RF power and the lepton beam energy fixed,  
it  remains  to  design  the  interaction region and a lepton accelerator lattice,
 which provides proton beta functions at the IP of at most 
 $\beta_{xp} \beta_{yp} = $1\,m$^2$,  
and  which allows to match the cross section of the lepton beam while providing sufficient dynamic aperture. Since the electron beam has naturally unequal emittances $\epsilon_x$, $\epsilon_y$, whereas the proton beam has naturally equal emittances in the transverse planes, the ratio of the proton beta function at the IP $r_{\beta}$ is chosen close to four ($r_{\beta} = 3.6$) to ease matching the beam cross sections at the IP. 

The lepton emittance determines the lepton beta functions at the IP since the beam 
cross sections need to be matched $\beta_e \epsilon_e=\beta_p \epsilon_p$.
The choice of the lepton beam  emittance is constrained to a window limited by dynamic and 
geometric aperture considerations, beam-beam tune shift limitations, by
the so-called hourglass effect due to the long proton bunches 
(the rms proton bunch length is $\sigma_p=7.5$\,cm), and by the necessity of small beam divergence at the IP to ease the beam separation and to avoid parasitic beam-beam effects.

The hourglass effect imposes a soft limit on the minimum beta function and the corresponding value of the emittance: 
$\beta_{xe} \gtrsim \sigma_p \rightarrow 
\epsilon_{xe} \lesssim \epsilon_{pN}\beta_{xp}/(\gamma_p\sigma_p) = 13$\,nm,
$\beta_{ye}\gtrsim \sigma_p \rightarrow 
\epsilon_{ye}\lesssim \epsilon_{pN}\beta_{yp}/(\gamma_p\sigma_p) = 3.8$\,nm.

The lepton emittance should further be sufficiently small to avoid a too large maximum beta-function in the low beta quadrupoles. The beta function at the IP, $\beta^*$, also determines the natural chromaticity contributions from the interaction region 
\begin{equation}
\label{eq:xsi}
\xi_{IP}= \frac{1}{\pi}\cdot \frac{D_{eff}}{\beta^*},
\end{equation}                                                    
where $D_{eff}$ is the effective distance of the low beta quadrupoles from the IP
and the chromaticity contributions for the two transverse planes
are assumed to be balanced.
If the chromaticity contributions from the IR exceed a certain fraction $f$ of the contributions from the arc 
\begin{equation} 
\label{eq:xsiarc}
\xi_{arc} = \frac{N_{cell} \cdot \tan{(\Phi_{cell}/2)}}{\pi}
\end{equation}  
\cite{ferdi8}  the dynamic aperture might become a critical issue \cite{ferdi9} according to experience at 
HERA, where the critical fraction is about $f=\xi_{IP}/\xi_{arc}=0.6$ \cite{ferdi10} (c.f. LEP: $f=0.5$
\cite{ferdi11}).   Here $N_{cell}$ is the number of FODO cells and $\Phi_{cell}$
is the betatron phase advance per FODO cell.

With these considerations  the limiting value of the horizontal lepton emittance
$\epsilon_{xe}$ for matched beam  cross sections is  given by 
\begin{equation}
\label{eq:horemlim}
\xi_{IP} \leq f \xi_{arc} \rightarrow \epsilon_{xe,ye} \leq 
\frac{f N_{cell} \cdot \tan{(\Phi_{cell}/2)} \cdot \beta_{xp,yp} \cdot \epsilon_{pN}}{D_{eff} \gamma_p}
\end{equation} 
We assume the fraction $f$ to be as large as the LEP value of 0.5 and assume that the chomaticity 
contributions from the non-interaction straights constitute $50\%$ of the off-arc chromaticity contributions so that $f=0.25$. 
With a LEP-like lattice and $N_{cell} = 290$, assuming an effective $D_{eff}  = 4$\,m, one obtains for the emittances limiting values of  $\epsilon_{xe} < 23$\,nm, $\epsilon_{ye} < 5$\,nm. 
As it turns out this small $\epsilon_{xe}$ value can just be achieved with the strongest focusing in a LEP-like structure with phase advances per FODO cell of 
108$^{\circ}$ and 90$^{\circ}$ in the horizontal and vertical plane respectively
\cite{ferdi11}. 

The other limitation on the emittance of the lepton beam is due to the beam-beam effect which is 
suffered by the lepton beam and which is parameterised by the linear tune shift parameter
\begin{equation}
\label{eq:tune}
\quad
\Delta \nu _{x,y}^e =\frac{\gamma _p r_0 N_p \beta _{xe,ye} }{2\pi \gamma _e 
\cdot \epsilon _{pN} \cdot \sqrt {\beta _{xp,yp} } (\sqrt {\beta _{xp} } 
+\sqrt {\beta _{yp} } )}=\frac{r_0 N_p \sqrt {\beta _{xp,yp} } }{2\pi \gamma 
_e \cdot \epsilon _{xe,ye} (\sqrt {\beta _{xp} } +\sqrt {\beta _{yp} } )}
\end{equation}
where $r_0$ is the classical electron radius. Tune shift parameters, which have been achieved at
HERA,  are $\Delta \nu^{He}_y=0.05$ and $\Delta \nu^{He}_x=0.03$ 
\cite{ferdi13,ferdi14}.  The corresponding limit on the lepton beam emittances depends on the ratio
of the proton beta functions at the  IP,  $r_{\beta}= \sqrt{\beta_{xp}/\beta_{yp}}$: 

%
\begin{eqnarray} \label{eq:emits}
 \epsilon _{xe} =\frac{r_0 N_p \cdot r_\beta }{2\pi \gamma _e \Delta \nu 
_x^{He} (1+r_\beta )} \\ 
 \epsilon _{ye} =\frac{r_0 N_p }{2\pi \gamma _e \Delta \nu _y^{He} 
(1+r_\beta )}  
\end{eqnarray}
%

For $r_{\beta} = 1.9$ one obtains emittance limits of $\epsilon_{xe}=12$\,nm and
$\epsilon_{ye} = 4$\,nm, respectively. However, since it will be difficult
to separate the beams sufficiently for the short bunch spacing of $25 ns$, a larger horizontal beam-beam tune shift will be taken into account which allows a smaller value of the horizontal emittance.  
Based on these considerations, the horizontal emittance for the lepton accelerator in LHeC 
is chosen to be $\epsilon_{xe}=  7.6$\,nm  while
the vertical emittance value is 3.8\,nm, assuming a coupling of 50\%.
The choice of the ratio of proton beta function at the IP  of $r_{\beta} = 3.6$ implies lepton $\beta$-functions  at  the
IP of $\beta_{xe}=12.7$\,cm and $\beta_{ye}=7$\,cm. In order to achieve these emittances without too strong focusing, the number of FODO cells in the arc of LHeC is increased from the LEP value of $290$ to $376$.

The main parameters which are implied by the above considerations are listed in 
Tab.\,\ref{tablhecmain}.

\begin{table}[htdp]
\caption{\it Main Parameters of the Lepton-Proton Collider}
\begin{center}
\begin{tabular}{|l|c|c|c|}
\hline
Property                                      &Unit       &      Leptons&      Protons  \\
\hline          
Beam Energies                             &   GeV    &          70      &     7000      \\
Total Beam Current                      &     mA     &          74    &     544    \\           
Number of Particles / bunch           &       $10^{10}$    &         1.04      &   17.0   \\                        
Horizontal Beam Emittance              &      nm           &    7.6  &       0.501     \\        
Vertical Beam Emittance                  &    nm           &    3.8          &  0.501       \\      
Horizontal $\beta$-functions  at IP          &      cm          &     12.7    &     180      \\         
Vertical $\beta$-function at the IP           &     cm          &     7.1     &    50          \\      
Energy loss per turn                     &    GeV         &     0.707    &   $ 6 \cdot10^{-6}  $    \\      
Radiated Energy                           &   MW          &     50         &  0.003      \\    
\hline   
Bunch frequency / bunch spacing   &      MHz  / ns &    \multicolumn{2}{|c|}{   40 / 25}          \\
Center of Mass Energy               &         GeV    &    \multicolumn{2}{|c|}{1400}         \\  
Luminosity                                   &   $10^{33}$cm$^{-2}$s$^{-1}$  & \multicolumn{2}{|c|}{1.1}   \\                    
\hline
\end{tabular}
\end{center}
\label{tablhecmain}
\end{table}
%
%
\subsection{Lepton Ring}
A detailed design of an electron ring has not been carried out. The 
parameters of the lepton storage ring of the lepton proton collider are 
chosen to be as close as possible to those of LEP \cite{ferdi12}, which was housed in 
the LHC tunnel.

The accelerator lattice consists of a FODO structure in the eight arcs with 
a betatron phase advance per cell of \textit{$\phi $}$_{cell }= 72^{\circ}$ in the horizontal and in the vertical plane. The cell length is $L_{cell}$ =  60.3 m and the bending radius is $\rho = 2997$\,m. The ring has a circumference of 
$C = $26658.876\,m. To further reduce the horizontal emittance, the damping 
distribution will be changed in addition to the strong focusing.
The ring will be operated with a relative shift of the RF 
frequency of \textit{$\Delta $f/f}$_{RF}$ = 2.5$\cdot $10$^{-7}$ which corresponds to an energy 
shift of $\Delta E/E= -1.8\cdot 10^{-3}.$ This changes the damping partition 
numbers in favour of horizontal damping to a value of $J_{x }$= 1.57 and 
yields a beam emittance \textit{$\epsilon $}$_{xe}$ of 7.6\,nm at 70\,GeV.
The corresponding lepton 
beam energy spread is \textit{($\delta $E/E)}$_{RMS}$ = 2.4$\cdot $10$^{-3}.$ 

The beam current of 70.7 mA is composed of 2800 bunches spaced by 25\,ns with $1.40 \cdot 10^{10}$ particles 
in each bunch. 

With the bend radius of 2997\,m, a 70\,GeV electron (positron) will suffer an 
energy loss of 706.8\,MeV per turn by the emission of synchrotron radiation with 
a critical energy of 254\,keV. The power loss of a beam of 70.7\,mA amounts to 
50\,MW. The linear power load to the vacuum system is 27\,W/cm. This is a large 
but feasible value. It exceeds the values of HERA (5.2\,MW dissipated over 
3818\,m yielding 13.5\,W/cm) by about a factor of two. The cooling system 
used at HERA is both conservative and conventional. The values for LEP were 
9\,W/cm with a critical energy of 522\,keV for a beam energy of 100\,GeV and 
a beam current of 12\,mA \cite{ferdi15}.

The RF system is assumed to consist of superconducting 1GHz RF resonators 
with an accelerating gradient of 12 MV/m. An active RF structure length of 
127\,m or 845 resonator cells is required to produce a total voltage of 1521 
MV. This voltage allows a synchronous phase of 28$^{\circ}$ and provides an RF 
bucket which accommodates ten times the RMS beam energy spread. This is 
expected to provide sufficient margin for a good beam lifetime. The 
parameters of the lepton ring are summarized in Tab\,\ref{tabfer3}.
\begin{figure}[htbp]
   \centerline{\includegraphics[width=0.8\textwidth]{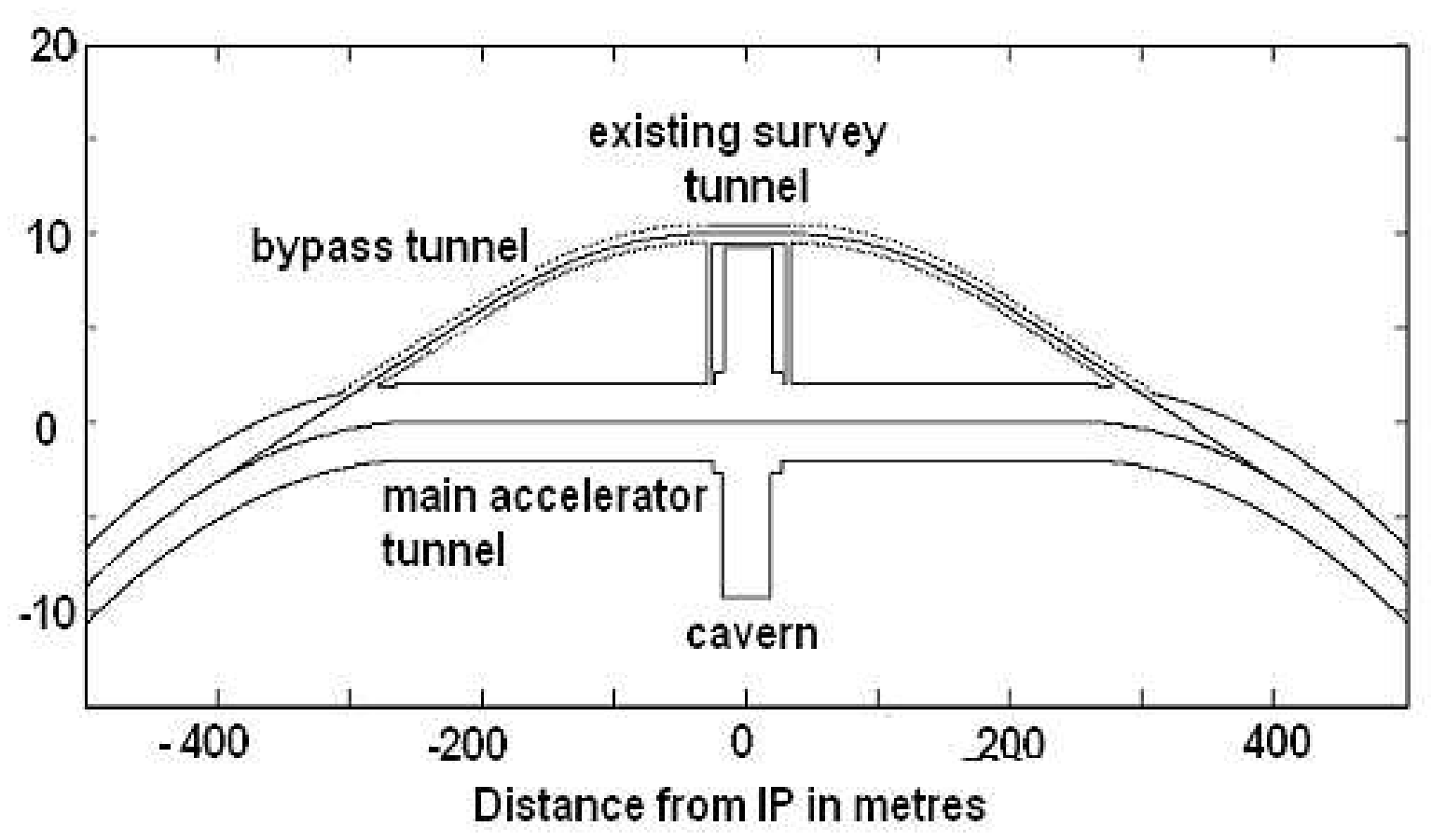}}
\caption{\it Top view (schematically) of straight section around IP1 (IP5) with an e-ring
 bypass around the experimental caverns of ATLAS and CMS. The scales are in meters.}
  \label{fig:bypass}
\end{figure}

Proven superconducting RF technology for lepton storage rings exists for 
only half the envisioned RF gradient \cite{ferdi16}. However, design efforts for 
superconducting RF for continuous operation (CW), for large beam currents 
and high bunch intensities are underway for ERL and CW-LINAC applications 
(see for example \cite{ferdi17}). The corresponding designs could serve as a 
starting point for an appropriate layout of a LHeC RF system. More 
development is required in the design of input couplers and higher order 
mode couplers. 


Geometrical considerations need to be taken into account in accommodating an 
additional lepton ring in the existing LHC tunnel. In the arc sections there 
appears to be sufficient space to place the lepton beam line above the LHC 
magnets. In the straight section, it may be cumbersome to accommodate the 
large RF systems of the lepton ring.

Since the ATLAS and CMS detectors are assumed to remain active at their 
locations when the lepton-proton collider is operated, a bypass must be 
provided round them. There exist survey tunnels 
which are  parallel with the LHC straight sections 1 and 5 which could be 
used for a bypass of the caverns which house the experimental detectors. They have 
a distance of about 10\,m from the LHC beam axis and a length of about 100\,m. 
Two about 250\,m long, up to 2\,m diameter connection tunnels would have to be 
drilled from the end of the arcs to connect to these tunnels (see sketch in 
Fig.\,\ref{fig:bypass}).

Lepton beam instabilities are not expected to become an important 
performance limitation given that the design single bunch currents are 
relatively modest. The expected total impedance, roughly estimated, is less 
than the impedance of LEP. The bunch population is much lower than 
that of LEP. 
For this reason, there should be no single bunch beam current limitations. A 
conventional active damper system could be used to damp coupled bunch 
oscillations if needed. 
\begin{figure}[htbp]
\centerline{\includegraphics[width=0.6\textwidth]{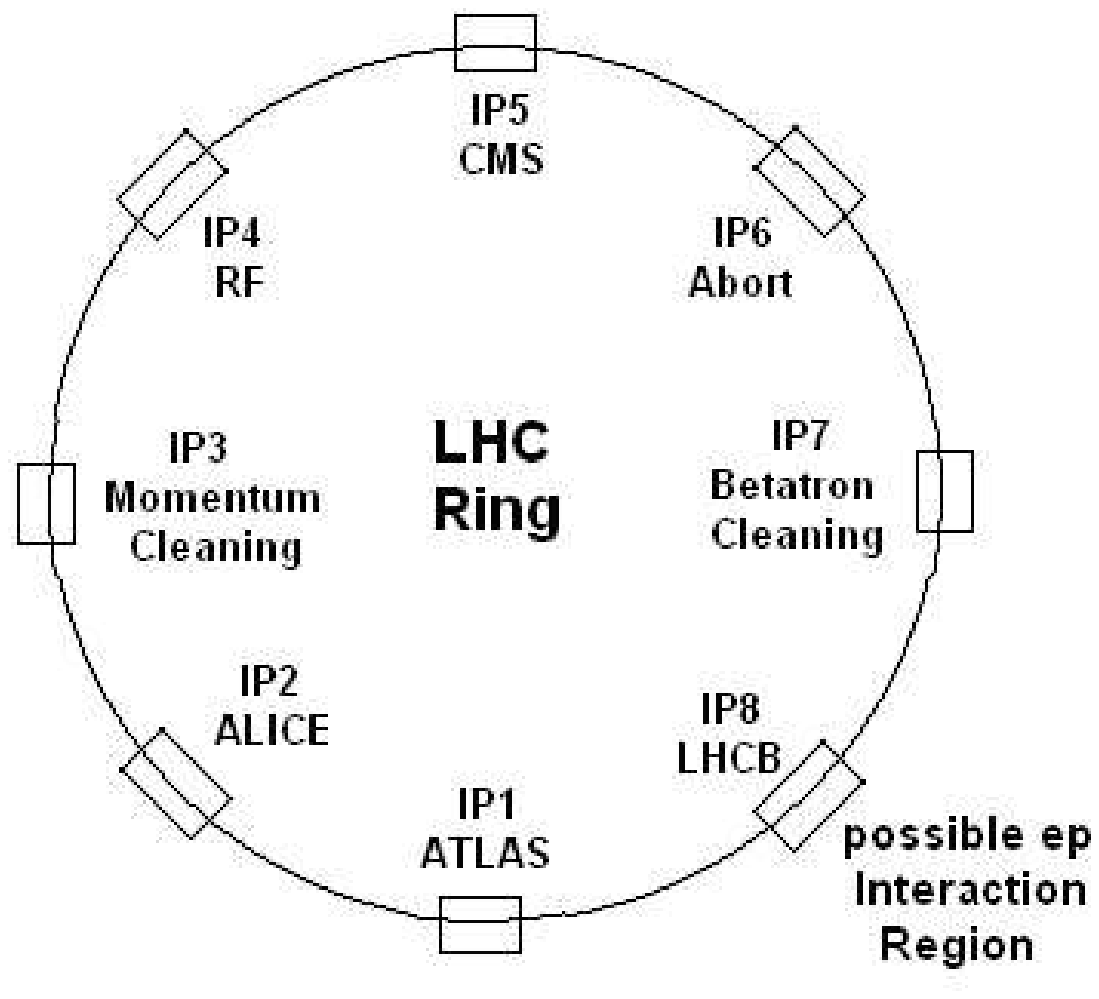}}
\caption{\it Possible location for the lepton-proton interaction region assuming
that the B physics programme is finished when the LHeC is installed.}
\label{fig:ip8}
\end{figure} 

\begin{table}[htbp] 
\caption{\it Parameters of the electron ring accelerator}
\label{tabfer3}
\begin{center}
\begin{tabular}{|p{243pt}|p{63pt}|p{122pt}|}
\hline
\multicolumn{3}{|p{428pt}|}{\textbf{Electron Ring Parameters}}  \\
\hline
Parameter& 
Unit& 
Value \\
\hline
Circumference C& 
m& 
26658.86 \\
\hline
Beam Energy E$_{e}$& 
GeV& 
70 \\
\hline
Arc Focusing& 
& 
FODO \\
\hline
Cell length L$_{c}$& 
m& 
60.3 \\
\hline
Bending radius $\rho $& 
m& 
2997 \\
\hline
Horizontal betatron Phase Adv./cell $\Delta \phi _{x}$& 
degree& 
72 \\
\hline
Vertical betatron Phase Adv./cell $\Delta \phi _{y}$& 
degree & 
72 \\
\hline
Number of FODO cells in the Arcs N$_{cell}$ & 
& 
376 \\
\hline
Arc Chromaticity (hor/vert.) $\xi _{x,y}$& 
& 
94/120 \\
\hline
Beam Current I$_{e}$& 
mA& 
70.7 \\
\hline
Bunch spacing $\tau _{b}$& 
ns& 
25 \\
\hline
Number of bunches n$_{b}$& 
& 
2800 \\
\hline
Number of particles per bunch N$_{e}$& 
10$^{10}$& 
1.4 \\
\hline
Momentum compaction factor $\alpha $& 
10$^{-4}$& 
1.34 \\
\hline
Horizontal beam emittance $\epsilon _{xe}$& 
nm& 
7.6 \\
\hline
Vertical beam emittance $\epsilon _{ye}$& 
nm& 
3.8 \\
\hline
RMS energy spread $\sigma _{e}$& 
10$^{-3}$& 
2.4 \\
\hline
RMS bunch length& 
mm& 
7.1 \\
\hline
Particle Radiation energy loss per turn eU$_{loss}$& 
MeV/turn& 
706.8 \\
\hline
Beam Power loss P$_{loss}$& 
MW& 
50 \\
\hline
Circumferencial Voltage U& 
MV& 
1521 \\
\hline
Synchronous Phase $\phi _{synch}$& 
degree& 
27 \\
\hline
RF frequency f$_{rf}$& 
MHz& 
1000 \\
\hline
Bucket height h$_{b}$& 
$\sigma _{e}$& 
8.4 \\
\hline
RF frequency shift& 
Hz& 
250 \\
\hline
Synchrotron frequency f$_{s}$& 
f$_{rev}$& 
0.191 \\
\hline
\end{tabular}
\label{tab:ering}
\end{center}
\end{table}

An open issue concerns the dynamic aperture of the lepton ring taking into 
account the contributions of the chromaticity in the interaction region. In 
this design study the beam emittance has been derived using scaling laws. Dynamic 
aperture studies will be necessary to assure sufficient stability. 
%
\subsection{Interaction Region} 
The electron-proton interaction region is taken to be installed in one of 
the existing LHC straight sections. The main detectors CMS and ATLAS are 
assumed to remain active during the operation of the lepton proton collider 
thus excluding the possibility of lepton-proton collisions in the straight 
sections around IP5 and IP1. The straight sections around IP7 and IP3 are 
needed for beam cleaning, which cannot be compromised for LHC proton 
operation. IP4 and IP6 are occupied by the proton RF and beam dump systems. 
This leaves only IP2 which is occupied by the ALICE detector and IP8 which 
is occupied by LHCb. A lepton-proton physics programme at the 
LHC would be unlikely to begin before the $B$-physics programme is completed. 
In this case one may envisage an interaction region 
for the LHeC around IP8 (Fig.\,\ref{fig:ip8}).


In order to provide small beta, the low beta quadrupole magnets have to be 
reasonably close to the IP. This requires a quick separation of the two 
beams outside the collision region. Separation by strong magnetic fields 
produces high power synchrotron radiation which is problematic  
because of experimental backgrounds and heating of the vacuum system. The 
alternative, a large crossing angle, reduces the luminosity. The IR design 
has to compromise between these difficulties. 

In order to allow for simultaneous lepton-proton and proton-proton 
operation, the following collision scheme is proposed. The lepton bunch 
spacing is 25\,ns, which is  the nominal LHC proton bunch spacing. 
The two proton beams traverse the low beta quadrupoles of the lepton 
beam and  the IP with a relative angle of $\simeq$ 6\,mrad in the 
horizontal plane. The non-colliding proton bunch is vertically displaced, 
thus providing a vertical separation to the lepton and colliding proton 
beam. The lepton beam passes the IP with an angle of 
\textit{$\theta $}$_{c }$= 2\,mrad 
in the horizontal plane relative to the colliding proton beam. 

The interaction region design is thus based on a moderate horizontal crossing 
angle of 2 mrad. This implies that the interaction geometry is 
anti-symmetric around the interaction point which leads to a narrower 
synchrotron radiation fan. This helps with the collimation scheme. The 
crossing angle avoids large synchrotron radiation power dissipation in 
the cold beam pipes of the proton low-beta quadrupoles. To regain the 
associated luminosity reduction by a factor of 3.5 caused by the crossing 
angle, so-called crab-crossing is assumed, which requires a tilt of the 
proton bunches around a vertical axis. This is discussed further below. 

At the first parasitic electron-proton collision point at 3.72\,m from the IP, the lepton 
and the proton beam are separated by about 7.8\,mm or 8.4\,$\sigma $ of the 
horizontal lepton beam size. This separation is considered sufficient to 
avoid potentially harmful interactions due to so-called parasitic crossings.

The length of the magnet-free space for the detector beam-pipe is assumed to 
be $\pm$1.2 m. The space requirement for the first low beta quadrupole of the 
lepton beam is 400\,mm in diameter. This provides a detector acceptance angle 
of 9.4$^{\circ}$ around the longitudinal axis. 

The low beta quadrupole magnets for both protons and leptons are assumed to 
be superconducting. The focusing of the electron beam could be accomplished 
using a low-beta quadrupole triplet located at a distance of 1.2\,m from the 
IP followed by an 50\,m long drift without focusing elements. The 
superconducting low beta quadrupole magnets have a gradient of 93\,T/m, 
lengths of 96\,cm, 204\,cm and 114\,cm, respectively, and they provide half 
apertures of 30\,mm, 40\,mm and 55\,mm. 

The beam separation is accomplished as follows. The low beta quadruples are 
displaced by 0.25\,mm from the beam axis which constitutes a 4.15\,m long soft 
separating magnetic field which provides a deflection of 0.4\,mrad. The low beta 
triplet is followed by a long soft separator dipole magnet with a field of 
only 0.023\,T and a length of 15\,m. The separation provided by this 
magnet is 1.5\,mrad. This arrangement, together with the crossing angle of 
2\,mrad provides a beam separation of 62\,mm at 22\,m, where the first low beta 
magnet for the proton can be located. 

These magnets can be built using standard superconducting and normal conducting 
magnet technology. The larger aperture of the third quadrupole magnet is 
needed to provide sufficient aperture for the second, non-colliding, proton beam. 

The colliding proton beam passes off-centre through the lepton low beta 
triplet before entering the first magnet of the proton low-beta triplet at 
22\,m. The first magnet must ensure an integrated strength of 1564\,T, the 
second lens requires 2070\,T and the third lens has an integrated gradient of 
1058\,T. The total length is about 45\,m. Gradients of the order of 115\,T/m 
are required and an aperture of 15\,mm is needed. All quadrupole 
magnets are assumed to have a cold beam pipe and cold iron to provide flux 
return. 

The first of these magnets is a septum half quadrupole as in the case of the 
HERA interaction region. The width of the septum is 12mm. The pole tip field 
is 2.79 T. The mirror-yoke is made of high quality magnetic steel. The 
large return yoke of the quadrupole magnets of the second proton low-beta 
lens accommodates a lepton beam pipe at room temperature which is
to be cooled. The 
separation at the third group of magnets of the proton low beta triplet is 
sufficiently large, for the lepton to pass outside the cryostat.
 
After the low-beta triplet of the proton beam, the lepton beam is deflected 
by 5\,mrad vertically using two 10m long dipoles which provide kicks of 1\,mrad 
and 4\,mrad respectively. At the end of the straight section, the lepton beam 
is about 80\,cm above the proton beam axis. The proton beam orbits  
diverge to 80\,cm separation. After the vertical deflection of the lepton 
beam, the protons are matched to their arc trajectory with three ten meter 
long superconducting dipole magnets. 

The non-colliding proton beam is assumed to cross the colliding proton beam 
at the IP with an angle of 6 mrad and a sufficient vertical separation. 
It is assumed to bypass the proton low-beta triplet. No attempts have been 
made to produce a layout of the lattice for this beam in the IR because it 
seems non-problematic.

This arrangement allows for a beta function of the lepton beam at the IP of 
$\beta_{xe}=$12\,cm and $\beta_{ye}=$ 7\,cm. The peak values of the vertical and horizontal 
lepton beta functions amount to 906\,m and 269\,m. The critical chromaticity 
contributions from the IR are quite modest with values of $\xi_{xIR}$= -7 and 
$\xi_{yIR }$= -39, which is about 21{\%} of the contributions of the arc. 
Correction of this chromaticity is not expected to result in any significant 
reduction of dynamic aperture. The horizontal and vertical peak beta 
functions of the proton beam are 2668\,m and 2637\,m, respectively,
and so are considerably smaller 
than at the proton-proton interaction points and are assumed to be 
non-problematic. The IR chromaticity contributions of the proton beam are 
$\xi_{pi r}$= -8.

The low beta quadrupole lenses provide sufficient aperture of at least 13.5 
times the RMS beam size in the case of protons and at least 20 times the RMS 
beam size in the case of leptons. According to experience at HERA \cite{ferdi13}, this 
would be sufficient for avoiding beam lifetime reductions or poor 
backgrounds.

\begin{figure}[htbp]
\centerline{\includegraphics[width=0.85\textwidth,angle=0.]{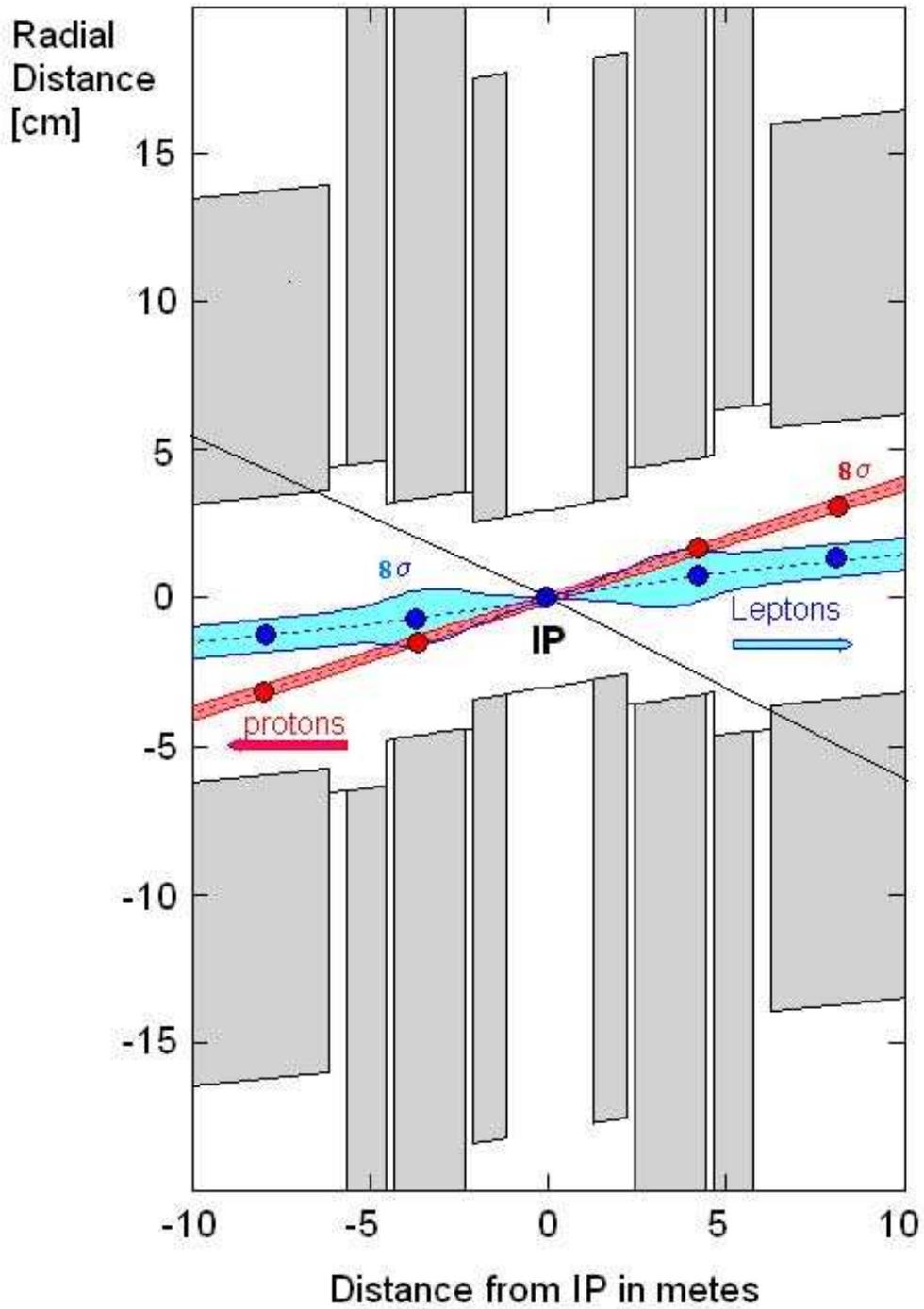}}
\caption{\it Top view (schematically) around the IP. The 8 $\sigma $ beam envelope 
of the proton and the lepton beams are shown. The non-colliding 
proton beam (dashed line) envelope is not shown.}
\label{fig:topir} 
\end{figure}

\begin{figure}[h]
\centerline{\includegraphics[width=\textwidth,angle=0.]{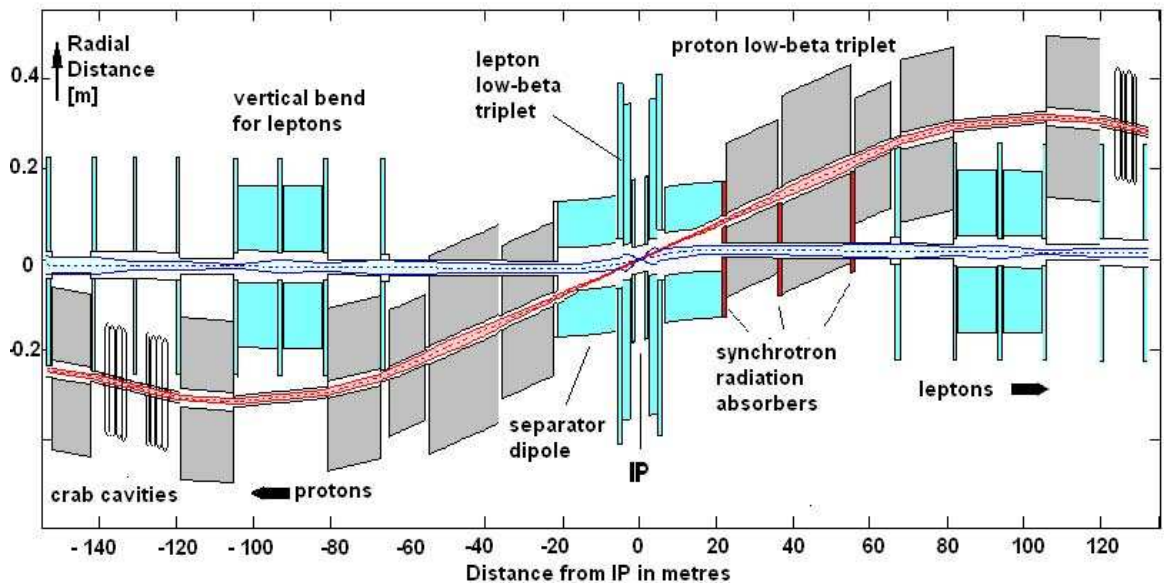}}
 \caption{\it IR conceptional overview (top view)}
 \label{fig:irover}
\end{figure}

\begin{figure}[h]
\centerline{\includegraphics[width=.7\textwidth]{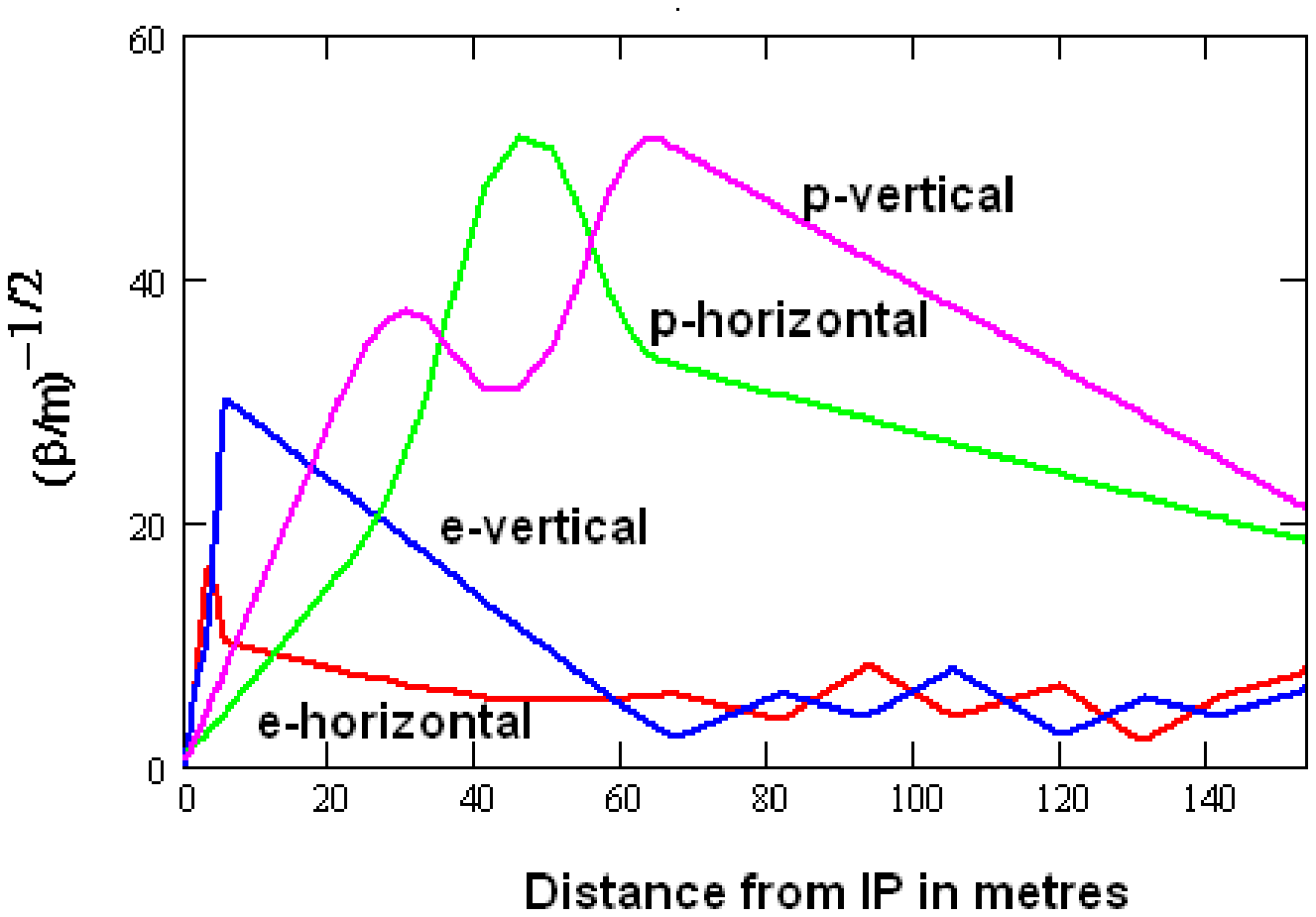}}
  \caption{\it Lattice functions for the proton and lepton beam in the lepton-proton IR.}
 \label{fig:latfun}
\end{figure}

The interaction region is sketched in Fig.\,\ref{fig:topir} and  Fig.\,\ref{fig:irover}.
The lattice functions  are plotted in Fig.\,\ref{fig:latfun}. The interaction region parameters
are shown in Table\,\ref{tab:IR}.
%

\begin{table}[h]
\caption{\it Parameters of the high luminosity interaction region for the LHeC}
\begin{center}
\begin{tabular}{|p{189pt}|p{49pt}|p{74pt}|p{90pt}|}
\hline
\multicolumn{4}{|p{403pt}|}{\textbf{Interaction region parameters}}  \\
\hline
\textbf{property}& 
\textbf{unit}& 
\textbf{leptons}& 
\textbf{protons} \\
\hline
Horizontal Beta function at IP& 
cm& 
12.7& 
180 \\
\hline
Vertical beta function at IP& 
cm& 
7.07& 
50 \\
\hline
Horizontal IR Chromaticity& 
& 
-7.5& 
-7.9 \\
\hline
Vertical IR chromaticity& 
& 
-29.7& 
-7.7 \\
\hline
Maximum horizontal Beta & 
m& 
131.7& 
2279 \\
\hline
Maximum vertical Beta & 
m& 
704.4& 
2161 \\
\hline
Minimum of available Aperture& 
$\sigma _{x}$& 
16& 
13.5 \\
\hline
Low beta quadrupole gradient& 
T/m& 
93.3& 
115 \\
\hline
Separation dipole field& 
T& 
0.033& 
- \\
\hline
Sychrotron Radiation Power& 
kW& 
9.1& 
- \\
\hline
Low beta quadrupole length& 
m& 
.96/2.43/1.14& 
16.5/18.6/11 \\
\hline
Low beta quadrupole apertures& 
mm& 
30/40/50& 
12/15/15 \\
\hline
Distance of first quadruple from IP& 
m& 
1.2& 
22 \\
\hline
Detector Acceptance Polar Angle& 
degree& 
\multicolumn{2}{|c|}{9.4}  \\
\hline
Crossing Angle& 
mrad& 
\multicolumn{2}{|c|}{2}  \\
\hline
\end{tabular}
\label{tab:IR}
\end{center}
\end{table}

The crossing angle of $\theta_c$ = 2\,mrad  reduces the luminosity by a 
factor of 3.5. One can recover from this reduction and can avoid any 
detrimental effects from the finite crossing angle if the proton beam is 
tilted around a vertical axis by $\theta_c/2$. This can be accomplished by RF 
resonators with a transverse deflecting field, so-called crab cavities. In 
order to obtain an almost linear kick with the distance from the centre of 
the proton bunch over the entire length of the bunch \textit{($\sigma $}$_{p }$= 75\,mm 
is the RMS  proton bunch length), the RF wave length has to be much larger than the 
bunch length. An RF frequency of 500\,MHz fulfils this requirement (the 
non-linearity at \textit{ $\sigma $}$_{p}$ from the centre amounts to 9\,\%).
With the beta function of 708\,m 
at the properly placed crab cavities, the required transverse kick at 
\textit{ $\sigma $}$_{p}$ from the centre is $\kappa = 2.1~\mu $rad. The RF phase angle corresponding to one $\sigma _{p}$ is 42$^{\circ}$. This means that the RF structure has to provide an integrated peak field \textit{of U}$_{crab }$\textit{= $\kappa $E}$_{p }= $20.8\,MV.With a gradient of $G_{crab }= $3.4\,MV/m, the two crab cavity systems 
around the IP must have an active length of 6.1m each. This could be 
accomplished by a three 6-cell resonators. A 12 $m$ long superconducting 
500 MH$z$ RF structure is not expected to affect the LHC impedance budget 
considerably if the higher order modes are damped below a Q$_{HOM}$ of 1000. 
Shorting of the resonators during p-injection and ramping may be required. 

The two crab cavities can be installed in the IR at the point of maximum 
separation of the proton orbits between 120\,m and 140\,m distance from the 
IR. The horizontal phase advance at 120\,m from the IP amounts to 25.6$^{\circ}$. 
Thus a third resonator may be needed to provide exact closure of the 
crab-orbit bump. In order to avoid blow-up of the proton beam, the 
tolerances on the RF system and the required precision of the field 
amplitudes in the presence of high beam loading are very demanding.
No attempt has been undertaken so far to specify an RF design for the 
crab-system. 

The horizontal amplitude of the 'crabbed' trajectories amounts to 0.3mm for particles
which are one standard deviation away from the centre. The corresponding increase 
in aperture requirement is negligible.
%

For the low $x$, low $Q^2$ part of the physics programme, the detector needs 
acceptance to polar angles of 1$^{\circ}$ with respect to the detector axis. This 
cannot be provided by the high luminosity IR as presented here. Since this 
part of the physics programme needs substantially less luminosity ($\sim$ 
one order of magnitude), the low beta magnets can be placed further away from 
the IP (by 3m). The beta function at the IP the increases by a factor of 4 and a 
larger crossing angle can be tolerated. The low beta quadrupole triplet for 
the lepton beam needs a larger aperture to accommodate all the beams. No 
further effort has yet been made with the low $x$ IR configuration, which is 
not expected to represent a design challenge.
%
\subsection{Synchrotron Radiation}
The synchrotron radiation produced by the beam separation magnets, by 
misalignment of the quadrupole magnets, and by closed orbit distortions in 
the IR, must be minimized by using beam-based alignment, by careful orbit 
control and by an effective collimation system. These issues have been 
discussed in the preparation and  commissioning of the HERA luminosity 
upgrade \cite{ferdi18,ferdi19}. The details will not be discussed here but it is important 
to notice that the synchrotron radiation produced in the proposed IR is 
somewhat less critical than in the case of HERA.

\begin{figure}[htbp]
%
\centerline{\includegraphics[width=0.7\textwidth,angle=0.]{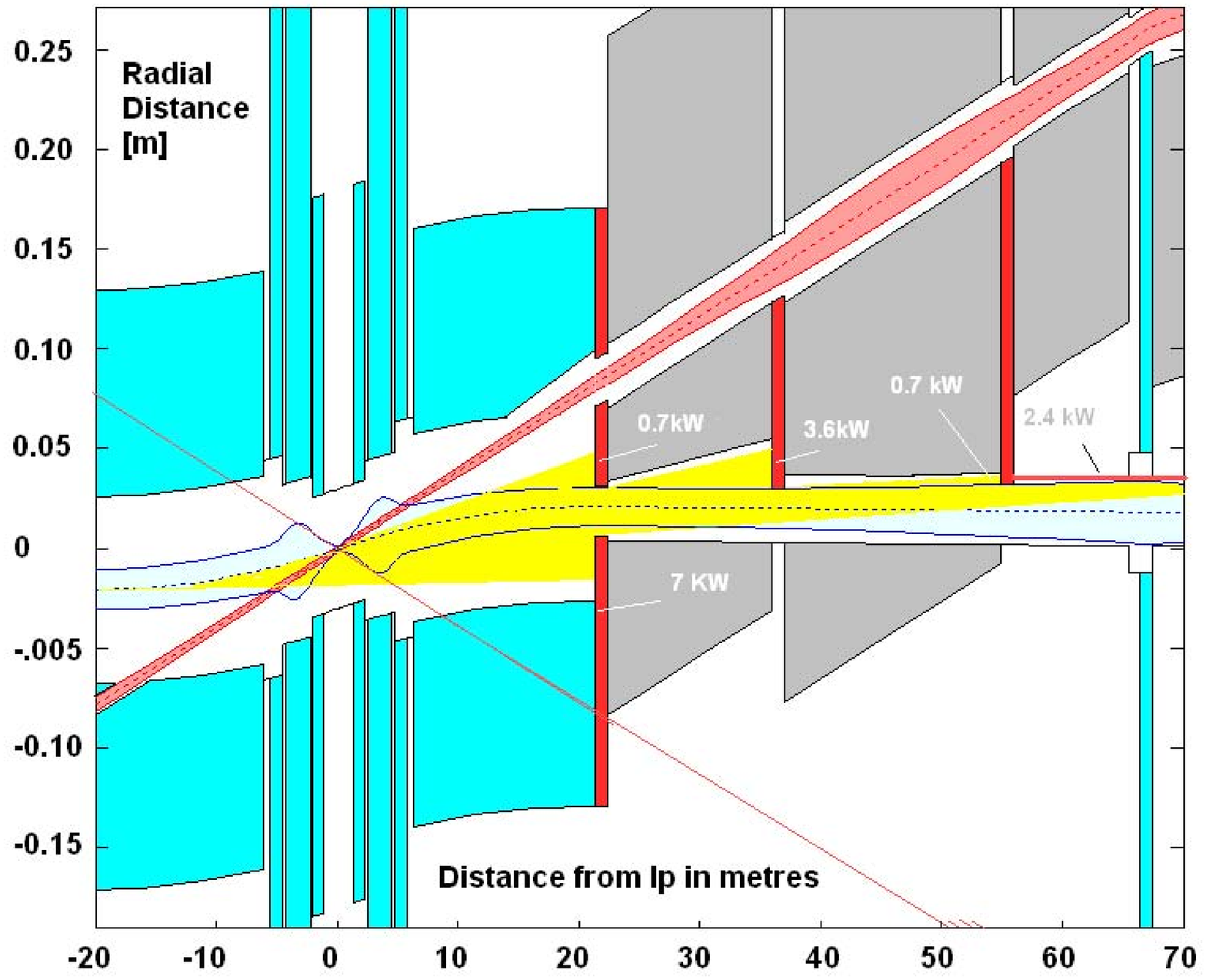}}
\caption{\it Deposition of Synchrotron radiation: Most of the power of 9.1\,kW in the synchrotron radiation fan (shown in yellow) is absorbed at the collimator-absorber at 22\,m from the IP (dark/red). No direct radiation is deposited inside the superconducting magnets.}
\label{fig:syn}
\end{figure}

The soft bending field with a bend radius of \textit{$\rho $}$_{ir}= $10,010\,m produces a 
radiation of 9.1\,kW and a critical energy of 76\,keV, which is transported 
within a fan of 1.9\,mrad horizontal opening angle and a vertical RMS 
thickness of $\approx $2\,mm. The narrow opening angle of the fan is due to 
the anti-symmetric arrangement of the IR. Due to the crossing angle in the 
IP, the whole synchrotron radiation fan is tilted by 2\,mrad away from the proton 
beam. Therefore, the high power synchrotron radiation does not penetrate 
the cold low-beta quadrupole magnets of the proton beam. Fig.\,\ref{fig:syn}
shows  the distribution of synchrotron radiation in the IR.

\begin{figure}[h]
\centerline{\includegraphics[width=0.6\textwidth,angle=0.]{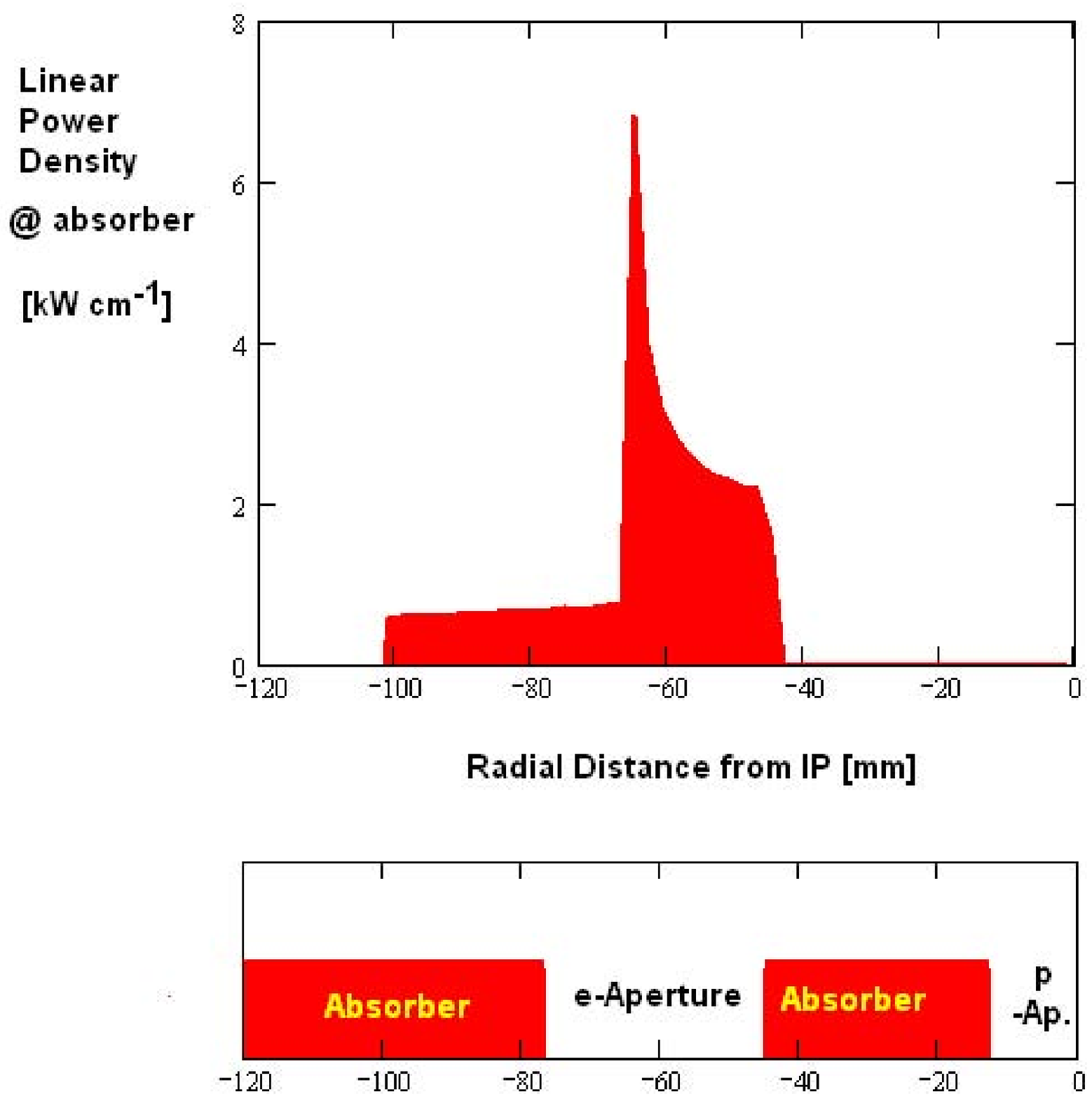}}
\caption{Synchrotron radiation linear power density at 21\,m from the IP,
 at the entrance of the absorber. The red boxes below  indicate the position of the absorber elements.}
\label{fig:pow}
\end{figure}

\begin{figure}[h]
\centerline{\includegraphics[width=0.5\textwidth,angle=0.]{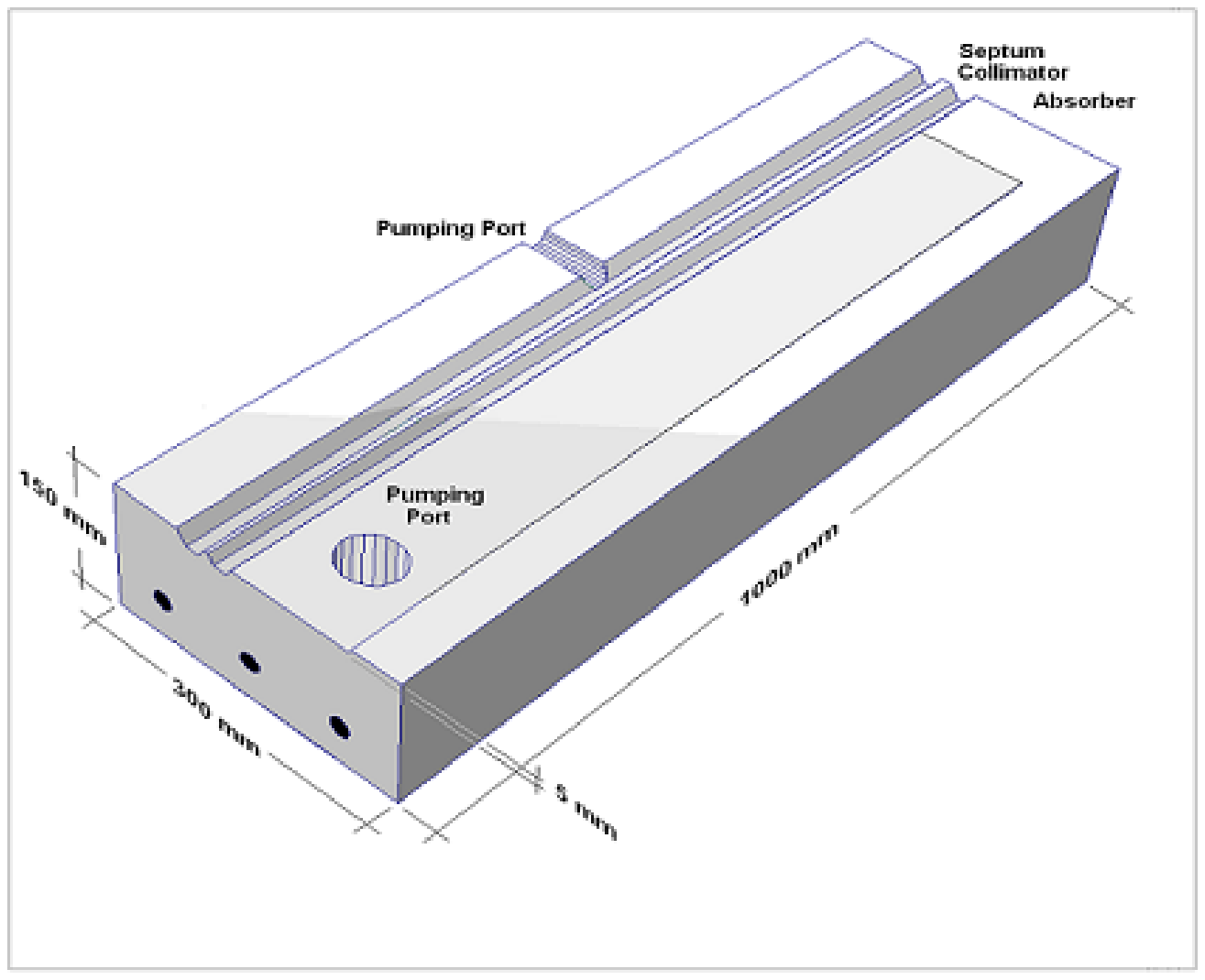}}
\caption{\it Sketch of the HERA-type, conical, 100\,cm long, 30\,cm $\times$ 30\,cm cross section, water-cooled synchrotron radiation absorber  made of copper. The radiation is absorbed over a length of 400\,mm and the full width of 90\,mm of the absorber (only the lower half of the absorber is shown).}
\label{fig:abs}
\end{figure}

The radiation is stopped at the 1m long absorber placed at 21\,m from the IP 
before the first proton low-beta quadrupole. The linear power density 
reaches maximum values of 4.0\,kW/cm (Fig.\,\ref{fig:pow}). The absorbing surface is 
tilted by $(\pi $/2-5mrad) so that the surface power density is 
reduced to approximately 100\,W/mm$^{2}$. A corresponding collimator exists at 
HERA \cite{ferdi19,ferdi21}. It absorbs about 3 to 4\,kW of radiation with a power density of 
40\,W/mm$^{2}$. The absorber presented here (Fig.\,\ref{fig:abs})
is a realistic extrapolation of the 
HERA absorber. A sketch of a possible absorber for LHeC is shown in
Fig.\,\ref{fig:abs}.
 A second absorber is needed between the first and the second low beta 
quadrupoles of the proton beam optics. According to HERA experience \cite{ferdi20} this 
collimation system should be adequate for both elimination
of experimental 
background and protection of accelerator components.
%
\subsection{Magnetic Elements of the Interaction Region}
The superconducting low beta quadrupoles for the electron beam can be built 
using standard superconducting technology. The first two low beta quadrupoles 
for the protons are more challenging, because the lepton beam pipe has to pass 
through the cryostat of these magnets.

\begin{figure}[h]
  \centerline{\includegraphics[width=0.6\textwidth]{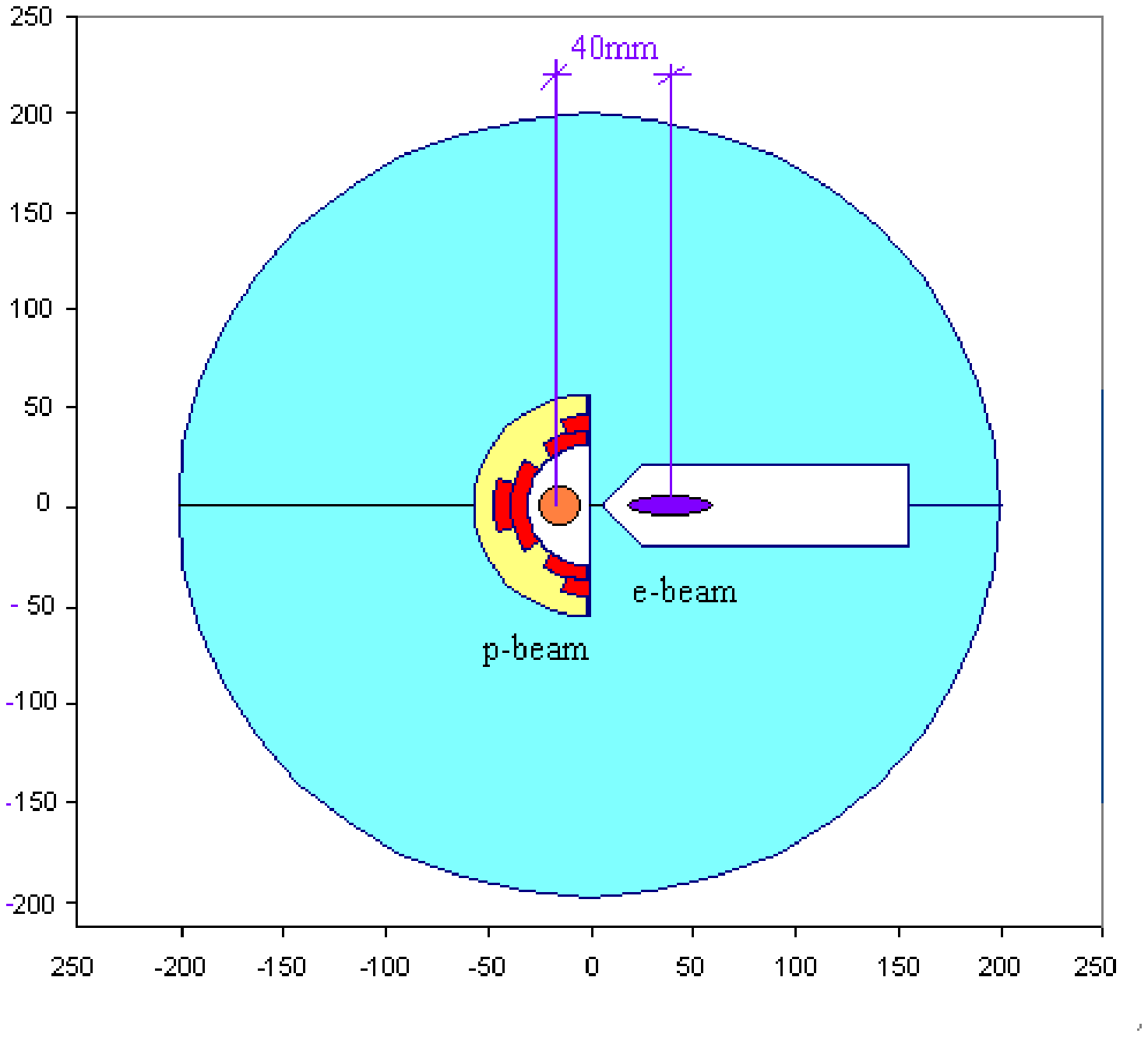}}
  \caption{\it Sketch of the cross section of the 30\,mm aperture septum 
  quadrupole used for the vertical low beta lens of the proton beam.}
 \label{fig:septquad}
\end{figure}

\begin{figure}[h]
  \centerline{\includegraphics[width=0.4\textwidth,angle=-90.]{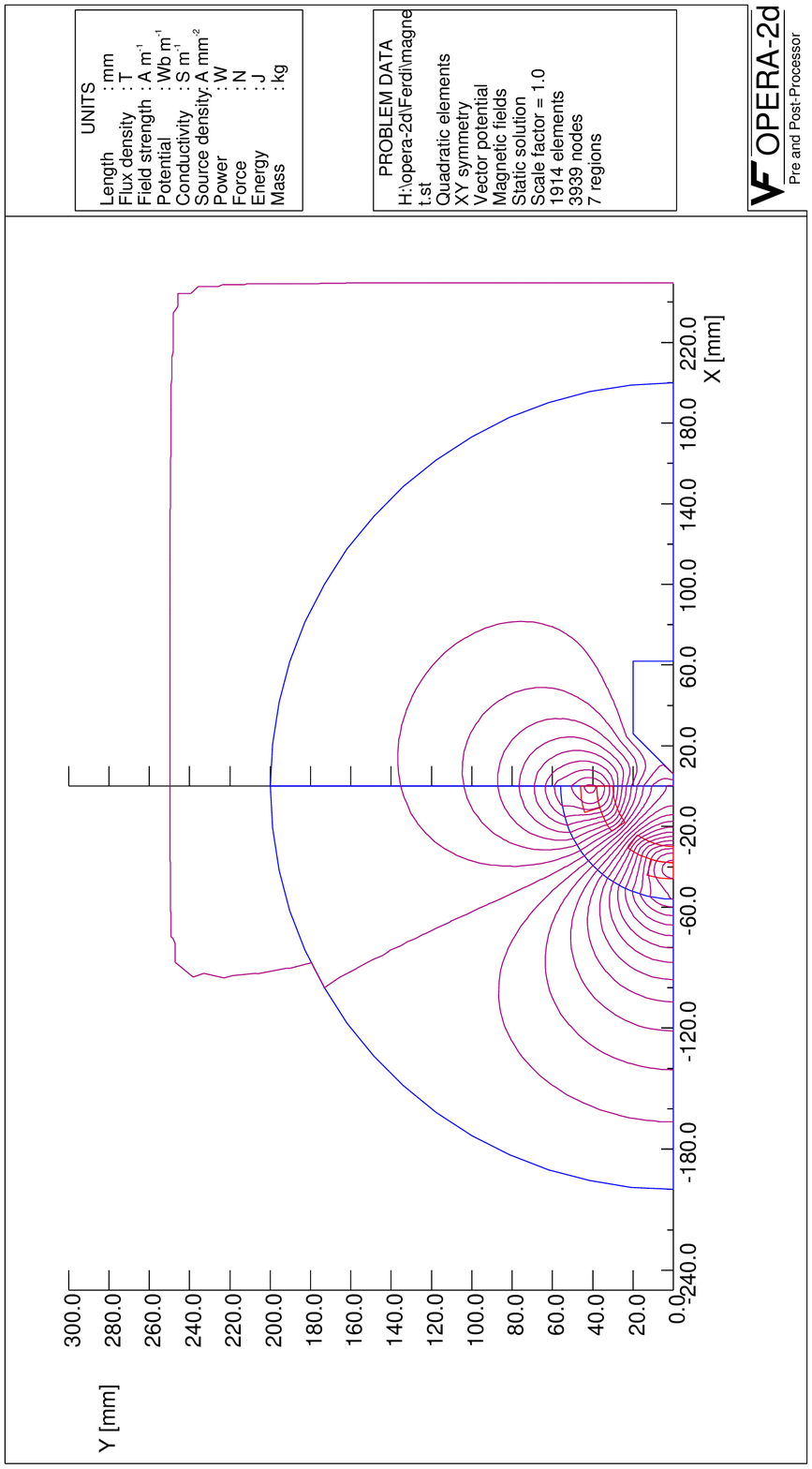}}
  \caption{\it Field lines of the septum quadrupoles as calculated with OPERA-2D [22].}
 \label{fig:field}
\end{figure}

\begin{figure}[h]
 \centerline{\includegraphics[width=0.4\textwidth,angle=-90.]{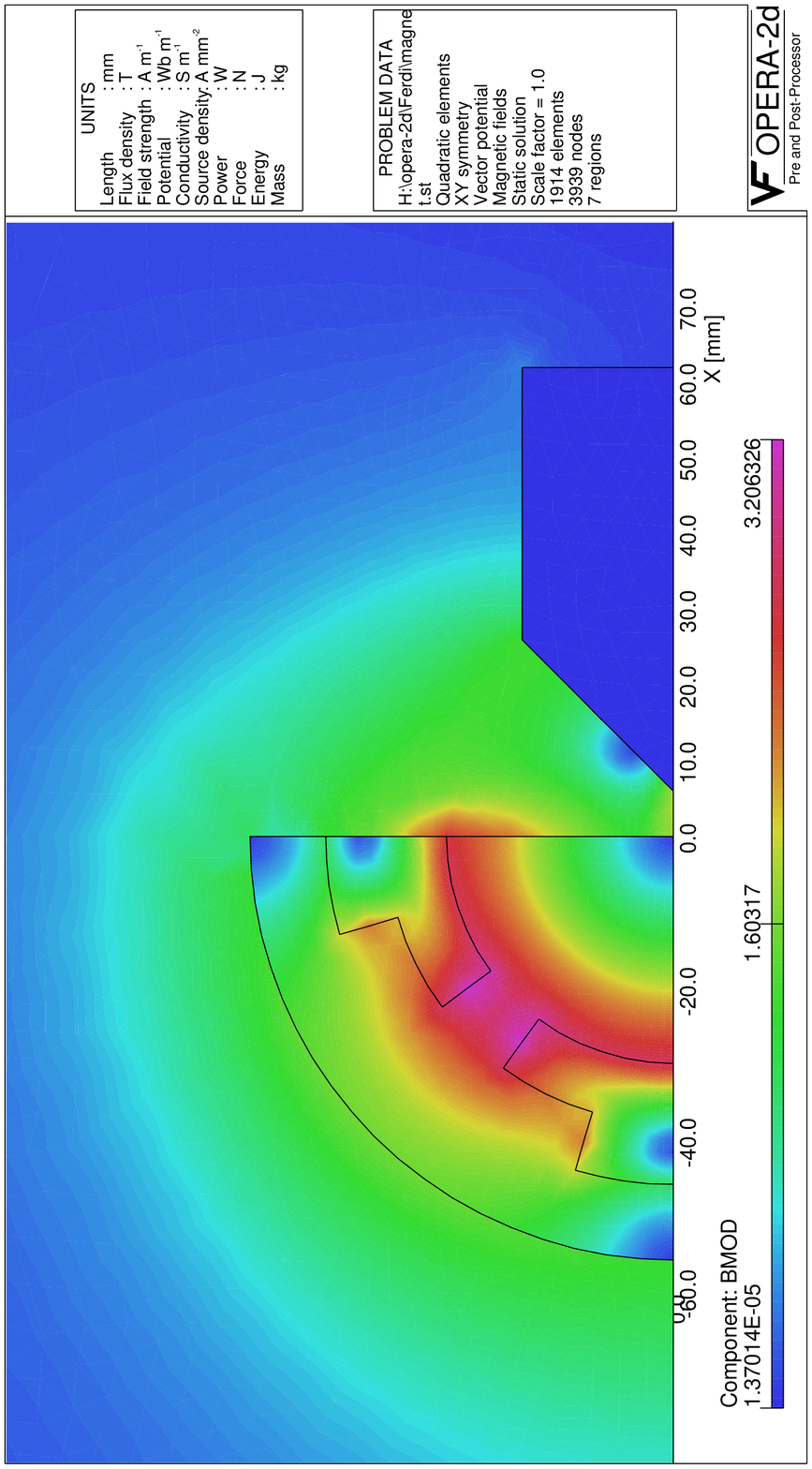}}
  \caption{\it Field map of the septum quadrupole obtained with OPERA-2D [15]. The magnetic induction does not exceed values of 1.6 T and the mirror plate functions well for this magnet. The magnetic field inside the septum gap is only a few Gauss.}
 \label{fig:fmap}
\end{figure}

The first lens is a septum quadrupole laid out as a superconducting half 
quadrupole. The radius of the half aperture is 30\,mm which provides a 15\,mm 
aperture for the beam. The left side half of the magnets looks like a 
standard superconducting quadrupole (Fig.\,\ref{fig:septquad}). The other half consists 
of magnetic iron with a gap for the lepton beam. The achieved gradient is 
93\,T/m. The magnetic mirror plate works well up to a magnetic induction of 
2.79\,T near the coil. The magnet has a reasonable field quality and has no 
field in the gap for the lepton beam. This has been confirmed by magnetic 
field calculation using the code Opera-2D \cite{ferdi22}. 
Figures\,\ref{fig:septquad}, \ref{fig:field}  and \ref{fig:fmap} show a
cross section with the conceptual layout and the results of the field 
calculation. A possible issue with this magnet is the mechanical stability of 
the coil.

At the  second proton low beta quadrupole which needs 
the full horizontal aperture, the two beams are separated by 85mm so that 
the lepton beam pipe can pass outside the regular coil through the flux 
return yoke. This magnet needs a wide flux return. The coil is a standard 
superconducting quadrupole coil with an aperture of 30mm.

\subsection{Beam-Beam Effects and Luminosity}

In the following, the luminosity will be recalculated from the design 
parameters. The large bunch length of the proton beam of 7.5 cm will cause 
a luminosity reduction since it enhances the effective beam size of the 
protons experienced by the electron beam. The effect of the short lepton 
bunch length can be neglected and the 
hour glass luminosity 
reduction factor is
\begin{equation}
\label{eq6}
R=\frac{2\cdot \beta _{yp} \beta _{ye} }{\sqrt \pi \sigma _p \sqrt {\beta 
_{yp} ^2+\beta _{ye} ^2} }\cdot \exp \left( {2\cdot \frac{\beta _{yp} \beta 
_{ye} }{\sigma _p \sqrt {\beta _{yp} ^2+\beta _{ye} ^2} }} \right)\cdot K_0 
\left( {\left( {\frac{\beta _{yp} \beta _{ye} }{\sigma _p \sqrt {\beta _{yp} 
^2+\beta _{ye} ^2} }} \right)^2} \right)=0.95
\end{equation}
The crossing angle should not reduce the luminosity if it is properly 
compensated by the crab-tilt of the proton bunches. 
Another effect which 
influences the luminosity is the so-called dynamic beta, the distortions of 
the beta functions in the core of the beam by the beam-beam interaction. 
This distortion decreases the lepton beta functions at the IP in the 
electron-proton case if the tunes are above the integer and below the quarter 
integer resonance 
\begin{equation}
\label{eq7}
\frac{\Delta \beta _{x,y} }{\beta _{x,y} }=- \frac{r_0 \cdot N_p \cdot \beta 
_{x,y} \cdot \cot (2\cdot \pi \cdot Q_{x,y} )}{\gamma _e \cdot \sigma 
_{x,y}^p \cdot (\sigma _x^p +\sigma _y^p )}.
\end{equation}
This does not affect the beam tails and has no influence on the aperture 
requirements for the beam. It does affect, however, the beam matching and the 
chromaticity. Dynamic focusing can therefore not be used to increase the 
luminosity without taking  into account the considerations chromaticity 
limitations as discussed before. The dynamic beta depends strongly on the 
choice of the tunes. For very strong beam beam forces, the beta functions
might be increased by higher order effects. For HERAe-like tunes Q$_{xe}$= 0.10 and Q$_{ye}$ = 0.11 
one obtains a  strong reduction of the horizontal electron beta functions from \textit{$\beta $}$_{xe 
}$= 12.7\,cm to \textit{$\beta $}$_{xe }$= 6.9\,cm, and a small increase for the vertical beta function from $\beta_{ye} $ =\,7cm 
to $\beta _{ye}$ = 7.3\,cm. Note that for positron rather than electron running, 
the tunes should be taken below the integer resonance.

The beam-beam tune shift values which result from the parameters are
\begin{eqnarray}
\label{eq8} 
 \Delta \nu _x^p =\frac{r_p N_e \beta _{xp} }{2\pi \gamma _p \sigma _{xe} 
(\sigma _{xe} +\sigma _{ye} )}=0.83\cdot 10^{-3} \\ 
 \Delta \nu _y^p =\frac{r_p N_e \beta _{yp} }{2\pi \gamma _p \sigma _{ye} 
(\sigma _{xe} +\sigma _{ye} )}=0.32\cdot 10^{-3} \\ 
 \Delta \nu _x^e =\frac{r_0 N_p \beta _{ye} }{2\pi \gamma _e \sigma _{xp} 
(\sigma _{xp} +\sigma _{yp} )}=48\cdot 10^{-3} \\ 
 \Delta \nu _y^e =\frac{r_0 N_p \beta _{ye} }{2\pi \gamma _e \sigma _{yp} 
(\sigma _{xp} +\sigma _{yp} )}=51\cdot 10^{-3}.
\end{eqnarray}
The vertical tune shift value is within the range of HERA operational parameters. However, large tune shifts in the order of 0.05 in both planes are beyond HERA experience.  Given the larger beam energy of 70GeV and the shorter damping time, these tune shifts may be considered tolerable. More studies are needed to verify whether this choice is acceptable or whether the design should be based on a larger crossing angle, larger electron emittance and smaller beta.
 
The long-range beam-beam tune shift parameters $\Delta \nu 
_{x}^{par}$, $\Delta \nu _{y}^{par}$ of the lepton beam at the 
parasitic crossing points are given by \cite{ferdi23}
\begin{eqnarray}
\label{eq9}
 \Delta \nu _x^{par} =\frac{N_p r_0 \beta _x^{par} }{2\pi \gamma _e \Delta 
x^2} \\ 
 \Delta \nu _y^{par} =\frac{-N_p r_0 \beta _y^{par} }{2\pi \gamma _e \Delta 
x^2} 
\end{eqnarray}
where$\beta _{x,y}^{par}$ are the beta functions, $\Delta x$ is the 
horizontal beam separation at the parasitic crossings and the vertical beam 
separation is zero. The corresponding values for the first parasitic 
crossing is quite large for a bunch spacing of 25ns (see Tab.\,\ref{tabtune} which 
summarizes the evaluation of these formulae). The operational experience 
with long-range beam-beam effects in LEP \cite{ferdi24} (at four instead of at one 
interaction point) indicates, that one might have to expect problems mainly due to beam-beam orbit effects. For larger  bunch 
spacing of 50\,ns and 75\,ns the long range beam-beam parameters are sufficiently small. 
A somewhat larger crossing angle might be necessary relax 
the long-range beam-beam effects sufficiently.  Further study is needed 
to confirm whether a crossing angle of 2\,mr is sufficient for 25\,ns
bunch spacing. 
\begin{table}[h]
\caption{Parasitic beam-beam tune shifts of the lepton beam}
\begin{center}
\begin{tabular}{|p{50pt}|p{51pt}|p{63pt}|p{61pt}|p{67pt}|p{66pt}|}
\hline
Bunch spacing&  
Crossing angle& 
Separation& 
Separation&
Horizont. parasitic beam-beam tune shift& 
Vertical parasitic beam-beam tune shift \\
\hline
{\small [ns]}&  
{\small [mrad]}& 
{\small [mm]}&  
{\small [$\sigma_{ex}$]} &
{\small }& 
{\small } \\
\hline
25& 
2.0& 
7.7& 
8.3&
0.0011& 
-0.0015 \\
\hline
50& 
2.0& 
15& 
23 & 
0.0001&
-0.002\\
\hline
75& 
2.0& 
27&
50& 
0.00003& 
-0.0004\\
\hline
\end{tabular}
\label{tabtune}
\end{center}
\end{table}

Taking these effects into account the luminosity for $I_{e }$= 71mA, $N_{p }$= 
1.68 10$^{11}$, \textit{$\epsilon $}$_{p }$= 0.5 nm,
\textit{$\epsilon $}$_{xe}$= 7 nm, \textit{$\epsilon $}$_{ye}$ = 5 
nm and $R$ = 0.94 amounts to
\begin{equation}
\label{eq10}
L=\frac{I_e \cdot N_p \cdot R}{2\cdot \pi \cdot e\cdot \sqrt 
{\epsilon _p \beta _{xp} +\epsilon _{xe} \beta _{xe} } \cdot \sqrt 
{\epsilon _p \beta _{yp} +\epsilon _{ye} \beta _{ye} } }=1.11\cdot 
10^{33}cm^{-2}s^{-1}\quad ,
\end{equation}
according to the design goal.


%
\section{Summary}
The physics and the design of a Large Hadron Electron Collider (LHeC) are sketched. 
It is illustrated that a unique and important
programme of physics is possible with a 70~GeV electron/positron beam in collision with
 the 7 TeV LHC proton (and ion) beam at a luminosity of $10^{33}$cm$^{-2}$s$^{-1}$.

Experiments at such a collider probe electron-quark and positron-quark physics in an 
unparalleled manner, thereby enabling a substantial extension of the discovery potential 
at the LHC and making possible measurements of a precision characteristic
of  lepton-hadron measurements. Highlights include

\begin{itemize}
\item{
      Observation and precision measurement of new physics in the lepton-quark spectrum at the TeV scale,
      which could reveal unexpected and new leptoquark phenomena. The LHeC 
      will provide precision measurements which are important to the interpretation and quantification
      of this new physics.}
      \item{
     Discovery and precision measurement of new physics in proton structure at very low Bjorken-$x$, which will be   crucial to superhigh energy neutrino physics, to forward physics at the LHC
     and basically to
      the development of our understanding of  QCD in the high parton density, low coupling
limit, and thereby the phase equilibria of chromodynamics in a variety of
      hadronic systems at the TeV energy scale.}
      \item{
    A new level of precision measurements and precision tests of the validity of QCD at a new distance scale,
    corresponding to substructure dimensions of $10^{-19}$\,m, which
      promise to have a direct bearing on the overall consistency of the Standard Model and its underlying physics as one moves towards
      the unification scale.}
    \item{
     Measurements which will make possible the determination of parton distribution functions of nucleons and nuclei over a
      hitherto inaccessible kinematic range in probe scale ($Q^2$) and longitudinal momentum fraction (Bjorken-$x$), and which are essential
      if the sensitivity at the LHC to new and rare physics in both $pp$ and $ep$ physics is to be optimised.}
 \item{
    The immense energy densities achieved in an $AA$ interaction at the LHC. 
To fully explore the nature of the
interactions will require comparable data in $pA$, $pp$, and $eA$ collisions. 
LHeC and the LHC will thus constitute an
experimental tool unparalleled in the history of hadron physics in that nowhere else
has there ever been such a range of possible measurements at such an energy scale.}
 \end{itemize}

A conceptual design of a high luminosity Large Hadron Electron
Collider, the LHeC, is presented. The approach takes advantage of developments in
technology which are now well advanced, and which are necessary for future electron/positron
linear accelerators, to achieve
an electron/positron storage ring 
of for example 70\,GeV energy in the LHC tunnel. It is shown how, with the careful design of 
the RF structure and
the interaction region, it is possible to achieve a luminosity of $10^{33}$\,cm$^{-2}$s$^{-1}$
 in collisions with one of the LHC hadron beams.
 The solution is based on 25\,ns 
bunch spacing with a small crossing angle of 2.0\,mrad  which requires 
crab-crossing for the proton beam.
 The concept is aimed at the simultaneous operation of LHC and LHeC
 and first considerations of how this can be achieved are stated. Wherever possible,
realistic constraints are included based on past operation of electron storage rings 
and on the operation of the HERA
electron and proton storage rings at DESY Hamburg.

Further work is needed to address a number of issues which have yet to be resolved 
concerning the feasibility of such a concept.  Nevertheless, to date it appears not unreasonable
to continue to contemplate a major and important $ep$ physics programme at the TeV scale 
as part of the  physics programme of the LHC. 


\newpage
\noindent
{\bf Acknowledgement} 

\noindent The authors wish to thank G.\,Altarelli, W.\,Bartel, J.\,Bl\"umlein, 
S.\,Brodsky, A.\,Bruell, 
A.\,Caldwell, A.\,DeRoeck, 
M.\,Diehl, J.\,Ellis, 
R.\,Engel, J.\,Engelen, J.\,Jowett, H.\,Jung,  J.\,Feltesse, S.\,Forte,  
A.\,Glazov,  T.\,Greenshaw, 
R.\,Heuer, E.\,Kabuss, U.\,Katz, H.\,Kowalski, T.\,La\v{s}tovi\v{c}ka, A.\,Levy, M.\,Mangano, 
M.\,Marx, S.\,Myers, C.\,Niebuhr, D.\,Pitzl, B.\,Reisert,  R.\,Rueckl,
D.\,Schlatter, S.\,Schlenstedt, M.\,Seidel, R.\,Schmitt,  U.\,Stoesslein, M.\,Strikman,
R.E.\,Taylor, W.\,K.\,Tung, R.\,Valbuena,  and A. Zarnecki for interesting discussions and 
providing valuable information.

The authors have worked for  two decades at the world's first high 
energy collider of different particle species,
HERA at DESY. It would be inappropriate not to record our thanks also to many
colleagues, too numerous to name here,   with 
whom we have worked throughout this time at HERA and who have developed
the physics and the techniques for experimentation at such a collider
on which much of what is sketched in this paper relies.
%

\newpage
\end{document}